\PassOptionsToPackage{square,comma,numbers,sort&compress,super}{natbib}
\documentclass{article}

\usepackage[preprint]{neurips_2024}

\usepackage[utf8]{inputenc}
\usepackage[T1]{fontenc}
\usepackage{amsfonts}
\usepackage{nicefrac}

\title{LMO-DP: Optimizing the Randomization Mechanism for Differentially Private Fine-Tuning (Large) Language Models}

\author{
  Qin Yang \\
  University of Connecticut\\
  \And
  Meisam Mohammady \\
  Iowa State University \\
  \AND
  Han Wang \\
  University of Kansas \\
  \And
  Ali Payani\\
  Cisco Research
  \And
  Ashish Kundu \\
  Cisco Research
  \AND
  Kai Shu\\
  Illinois Institute of Technology
  \And
  Yan Yan \\
    Illinois Institute of Technology
  \And
  Yuan Hong \\
  University of Connecticut
}

\usepackage{microtype}
\usepackage{graphicx}
\usepackage{subfigure}
\usepackage{booktabs}
\usepackage[noend]{algorithmic}
\usepackage{algorithm}
\usepackage{amsmath}
\usepackage{amssymb}
\usepackage{mathtools}
\usepackage{amsthm}
\usepackage{multirow}
\usepackage{wrapfig}
\usepackage{lipsum}
\usepackage{tcolorbox}
\usepackage{subfigure}
\usepackage{colortbl}
\usepackage{xcolor}
\usepackage{url}
\usepackage[thinc]{esdiff}
\usepackage{graphicx}
\usepackage{dsfont}
\usepackage{xcolor}
\usepackage{soul}
\usepackage[normalem]{ulem}
\usepackage{hyperref}
\usepackage[capitalize,noabbrev]{cleveref}

\newcommand{\RN}[1]{%
\textup{\uppercase\expandafter{\romannumeral#1}}%
}

\theoremstyle{plain}
\newtheorem{theorem}{Theorem}[section]
\newtheorem{lemma}[theorem]{Lemma}

\theoremstyle{definition}
\newtheorem{definition}[theorem]{Definition}

\theoremstyle{remark}

\begin{document}

\maketitle

\begin{abstract}
Differentially Private Stochastic Gradient Descent (DP-SGD) and its variants have been proposed to ensure rigorous privacy for fine-tuning large-scale pre-trained language models. However, they rely heavily on the Gaussian mechanism, which may overly perturb the gradients and degrade the accuracy, especially in stronger privacy regimes (e.g., the privacy budget $\epsilon < 3$).\footnote{Most state-of-the-art (SOTA) methods have demonstrated high accuracy in case of relatively weaker DP guarantees, e.g., $\epsilon\geq 3$, but not small $\epsilon$.} 
To address such limitations, we propose a novel Language Model-based Optimal Differential Privacy (LMO-DP) mechanism, which takes the first step to enable the tight composition of accurately fine-tuning (large) language models with a sub-optimal DP mechanism, even in strong privacy regimes (e.g., $0.1\leq \epsilon<3$). Furthermore, we propose a novel offline optimal noise search method to efficiently derive the sub-optimal DP that significantly reduces the noise magnitude. For instance, fine-tuning RoBERTa-large (with 300M parameters) on the SST-2 dataset can achieve an accuracy of 92.20\% (given $\epsilon=0.3$, $\delta=10^{-10}$) by drastically outperforming the Gaussian mechanism (e.g., $\sim 50\%$ for small $\epsilon$ and $\delta$). We also draw similar findings on the text generation tasks on GPT-2. Finally, to our best knowledge, LMO-DP is also the first solution to accurately fine-tune Llama-2 with strong differential privacy guarantees. The code will be released soon and available upon request.
\end{abstract}

\section{Introduction}

Recently large language models (LLMs) have achieved breakthrough success by effectively processing and encoding huge volumes of text data from extremely large-scale training datasets. For instance, BERT  \cite{liu2019roberta} and GPT families \cite{yu2021differentially} have demonstrated state-of-the-art (SOTA) accuracy and enhanced performance in most of the learning tasks. In addition, such (open-sourced) language models are pre-trained on extremely large and generic datasets, and then fine-tuned to accurately support a wide variety of downstream tasks using a relatively smaller dataset in the task domain, e.g., sentence classification \cite{liu2019roberta}, text generation \cite{novikova2017e2e}, and code generation \cite{wang2018glue}. 

However, deep learning models have been proven to be vulnerable to privacy threats during training \cite{abadi2016deep, shokri2017membership, hayes2017logan}. Similar risks are also present in training or fine-tuning (large) language models, which could potentially lead to the leakage of sensitive information. A notable distinction in the case of training language models is that, since pre-trained datasets and models have been published, they have already rendered privacy leakage from the open-sourced datasets and models. Thus, it is desirable to privately fine-tune language models by \emph{protecting the sensitive information in the new dataset used for fine-tuning} \cite{li2021large, yu2021differentially, bu2022differentially, he2022exploring, bu2023differentially}.
 
To mitigate privacy risks in deep learning training and fine-tuning, differential privacy (DP) \cite{dwork2006differential} has been widely recognized as the de facto rigorous privacy model where adding or removing any data sample, or user in the (training) data would not cause significant leakage. In particular, the well-known Differentially Private Stochastic Gradient Descent (DP-SGD) \cite{abadi2016deep} method tightly balances the privacy and utility within the training, leveraging the privacy parameters $\epsilon$ and $\delta$ in the Gaussian mechanism. It achieves this by constraining the influence of individual examples through gradient clipping and adding Gaussian noise to the gradients within each batch, providing a DP guarantee for model training \cite{abadi2016deep}. To our best knowledge, the SOTA methods for privately fine-tuning language models (which are also DP-SGD variants) \cite{li2021large,he2022exploring,panda2023differentially,yu2021differentially,bu2022differentially,bu2023differentially} mainly focus on optimizing the gradient clipping mechanism to enhance utility and/or improve system performance (e.g., reducing memory) while maintaining privacy. 

Although Gaussian-based DP mechanisms can be tightly accounted with the Moments Accountant in DP-SGD, the magnitude of the noise itself is far from optimal, especially for small privacy budget (e.g., $0.1\leq \epsilon<3$). Then, it may overly perturb the gradients and degrade the accuracy. Therefore, the performance of DP-SGD and other variants (e.g., gradient clipping or memory-reducing mechanisms) can be significantly improved by \emph{designing an optimal or sub-optimal noise for the same DP guarantee}. Such novel and orthogonal noise-reduction method would make fine-tuning language models with strong DP guarantees practical, e.g., boosting the accuracy of fine-tuning RoBERTa-large (with 300M parameters) on the SST-2 dataset from $\sim50\%$ to $90\%+$ when $\epsilon=0.3$ and $\delta=10^{-10}$.

\subsection{Contributions}

To boost the tradeoff between privacy and accuracy for language models, we propose a novel Language Model-based Optimal Differential Privacy (LMO-DP) mechanism, which works as a plug-and-play model-agnostic module to strengthen private training performance. To this end, we design a novel LMO noise and adapt it to the LM training/fine-tuning process by replacing the Gaussian noise in the SOTA methods, e.g., DP-SGD and/or the variants (e.g., Ghost Clipping \cite{li2021large}). Our LMO noise is generated from a two-fold mixture distribution derived from an optimal differential privacy framework (R$^2$DP~\cite{mohammady2020r2dp}) in which the first fold is a Laplace distribution and the second fold is a combination of possible linear weighted probability density functions (PDFs). 
Specifically, we instantiate the second-fold mixture distribution as the scale parameter of Laplace distribution and apply R\'enyi Accountant \cite{wang2019subsampled} during the training/fine-tuning stage to realize a tight composition for the two-fold randomization mechanism.\footnote{Some recent tighter accountants \cite{koskela2022individual, zhu2022optimal} can also be integrated with the LMO-DP.
} 
Then, our major contributions are further discussed below. 

{\textbf{(1) First non-Gaussian DP mechanism for LM fine-tuning.} 
To our best knowledge, we propose the first non-Gaussian mechanism for privately fine-tuning language models. Recall that our approach ensures tight composition using the Rényi Accountant, which extends the Gaussian mechanism to universal DP mechanisms through the Rényi Differential Privacy \cite{mironov2017renyi}.

We establish a meticulously defined search space of PDFs while upholding a universal DP privacy guarantee (a subspace within the PDF space, incorporating randomization into the scale parameter of Laplace noise) \cite{mohammady2020r2dp}. 
Different from \cite{mohammady2020r2dp}, we instantiate this randomization as a linear combination of \emph{Gamma}, \emph{Exponential}, and \emph{Uniform} distributions due to their computational convenience and ability to approximate the entirety of the search space of PDFs. This also enables us to formulate privacy and utility within a unified framework for searching the noise parameters.

{\textbf{(2) First DP mechanism supports strong privacy guarantee for fine-tuning language models and LLMs.} To our best knowledge, we take the first step to fine-tune (large) language models with strong $(\epsilon, \delta)$-DP guarantees such as $\epsilon=0.3$ and $\delta=10^{-10}$.  
Meanwhile, we found that LMO-DP achieves superior convergence rates (empirically) in a diverse range of LM tasks (e.g., sentiment classification, and table-to-text generation) on models with parameters ranging from 300 million (e.g. RoBERTa, BERT) to 7 billion (e.g., Llama2-chat-7b)\footnote{Searching the optimal noise will be executed as a pre-processing procedure before the fine-tuning under the DP-SGD framework. Then, LMO-DP would only incur minor extra runtime during private fine-tuning.} 

\textbf{(3) Accuracy-boosting module to SOTA methods.} Recall that some SOTA methods focus on optimizing the gradient clipping to improve the system performance, e.g., low memory by Ghost Clipping \cite{li2021large}. Since LMO-DP mainly optimizes the randomization for DP mechanisms offline (pre-processing), it is orthogonal to those DP fine-tuning methods for LMs, e.g., \cite{bu2023differentially}. Thus, LMO-DP can inherit all the benefits of existing methods via integration with them, e.g., memory reduction via Ghost Clipping \cite{li2021large}. 

\section{Related Work}

\textbf{DP-SGD}~\cite{abadi2016deep} was initially devised for private neural network training. However, a significant challenge with traditional DP-SGD lies in compromised performance and the substantial time and memory overhead of private training. Researchers are actively addressing it by reducing online training costs and improving performances for LMs through more efficient parameter tuning. For instance, \cite{yu2021differentially} enhanced DP-SGD by exploring parameter-efficient tuning methods that focus on training only a fraction of the model parameters, resulting in improved utility, privacy, and reduced overheads. Another advancement comes from \cite{bu2022differentially}, introducing a model-agnostic DP bias-term fine-tuning (DP-BiTFiT) framework. It prioritizes optimizing the bias rather than model weights, and achieves efficiency by activating only the backward hook in PyTorch, thus saving time and space. 
Moreover, recent works have proposed clipping methods to reduce computational time and memory requirements for large language models. \cite{li2021large, bu2022scalable} introduced the ghost clipping method, significantly reducing memory usage during training and enhancing performance in text classification and generation tasks. \cite{he2022exploring} proposed adaptive group-wise clipping, encompassing per-layer, and per-device clipping techniques, suitable for deployment on multiple accelerators. 
\cite{bu2022scalable} proposed a mixed ghost clipping method on convolutional layers, that significantly eases the private training in terms of both time and space while maintaining the accuracy. \cite{bu2023differentially} introduces a novel book-keeping (BK) technique that enhances the computational efficiency by eliminating the need for a second back-propagation step in Ghost Clipping \cite{bu2022differentially}, while preserving the same accuracy.

\section{Preliminaries}
\label{sec:prem}

\subsection{DP-SGD}
\label{sec:DP-DPSGD}
We first provide the formal definition of differential privacy (DP)~\cite{dwork2006differential,dwork2014algorithmic}.
\begin{definition}[$(\epsilon, \delta)$-Differential Privacy]
Let $\mathcal{D}$ be the domain of datasets, and let $D, D'$ be two adjacent datasets in $\mathcal{D}$ such that $D'$ can be obtained from $D$ by adding or removing one sample. A randomization mechanism $\mathcal{M}: \mathcal{D} \times \Omega \to \mathbb{R}$ satisfies $(\epsilon, \delta)$-differential privacy if, for any $D$ and $D'$, and for all $O \subseteq \text{range}(\mathcal{M})$, $
\Pr[\mathcal{M}(D) \in O] \leq e^\epsilon \Pr[\mathcal{M}(D') \in O] + \delta$, where $\Omega$ is a sample space by randomization $\mathcal{M}$ to generate the output. For instance, Laplace mechanism is a commonly used technique for DP, relying on Laplace distributions.
\label{def:dp}
\end{definition}

DP-SGD \cite{abadi2016deep} generally refers to the DP during deep learning training with the Gaussian mechanism. This has been widely used to ensure differential privacy during deep learning training. Mostly it applies the Gaussian mechanism for DP, which first clips the gradients $\mathbf{g}_t(\textbf{x}_i)$ using a threshold $C$ for $\ell_2$-sensitivity $
\overline{\mathbf{g}}_t(\textbf{x}_i) = \mathbf{g}_t(\textbf{x}_i)/\max\left(1, \frac{\|\mathbf{g}_t(\textbf{x}_i)\|_2}{C}\right)$ and then injects the Gaussian noise $\mathcal{N}(0, C^2\sigma^2\mathbf{I})$ to these clipped gradients within each batch $
\tilde{\mathbf{g}}_t = \frac{1}{L}\left(\sum_i \overline{\mathbf{g}}_t(\textbf{x}_i) + \mathcal{N}(0, C^2\sigma^2\mathbf{I})\right)$ where $\textbf{x}_i$ is the $i$-th training sample.

The biggest challenge DP-SGD successfully tackles is bounding $\epsilon$ and $\delta$ during thousands of rounds of training, each of which accounts for spending a portion of the privacy budget. While existing literature relied on the strong composition theorem~\cite{dwork2010boosting}, which, although effective, can be imprecise and overlook the noise distribution specifics, DP-SGD introduces the \emph{moments accountant} which utilizes probabilistic insights to provably tighten the budget. In particular, the overall $\mathcal{O}(\epsilon.T,\delta.T)$ budget is reduced to $\mathcal{O}(\epsilon.\sqrt{T},\delta)$ in DP-SGD \cite{renyi}. 

\subsection{R\'enyi Accountant}
The R\'enyi Accountant \cite{wang2019subsampled} is one of such privacy accountant which accounts for R\'enyi Differential Privacy (R\'enyi-DP) \cite{renyi} (per step) and offers a tight composition of all the steps during the private training. R\'enyi-DP is defined as follows:

\begin{definition}
[$(\alpha, \epsilon_\alpha)$-Rényi Differential Privacy] A mechanism $\mathcal{M}$ is said to be $(\alpha, \epsilon_\alpha)$-Rényi DP with order $\alpha \in (1, \infty)$ if, for all adjacent datasets $D$ and $D'$, the $\alpha^{th}$ order Rényi divergence $\mathrm{D}_{\alpha}(\mathcal{M}(D) \Vert \mathcal{M}(D')) \leq \epsilon_\alpha$.

The $\alpha^{th}$ order R\'enyi divergence is defined as follows:
\[
D_\alpha(P \| Q) = \frac{1}{\alpha - 1} \log \sum_{r\in \mathbb{R}} \left( P(r)^\alpha Q(r)^{1-\alpha} \right)
\]
where $r\in \mathbb{R}$ and $P$ and $Q$ are some discrete probability distributions.

\end{definition}
\begin{lemma} [R\'enyi Accountant]
\label{lemrdp}
Given a sequence of $T$ random mechanisms $\mathcal{M}_1, \ldots, \mathcal{M}_j, \ldots, \mathcal{M}_T$ with associated $\alpha$-RDP budgets $\epsilon_{\alpha,1}, \ldots, \epsilon_{\alpha,j}, \ldots, \epsilon_{\alpha, T}$, the cumulative Rényi-DP privacy guarantee during training/fine-tuning process is  $\epsilon_{\alpha} =\sum_{j=1}^T \epsilon_{\alpha, j}$, where
\begin{align*}
\vspace{-0.1in}
\epsilon_{\alpha, j} &= \frac{1}{\alpha - 1} \cdot \log \left(\mathbb{E}_{\mathcal{M}_j(D)} \left[ \left(\frac{Pr[\mathcal{M}_j(D) = O_j]}{Pr[\mathcal{M}_j(D')=O_j]}\right)^{\alpha} \right] \right),\end{align*}
$T$ is the number of steps during the training/fine-tuning stage and 
$O_j$ $\left (O_j \subseteq \text{range}(M_j)\right )$ represents any potential output of $\mathcal{M}_j(D)$.
\end{lemma}

Then, tight privacy bounds for DP-SGD with Rényi DP can be derived, ensuring specified privacy ($\epsilon$) while controlling the privacy failure probability ($\delta$) \cite{wang2019subsampled}. Given a small $\delta$ (e.g., $10^{-5}$, $10^{-10}$), the total privacy loss $\epsilon$ can be computed using the privacy curve in \cite{gopi2021numerical}, given by:
\begin{align}
\vspace{-0.1in}
\epsilon = \mathcal{H}(\epsilon_\alpha, \delta) = \min_{\alpha > 1} \left\{\log\left (\frac{\log{\frac{1}{\delta}} + \epsilon_\alpha}{\alpha}\right)\right\}
\label{conversion}
\end{align}

The range of $\alpha$ can be specified empirically, e.g., 2,\dots,128.\footnote{Expanding the upper bound of $\alpha$ can facilitate the computation of R\'enyi privacy. However, larger values may result in too tiny moments for ineffective processing of floating point numbers.} Notice that many recent works \cite{pmlr-v151-zhu22c, koskela2023individual} provide tighter privacy accounting for DP, which could also be applied to formulate similar constraints. Then, we can replace the R\'enyi-DP with those new accountants with tighter bounds, though different solvers will be desirable.

\subsection{Problem Formulation: General Private Training Problem Definition}
\label{sec:problem}

We denote the fine-tuning dataset as $\mathcal{D} = {(\textbf{x}_i, y_i) | 1 \leq i \leq N}$, where $\textbf{x}_i \in \mathbb{R}^w$ and $y_i \in {1, 2, \dots, V}$. This dataset is used to fine-tune a (large) language model $\Theta: \mathcal{D} \to {1, 2, \dots, V}$ within $T$ steps. 
Let $\tilde{\textbf{g}}_{j}(D)$ denote the noisy-clipped model gradients after $j$ ($1\leq j \leq T$) steps of DP training/fine-tuning, $D$ and $D'$ are adjacent datasets.
Assuming the existence of an optimal noise generation mechanism $\mathcal{M}$ that enhances the tradeoff between privacy and accuracy during training/fine-tuning, we formulate an optimization problem for fine-tuning a (large) language model for a specific NLP task under the constraint of $(\epsilon, \delta)$-Differential Privacy (DP) as follows \cite{wang2020cat} (in Figure (\ref{fig:framework})):

\vspace{-0.2in}

\small

\begin{align}
& \min_{\forall j} \quad -\frac{1}{B_j}\sum_{i=1}^{B_j} \log \left (p^c_{i, j}\right )
\label{problem}
\end{align}

\vspace{-0.2in}
\begin{align}
& \text{s.t.} \quad \sup_{\forall{ D, D'} \in \mathcal{D}} \left \{ \mathcal{H} \left [ \sum^T_{j=1} D_{\infty}(\tilde{\textbf{g}}_j(D) \Vert \tilde{\textbf{g}}_j(D')), \delta\right ] \right \} \leq \epsilon \nonumber
\label{constraint}
\end{align}
\normalsize

Where $B_j$ is the batch size of Poisson-sampled indices $\mathcal{B}_j$, $p_{i,j}^c$ is the predicted probability of the correct class for sample $\textbf{x}_i$ in $\mathcal{B}_j$, $\tilde{\textbf{g}}$ is the noisy gradient, $D_{\infty}$ is the $\infty$-R\'enyi divergence and $\mathcal{H}$ is the RDP to $(\epsilon, \delta)$-DP conversion.

Given the noise searching space $\mathcal{S}$ for gradient, we denote the optimal noise parameters of the mechanism $\mathcal{M}$ for the above problem~\eqref{problem} as $\mathcal{S}_{\text{opt}}$. Solving this optimization problem is challenging since the noisy gradients $\tilde{\textbf{g}}$ are optimized or computed by the  randomly sampled input data $\mathcal{B}_j$, the model parameters at the previous step $\Theta_{j-1}$, and the randomization mechanism $\mathcal{M}$. Meanwhile, the optimal randomization mechanism $\mathcal{M}$ searches $\mathcal{S}_{\text{opt}}$ from the search space $\mathcal{S}$ to improve the NLP task-dependent utility metrics such as accuracy (e.g., minimizing the cross-entropy loss).

\begin{wrapfigure}{r}{0.6\textwidth}
  \centering
  \includegraphics[width=0.48\textwidth]{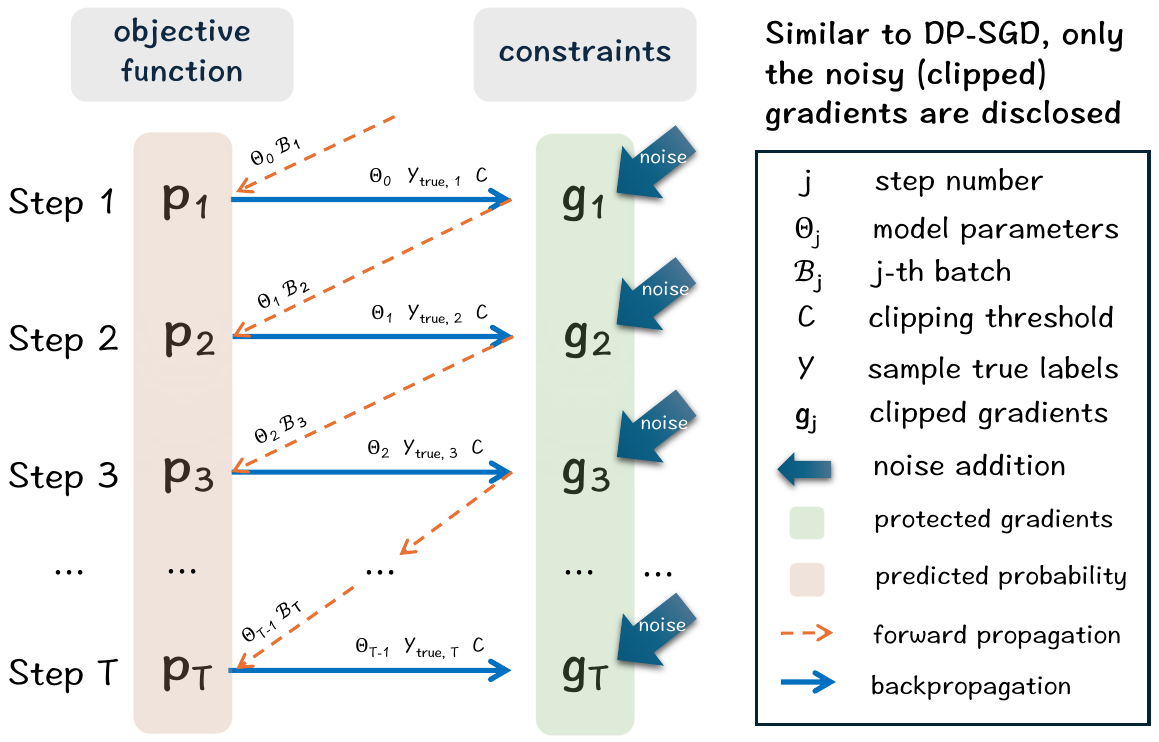}
  \caption{Private fine-tuning/training with optimal noise (similar to DP-SGD). $p_j$ is drawn from the batch. } 
  \label{fig:framework}
  \vspace{-0.8mm}
\end{wrapfigure}

Updating the noise parameters during the training or fine-tuning process seems like an intuitive proposal. However, we have found that simultaneously optimizing both the model parameters and the noise creates a complex interdependence where each optimization process relies on the prerequisite of the other. This may require significant computation during fine-tuning/training or initially introduce a large noise that causes vanishing or exploding gradients \cite{glorot2010understanding}, extending the training or fine-tuning time. Thus, we propose to optimize the noise \emph{offline} (while satisfying ($\epsilon,\delta$)-DP from end to end), and inject the noise to the clipped gradients online, detailed in Section~\ref{sec:lmo-dp}.

\section{Private Fine-Tuning with LMO-DP}
\label{sec:lmo-dp}

In this section, we first define the constraints for the gradients of fine-tuning/training with Rényi Accountant. Then, we define our LMO randomization mechanism and propose the Language Model-based Optimal Differential Privacy (LMO-DP) mechanism to solve the optimization problem. Specifically, we utilize the characteristics of Rényi divergence and an optimal noise generation mechanism to disentangle the complex optimization problem \ref{problem} into two phases: (1) offline noise optimization, and (2) model fine-tuning (parameters optimization) with the optimized noise injected to the clipped gradients. Although this two-phase optimization approach relaxes the global optimum,\footnote{In practice, such effect can be relieved since the gradients $\textbf{g}$ are clipped by $C$ for any model parameters $\Theta$ during fine-tuning. Then, the noise optimization and model parameter optimization can be relatively separated. } it eliminates the dependence on training data $D$, $D'$, and model parameters $\Theta_j$. Meanwhile, it can ensure $(\epsilon, \delta)$-Differential Privacy during the fine-tuning process while promoting the performance. 

\subsection{Phase (1): Rényi Accountant based Offline Noise Optimization}

Specifically, we leverage the fine-tuning/training dataset $\mathcal{D} = \{(\textbf{x}_i, y_i) | 1\leq i\leq N\}$, where $\textbf{x}_i \in \mathbb{R}^w$ and $y_i \in \{1,2,\dots, V\}$, to fine-tune a language model $\Theta: \mathcal{D} \to \{1,2,\dots,V\}$ within $T$ steps. When we apply a randomization mechanism $\mathcal{M}$ to model fine-tuning under $(\epsilon, \delta)$-Differential Privacy, assuming we have the optimal parameters $\mathcal{S}_{\text{opt}}$ searched from the given space $\mathcal{S}$ with a defined objective (e.g., usefulness \cite{mohammady2020r2dp}), the constraint in Problem \ref{problem} can be simplified to the following noise searching problem:

\small
\begin{equation}
    \mathcal{H}\left \{\sum^T_{j=1} \epsilon_{\alpha, j}\left (C, \mathcal{S}_{\text{opt}}\right), \delta \right \} \leq \epsilon
    \label{new}
\end{equation}
\normalsize
where $\epsilon_{\alpha, j}$ is Rényi-DP of $\mathcal{M}$ in the order $\alpha$ and step $j$, and $C$ is the clipping threshold.

Then, in Phase (1), we solve a more manageable problem: search the optimal noise with the privacy constraints in \ref{new} (all the privacy leakage on the gradients can be tightly accounted via the Rényi Accountant).

\subsection{A Versatile Search Domain: \texorpdfstring{$\mathcal{S}_\text{LMO}$}{LMO} Space}

First, we define a versatile search domain by instantiating the randomization optimization in R$^2$DP (two-fold randomization for optimal DP, see more details in Appendix \ref{sec:r2dp}) \cite{mohammady2020r2dp} for language model fine-tuning. Specifically, given the Laplace distribution $\Lambda (C, b)$ as the first-fold randomization, we define the $\mathcal{S}_{\text{LMO}}$ space by randomizing the scale parameter ($b$) according to mixture distributions (as the second-fold randomization), which are defined by multiple positive-supported probability density functions (PDFs).

This dual randomization process creates a space where elements are the moment-generating functions (MGFs) of the second-fold PDFs. This construction offers versatility by enabling linear combinations of various positively supported PDFs in the second-fold PDF, thanks to the MGF composability (Appendix~\ref{sec:mgfdef}). 
Moreover, the $\mathcal{S}_\text{LMO}$ Space provides a universal R\'enyi-DP guarantee (as detailed in the Appendix in \cite{mohammady2020r2dp}):

\vspace{-0.2in}

\begin{align*}
    \forall X=\text{Lap}(x) \in \text{$\mathcal{S}_\text{LMO}$}: e^{\epsilon}_\alpha(x) \propto \mathcal{O}\left(\frac{dM}{dx}\right).
\end{align*}

These classes of PDFs, known to be high-entropy PDFs ~\cite{Walker1965ProbabilityTA}, are characterized by two essential (appealing) properties: (1) comprehensiveness in (approximately) covering all $\mathbb P \in \Omega$ (demonstrated empirically), and (2) a \emph{universal} DP guarantee function for all $\mathbb P \in$ LMO, to support/facilitate solving the optimization problem~\ref{new}. 

Regarding comprehensiveness, we empirically assess LMO-DP noise comprehensiveness compared to a universally simulated space via a novel quantification test defined using several well-known metrics such as KL divergence, $\ell_2$ distance, and earth mover's distance (EMD). The quantification is detailed in Algorithm \ref{alg:quantify-lmo-space} in Appendix \ref{sec:proposed-approach}. Figure~\ref{nearoptim} in Appendix \ref{sec:quantify} shows that the $\mathcal{S}_\text{LMO}$ Space aligns closely with the simulated space. We note that in general, quantifying the comprehensiveness of a subset of probability functions lacks a universally accepted measure.  While this search space may not encompass the entire space, we will show its sufficiency for near-optimal accuracy through numerical results (see Figure~\ref{fig_entropy_variance} in Appendix \ref{sec:quantify}) and experiments in various learning settings (see Section~\ref{sec:exp}). 

\subsection{LMO-DP Mechanism}
From the constructed search space $\mathcal{S}_{\text{LMO}}$, we define the LMO-DP mechanism, which is a sub-optimal noise generation mechanism that is instantiated from the R$^2$DP mechanism \cite{mohammady2020r2dp} (\emph{it optimizes the DP mechanism for generic queries with different mixture of randomization mechanisms}) and adapted to the domain of language model fine-tuning. In our context, LMO-DP refers to the DP during language model fine-tuning/training with the LMO randomization mechanism $\mathcal{M}_{\text{LMO}}$. Considering the balance between computation usage and evaluation precision, we choose Gamma distribution, Exponential distribution, and Uniform distribution to formulate our LMO-DP mechanism as follows:

\begin{definition} [LMO-DP mechanism]
    Denote the Laplace scale parameter $(b)$ in the LMO-DP mechanism is modeled as a random variable $(Y)$, which follows by a linear weighted distribution built by a mixture of the Gamma distribution ($Y_1 \sim {\displaystyle \Gamma \left (k,\theta \right )}$), the Exponential distribution ($Y_2 \sim {\displaystyle Exp \left (\lambda \right )}$), and the Uniform distribution ($Y_3 \sim {\displaystyle U} \left (a, b \right )$); $Y=\sum_{k=1}^{3} {a_k\cdot Y_k}$.
\end{definition}

The LMO noise parameters $\mathcal{S}_{\text{opt}}$ are searched from the $\mathcal{S}_{\text{LMO}}$ space. 
Given the LMO noise parameters $\mathcal{S}_{\text{LMO}}$, for order $\alpha > 1$, the moment generating functions (MGFs) $M_{\text{LMO}}$ are defined as follows (detailed in Appendix \ref{sec:mgfdef}): $M_{\text{LMO}}(t) = M_{\Gamma} \left (a_1\cdot t\right ) + M_{Exp} \left (a_2 \cdot t\right ) + M_{U} \left (a_3 \cdot t\right )$ where $a_1 (\alpha-1) < 1/\theta,~
 a_2 (\alpha-1) < \lambda,~
 b>a,~
 k>0,~
 \theta>0$, 
 $\lambda>0$ for each $\alpha$; $a_1$, $a_2$, and $a_3$ are the weights for the Gamma, Exponential, and Uniform distributions, respectively; 
 $M_{\Gamma} (t)$, $M_{Exp} (t)$, $M_{U} (t)$ are the MGFs of Gamma distribution, Exponential distribution, and Uniform distribution. Specifically, 
$M_{\Gamma} (t) = \frac{1}{\left [ 1-t \cdot \theta \right]^k}$, $M_{Exp} (t) = \frac{1}{\left [ 1- t \cdot \lambda^{-1} \right]} $ and $M_{U} (t) = \frac{e^{t \cdot b}-e^{t \cdot a}}{t\cdot \left (b - a\right )}$.

Following the definition in Lemma \ref{lemrdp}, we can derive that the R\'enyi-DP of LMO Noise in each step is as follows (proven in Theorem (\ref{q_not_1})):

\begin{theorem}[R\'enyi-DP of LMO Noise] 
Given the LMO mechanism with its parameters $\mathcal{S}_{\text{LMO}}$ defined above and a real-valued query $C$, we have its MGF $M_{\text{LMO}}(C, \mathcal{S}_{\text{LMO}})$ to compute the R\'enyi-DP guarantee of an LMO noise (in each step in the training) $\epsilon_{\alpha}^{\text{LMO-DP}}$ can be derived as:
\small
\begin{align}
    \epsilon^{\text{LMO-DP}}_\alpha = \frac{1}{\alpha-1} \log \left \{ \frac{\alpha}{2\alpha-1} M_{\text{LMO}} \left (\alpha-1\right )  + \frac{1}{2}M_{\text{LMO}}\left (1-C-\alpha\right)  \right. \nonumber \\
    + \left. \frac{1}{2(1-2\alpha)}M_{\text{LMO}} \left (\left ( 1-2C\right )\alpha + \left ( C-1 \right )\right )\right \}
\end{align}
\end{theorem}

\normalsize

With the defined search space (with strict ($\epsilon,\delta$)-DP guarantee), we also provide an algorithm to search the optimal noise for the problem \ref{new} in Algorithm \ref{alg:rdp-optimization} (as detailed in Appendix \ref{sec:algmapp}). In addition, Appendix \ref{sec:vsguassian} demonstrates a significant improvement over the Gaussian noise (see Figures \ref{fig_compare_noises2} and \ref{fig_entropy_variance}). Note that the ablation study for the mixture distribution (guaranteeing the same DP) is given in Appendix \ref{sec:ablation}.

\subsection{Phase (2): LMO-DP based Private Language Model Fine-tuning}
\label{sec:frameworks}

After searching the optimal noise offline with the support of R\'enyi Accountant in Phase (1), the private fine-tuning in Phase (2) is similar to DP-SGD (supported by R\'enyi Accountant), e.g., clipping gradients, and injecting the optimized noise to the clipped gradients (detailed in Algorithm \ref{alg:lmo-dp}).

\begin{algorithm}[tb]
\label{LMO-DP}
\small
\caption{LMO-DP} 
\label{alg:lmo-dp}
\begin{algorithmic}[1]
\REQUIRE $A_1$: privacy budget $\{\epsilon, \delta\}$; $A_2$: hyperparameters in NLP tasks $\{\mathcal{D}, \eta_j, T, B, C, \mathcal{L}, \Theta_0\}$ -- the training data $\mathcal{D}=\left \{ x_i \right \} ^{N}_{i=1}$, learning rate $\eta_j$, the number of steps $T$, batch size $B$, clipping threshold $C$, loss function $\mathcal{L}$ and model parameters $\Theta_j$; $A_3$: parameters related to noise computation -- selected PDFs for mixture, the max order $\alpha_{\text{max}}$ during R\'enyi-DP accountant computation, and searched ranges of distribution parameters and their weights $\mathcal{S}$)

\ENSURE $\Theta_T$

\STATE $\text{\textbf{Call Algorithm \ref{alg:rdp-optimization} (R\'enyi Accountant Optimization) as $F_1$:}}$\\$\mathcal{S}_{\text{opt}} = F_1 (A_1, A_3)$ 

\FOR{$j$ in [1, T]} 
   \STATE draw a batch $\mathcal{B}_j$ via Poisson sampling
 
    \FOR{$x_i$ $\in$ $\mathcal{B}_{j}$} 
     \STATE  $\textbf{g}_{j} (x_i) = \bigtriangledown _{\Theta_{j}} \mathcal{L} \left (\Theta_{j}, x_i\right )$ \color{blue}\COMMENT{computing gradient $g_{j} (x_i)$}\color{black}
    \ENDFOR
    
   \STATE $\bar{\textbf{g}}_{j} \left ( x_i\right ) \gets \textbf{g}_{j} \left ( x_i\right ) /\max \left ( 1, \frac{\left \| \textbf{g}_{j}\left ( x_i \right )  \right \| _2}{C} \right )$  \color{blue}\COMMENT{clipping}\color{black}
    
\STATE $\tilde{\textbf{g}}_{j} \gets \frac{1}{B_j} \left (  {\textstyle \sum_{i=1}^{B_j}} \bar{\textbf{g}}_{j}\left (x_i\right ) + \mathcal{M}_{\text{LMO}}(C, \mathcal{S}_{\text{opt}})  \right )$ \color{blue}\COMMENT{adding LMO noise}\color{black}

\STATE $\Theta_{j+1} \gets \Theta_{j} - \eta_j \tilde{\textbf{g}}_{j}$ 
\ENDFOR

\STATE \textbf{Return} $\Theta_T$

\end{algorithmic}
\end{algorithm}

\section{{Experiments}}
\label{sec:exp}

Recent SOTA methods achieve high accuracy (e.g., $93\% \sim 94\%$) on large privacy budgets ($\epsilon>3$). However, it remains unclear how these models perform when subjected to stronger DP guarantees, characterized by a privacy budget $\epsilon<3$ (with $\delta$ ranging from $10^{-6}$ to $10^{-5}$). 
Following \cite{li2021large}, we use the full datasets and all the classes in the original datasets for fine-tuning. Our results highlight the versatility (higher accuracy with fewer steps), universality (including sentence classification \cite{wang2018glue} and table-to-text generation \cite{yu2021differentially}) and other related NLP tasks in \cite{liu2019roberta, devlin2018bert, yu2021differentially, touvron2023llama}. In addition, we take the first step to privately fine-tune large language models (i.e., Llama 2) with strong privacy guarantees (see the accurate results in Section \ref{sec:llama2}). We focus on strong DP guarantees (e.g., $\epsilon < 3$) and set $\delta=10^{-10}$ unless otherwise specified.

\subsection{Performance on Sentence Classification}
\label{4.3}

\begin{figure*}[!ht]
	\centering
	\subfigure[MNLI-m dataset]{
    \includegraphics[angle=0, width=0.24\linewidth]{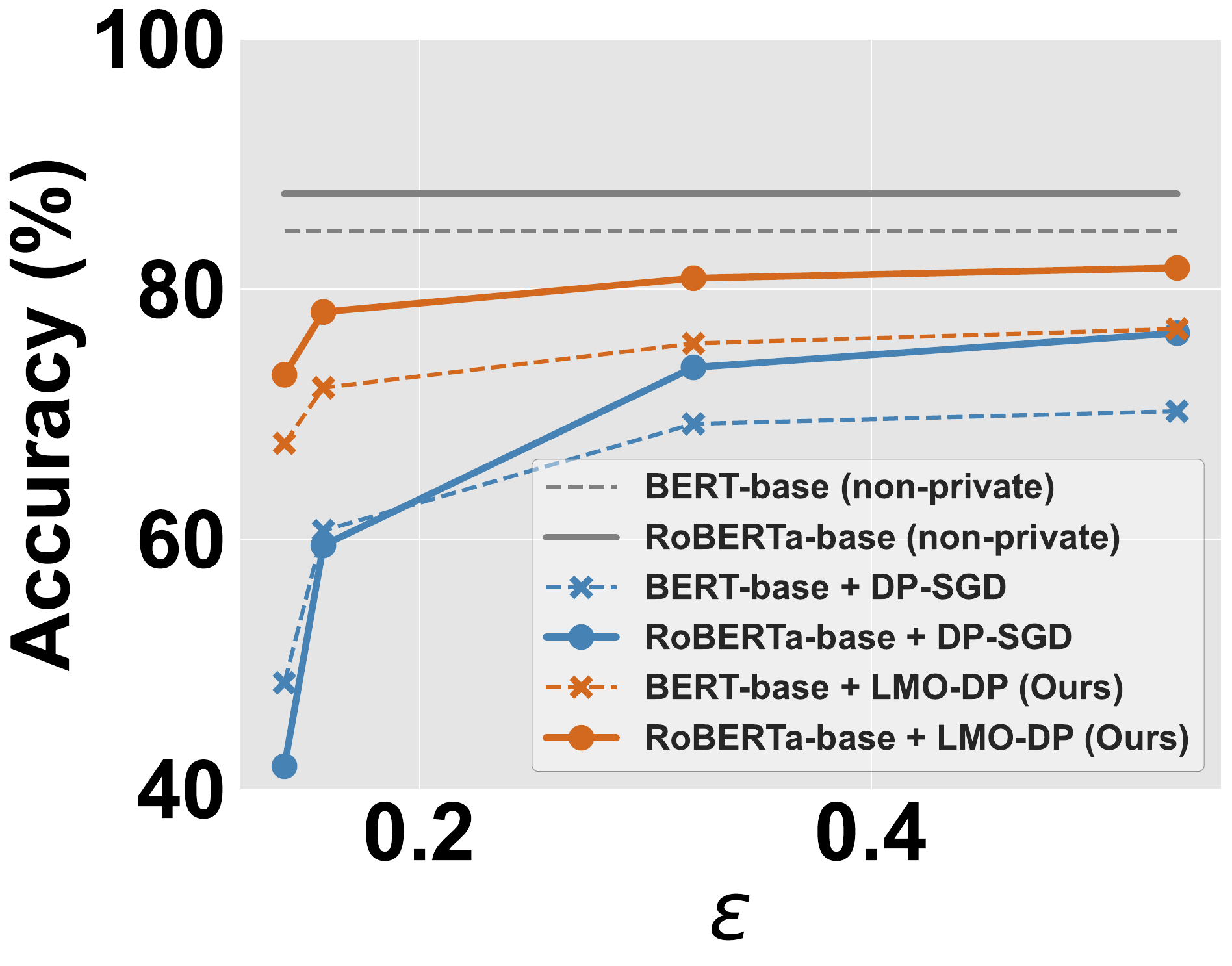}    \label{fig:task1_base_acc_mnli.pdf}}
    \hspace{-0.1in}
    \subfigure[SST-2 dataset]{
    \includegraphics[angle=0, width=0.24\linewidth]{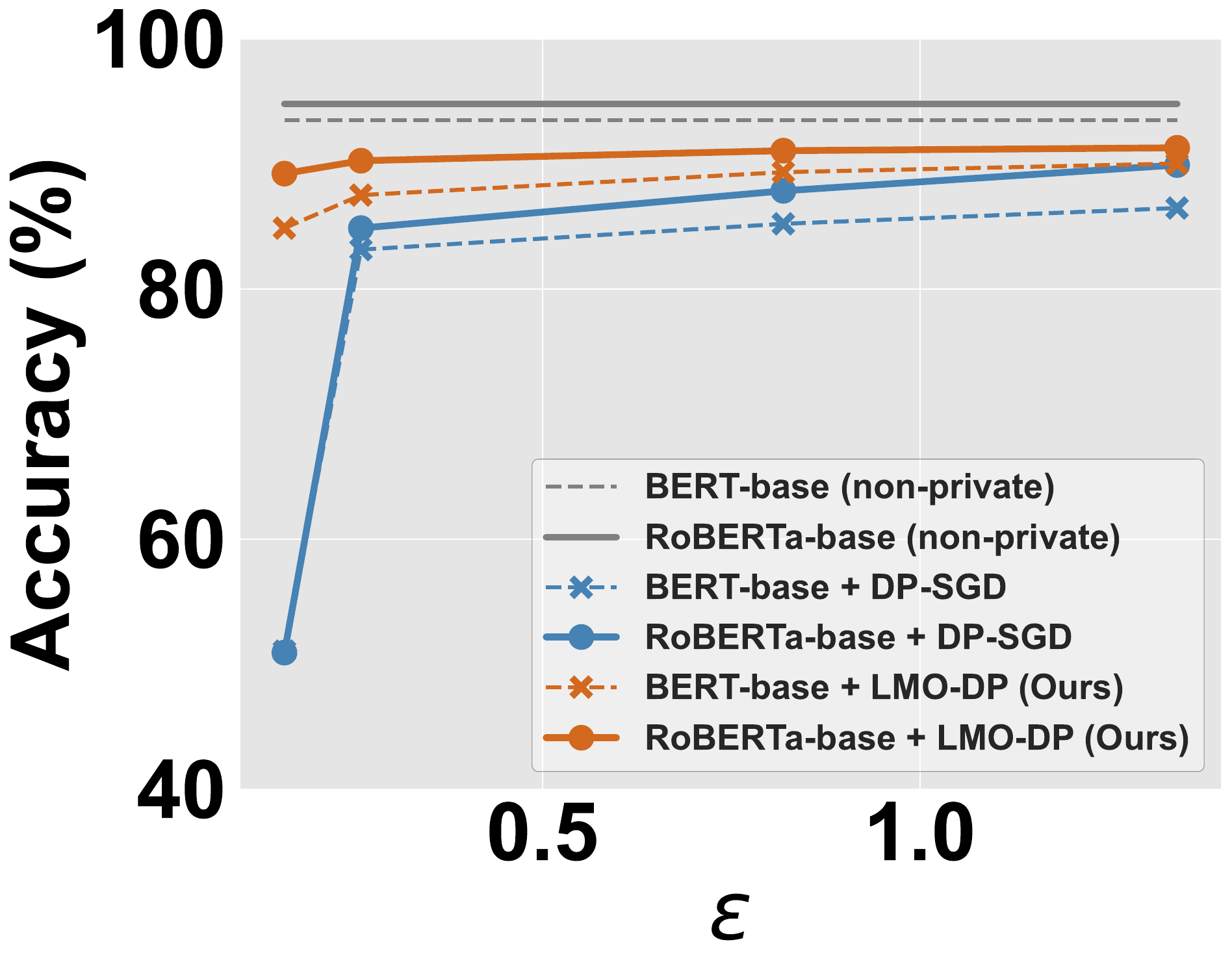}  \label{fig:task1_base_acc_sst2.pdf}}\vspace{-0.00in}
    \hspace{-0.1in}
	\subfigure[QNLI dataset]{
		\includegraphics[angle=0, width=0.24\linewidth]{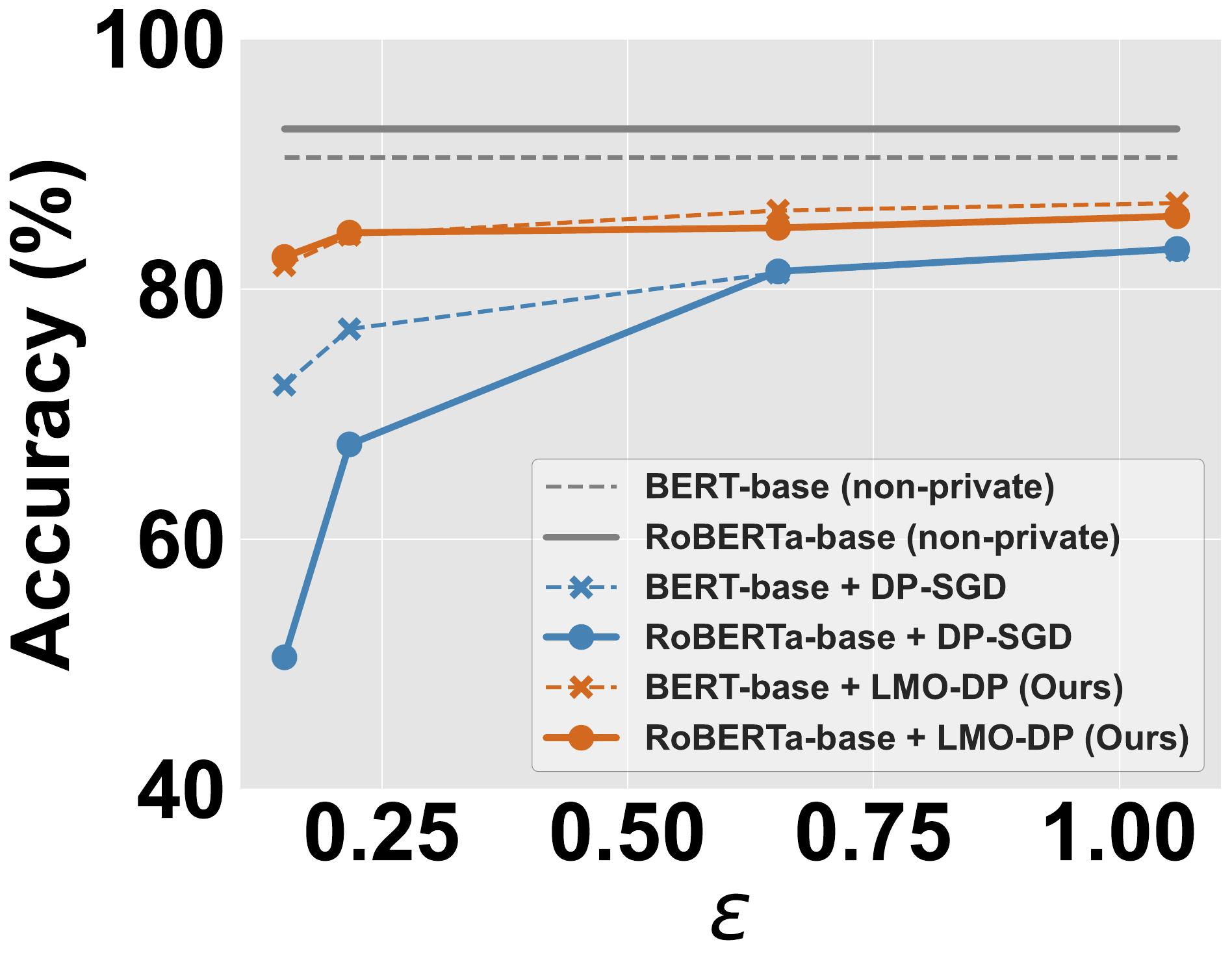}		\label{fig:task1_base_acc_qnli.pdf}}
	\hspace{-0.1in}
	\subfigure[QQP dataset]{
		\includegraphics[angle=0, width=0.24\linewidth]{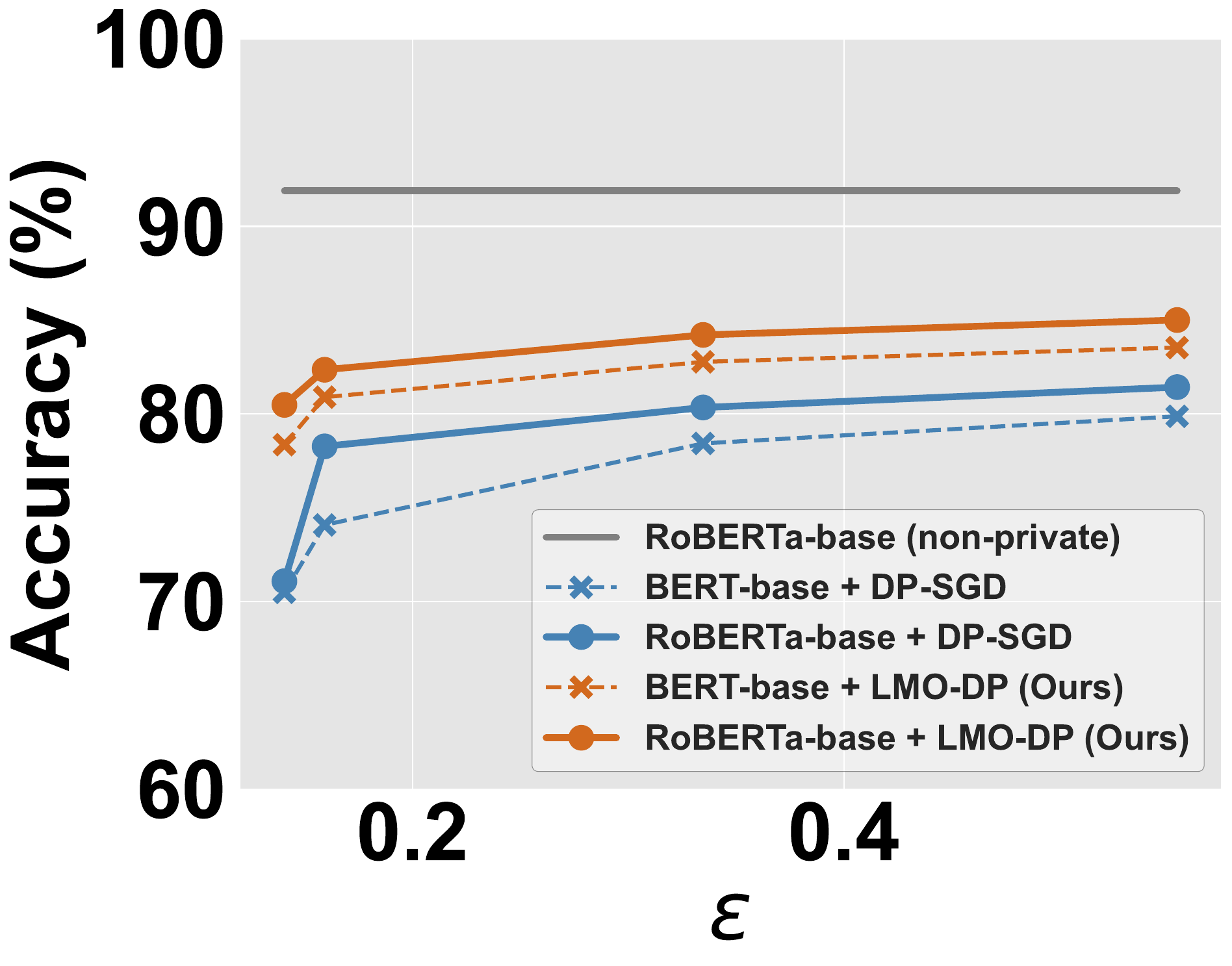}		\label{fig:task1_base_acc_qqp.pdf}}\vspace{-0.1in}
\caption[Optional caption for list of figures]
{$\epsilon$ vs. Accuracy of sentence classification task for BERT-base and RoBERTa-base models (100M parameters). For larger $\epsilon$, the results of baselines are approximating LMO-DP, and they are not plotted.}\vspace{-0.1in}
\label{classification_acc_basemodel}
\end{figure*}

{We evaluate the LMO-DP mechanism on sentence classification tasks aiming to distinguish between positive and negative emotions with GLUE benchmarks (MNLI-m, SST-2, QNLI, and QQP datasets) \cite{wang2018glue} with experiments conducted on RoBERTa-base, RoBERTa-large \cite{liu2019roberta}, BERT-base, BERT-large \cite{devlin2018bert} and Llama2-chat-7b \cite{touvron2023llama}, compared with the non-private training and baseline DP-SGD method \cite{li2021large}. In detail, SST-2 has more than 60k+ samples in the training set; QNLI has more than 100k+ samples; MNLI and QQP contain more than 350k but less than 400k samples for each dataset. SST-2, QNLI, and QQP include two classes each; MNLI includes three classes.}

\begin{figure*}[!ht]
	\centering
	\subfigure[MNLI-m dataset]{
    \includegraphics[angle=0, width=0.24\linewidth]{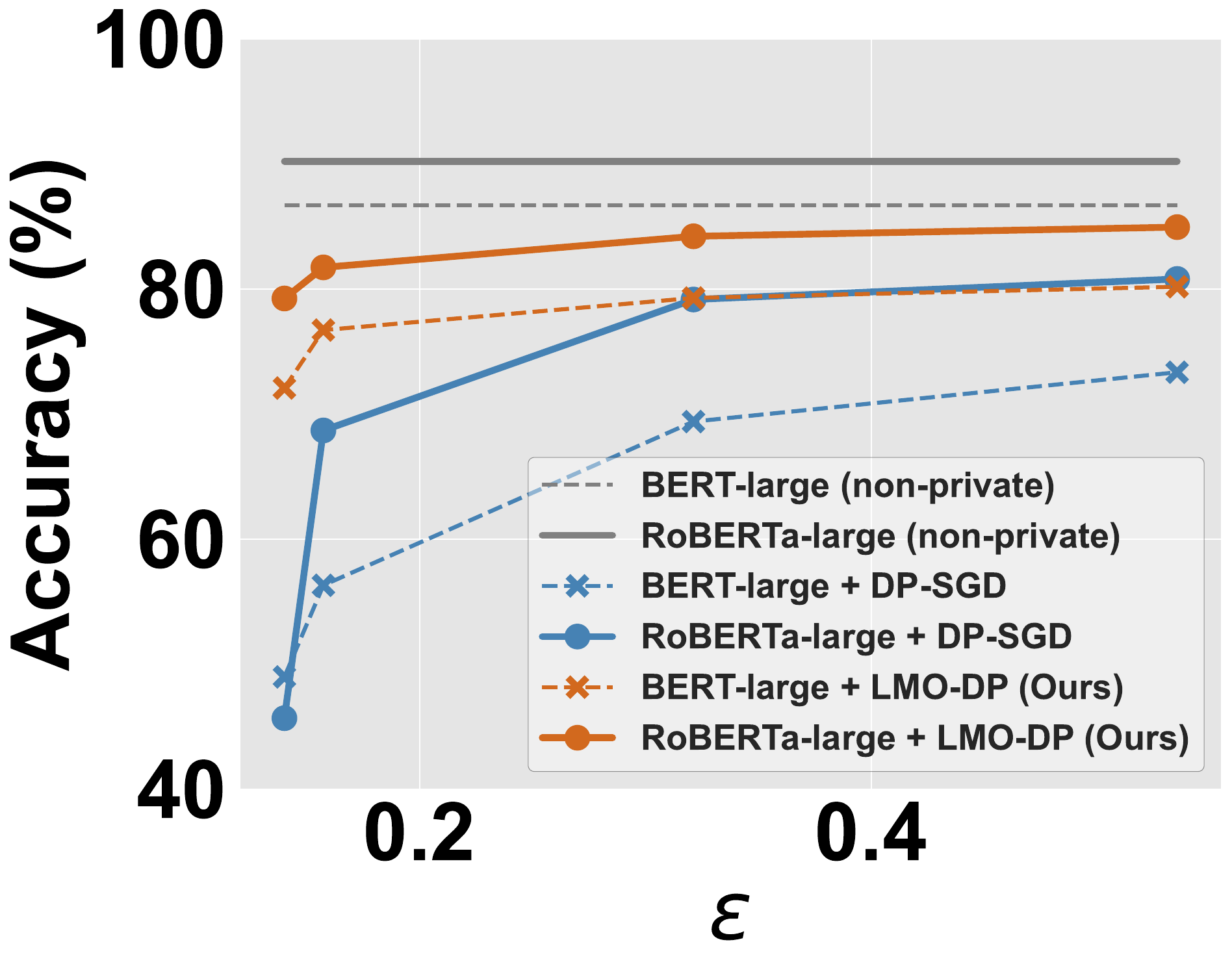}    \label{fig:task1_large_acc_mnli}}
    \hspace{-0.1in}
    \subfigure[SST-2 dataset]{
    \includegraphics[angle=0, width=0.24\linewidth]{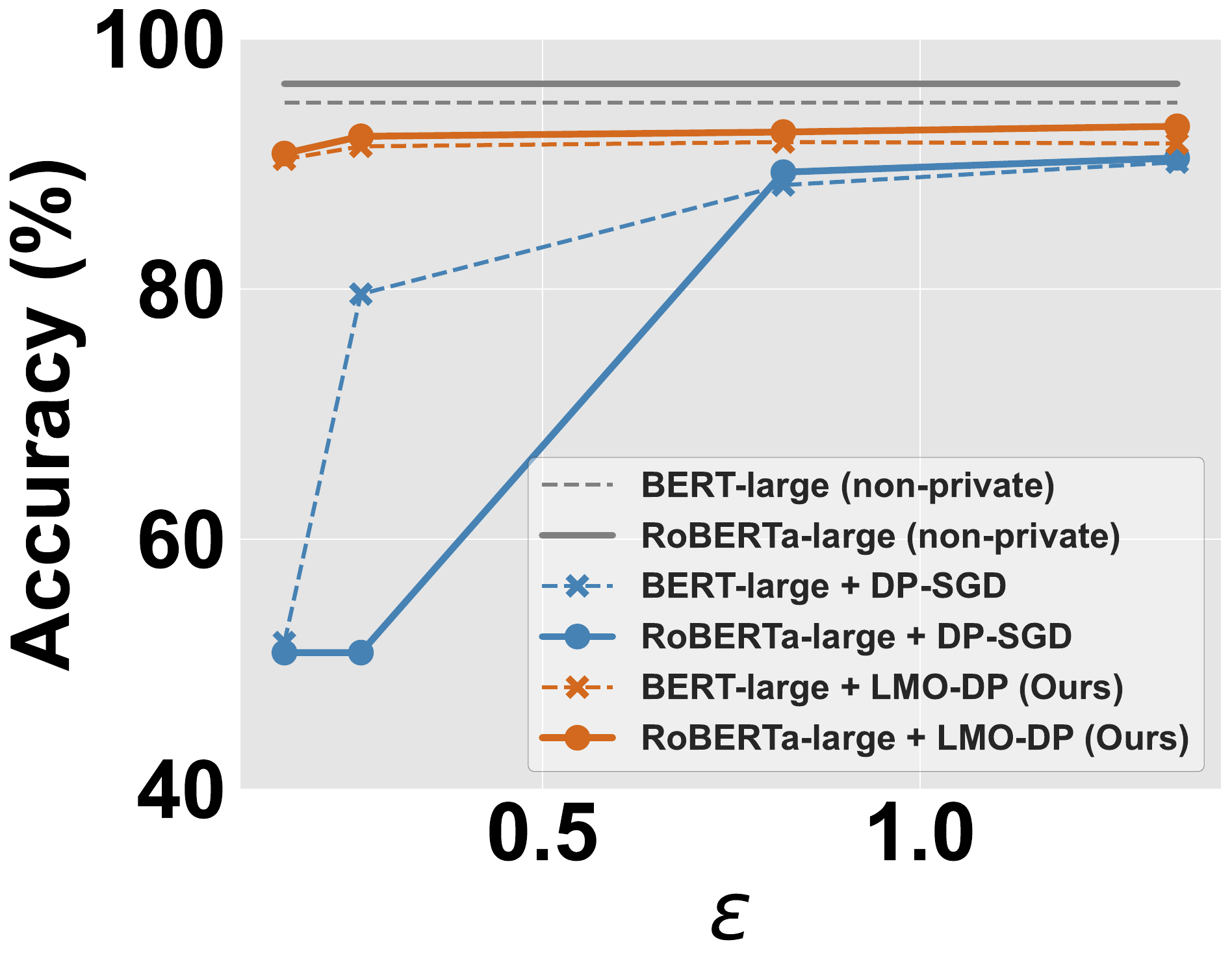}  \label{fig:task1_large_acc_sst2}}\vspace{-0.00in}
    \hspace{-0.1in}
	\subfigure[QNLI dataset]{
		\includegraphics[angle=0, width=0.24\linewidth]{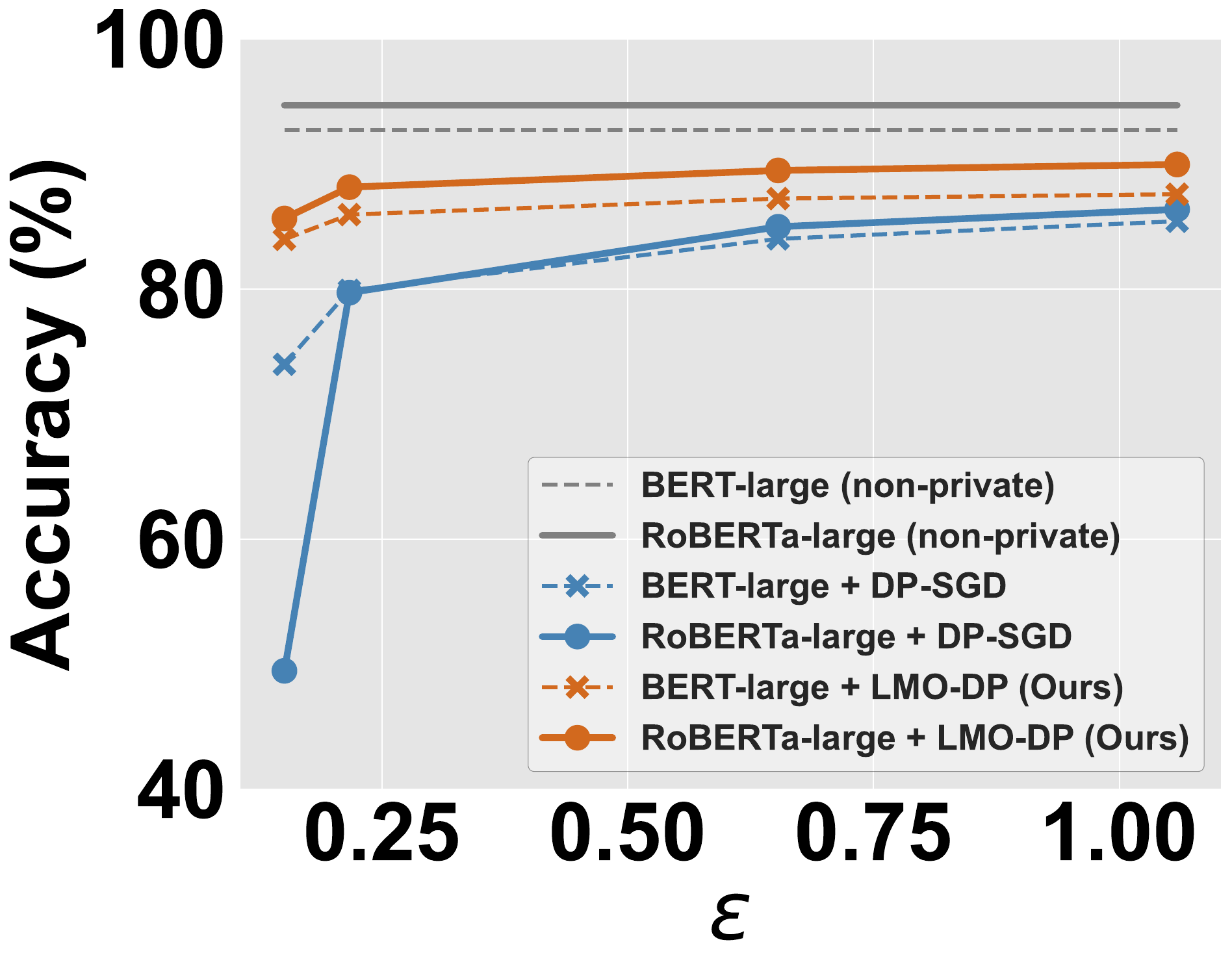}		\label{fig:task1_large_acc_qnli}}
	\hspace{-0.1in}
	\subfigure[QQP dataset]{
		\includegraphics[angle=0, width=0.24\linewidth]{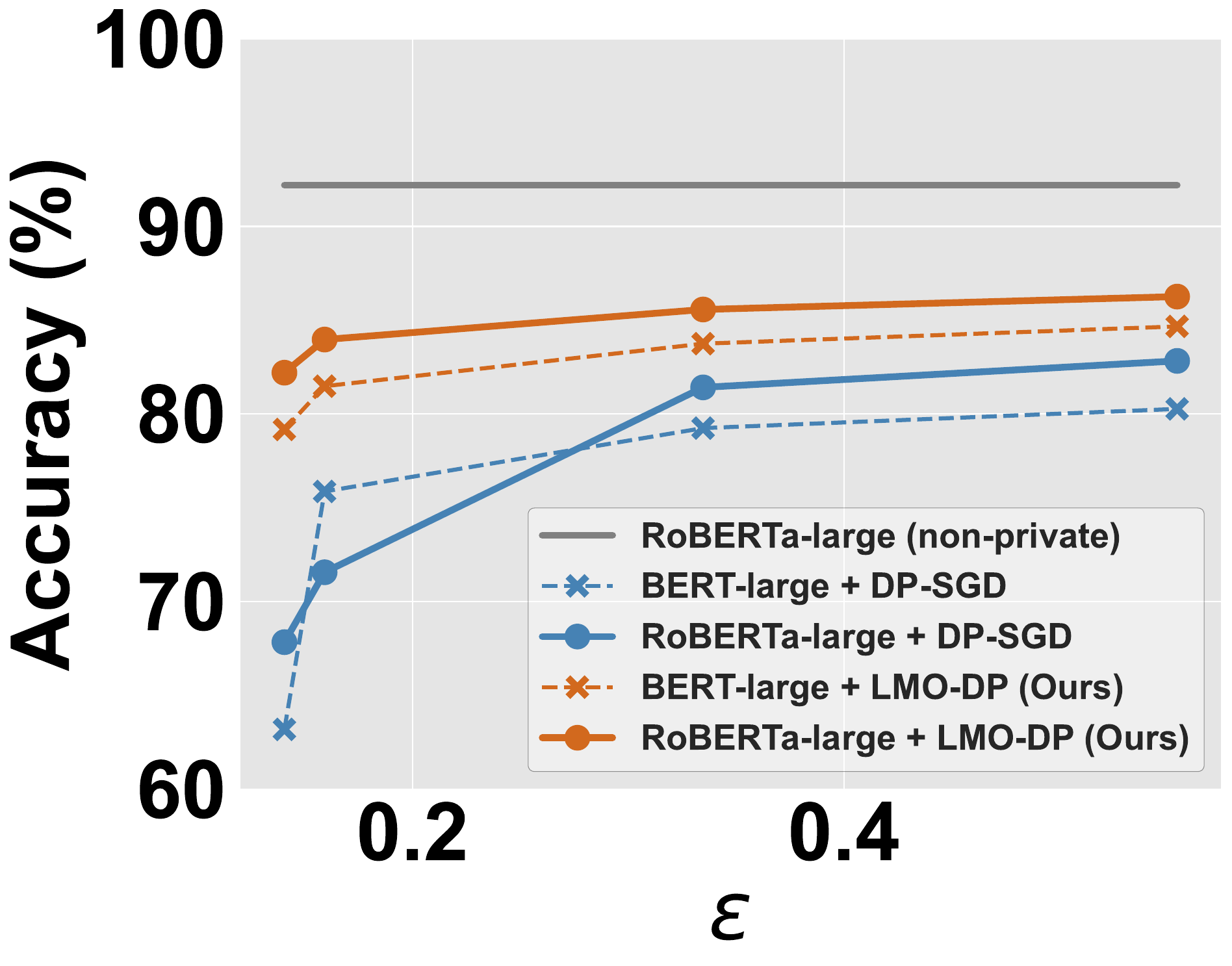}		\label{fig:task1_large_acc_qqp}}\vspace{-0.1in}
\caption[Optional caption for list of figures]
{$\epsilon$ vs. Accuracy of sentence classification task for BERT-large and RoBERTa-large (300M parameters). For larger $\epsilon$, the results of baselines are approximating LMO-DP, and they are not plotted.}\vspace{-0.1in}
\label{classification_acc_largemodel}
\end{figure*}

\begin{figure*}[!ht]
	\centering
	\subfigure[MNLI-m dataset (base)]{
    \includegraphics[angle=0, width=0.24\linewidth]{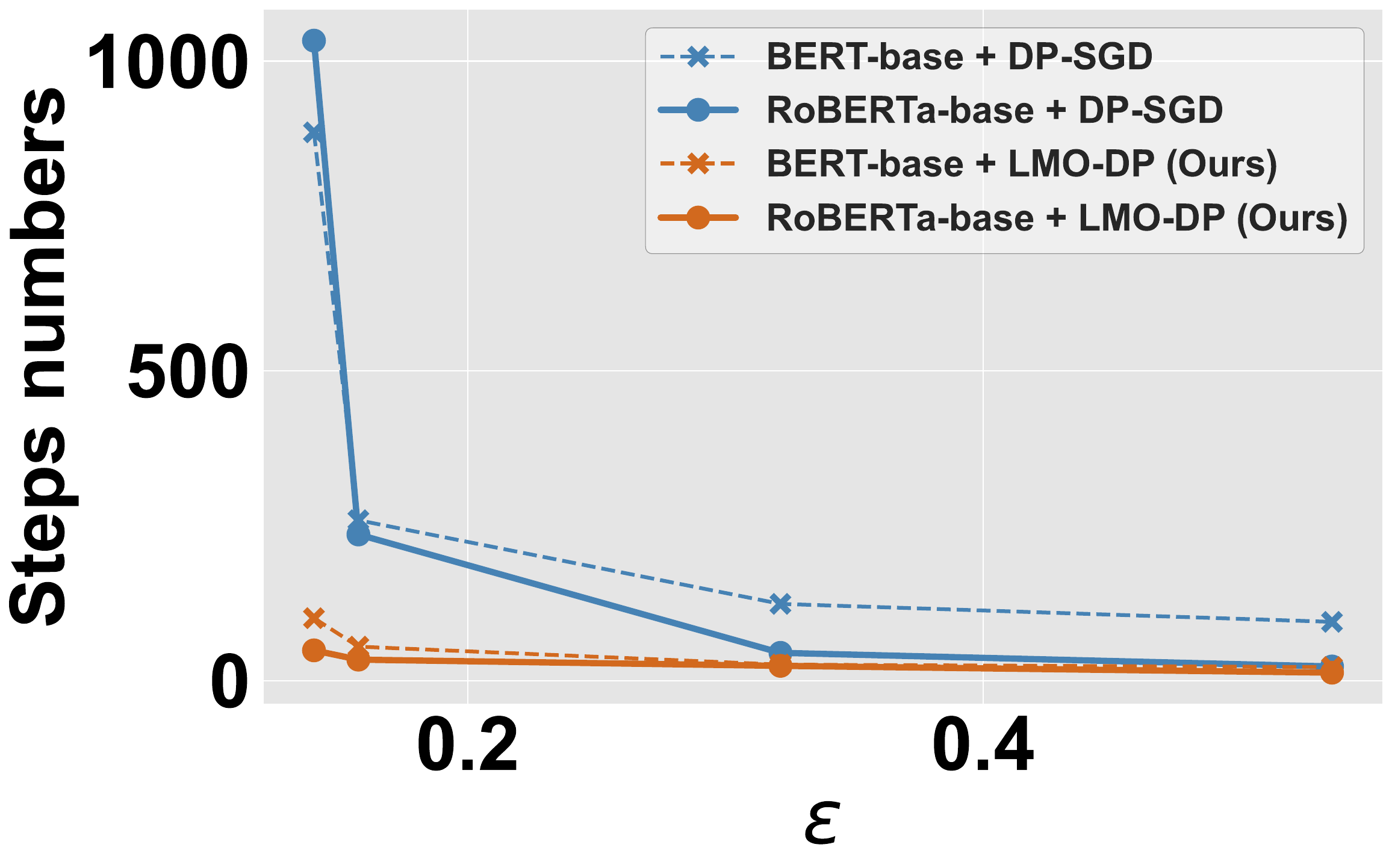}    \label{fig:task1_base_spd_mnli}}
    \hspace{-0.1in}
    \subfigure[QQP dataset (base)]{
    \includegraphics[angle=0, width=0.24\linewidth]{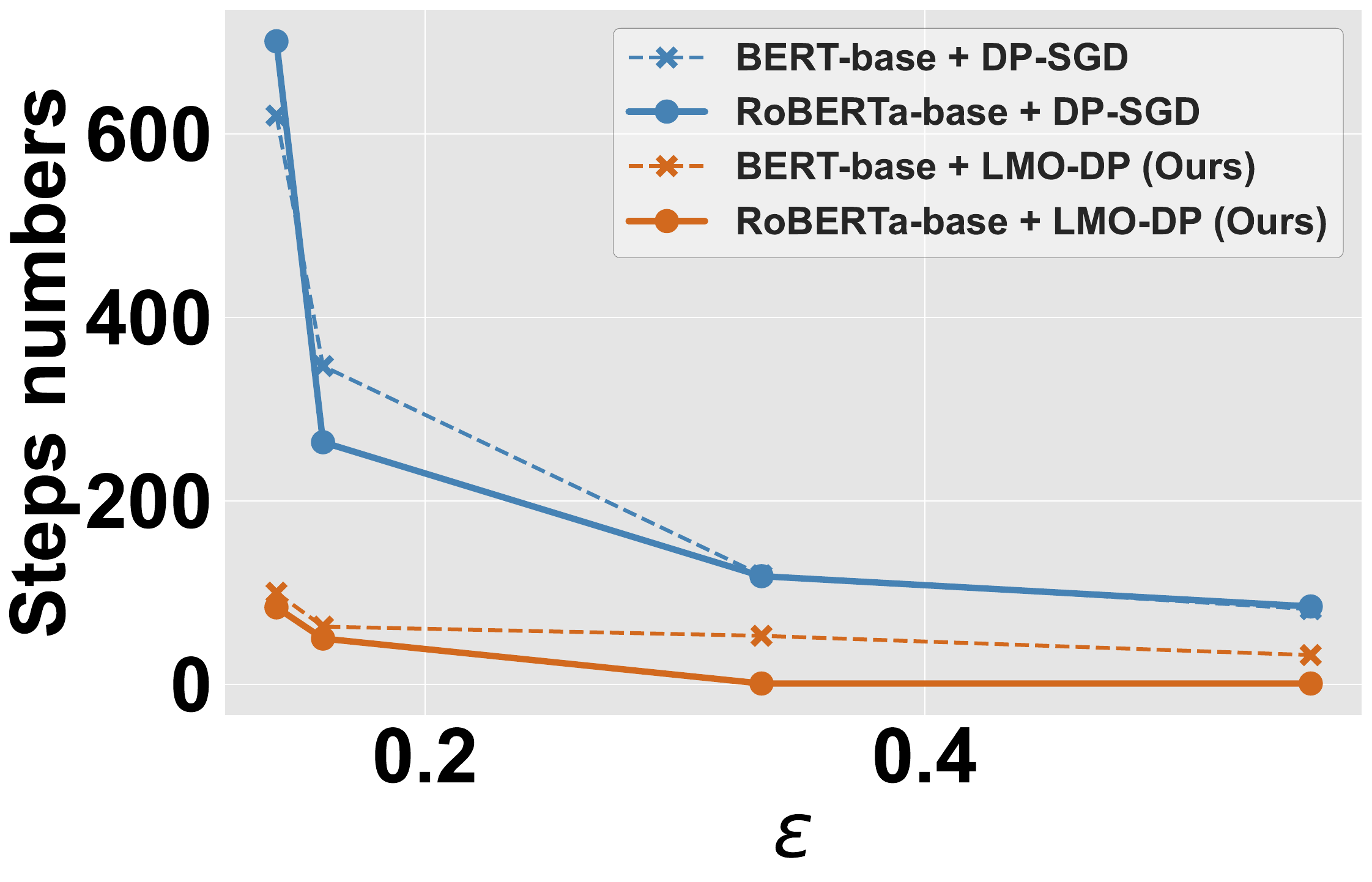}  \label{fig:task1_base_spd_qqp}}\vspace{-0.00in}
    \hspace{-0.1in}
	\subfigure[MNLI-m dataset (large)]{
		\includegraphics[angle=0, width=0.24\linewidth]{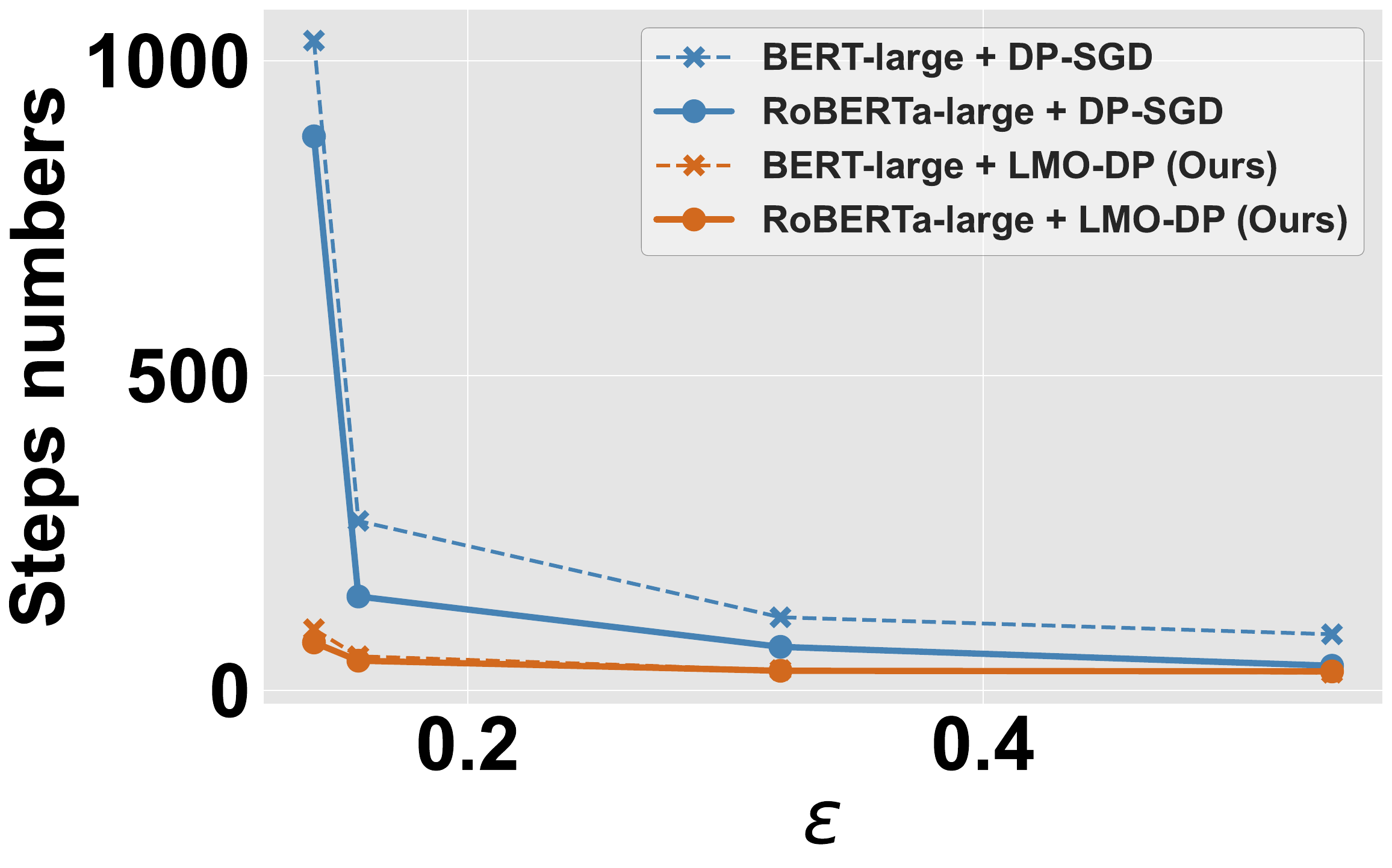}		\label{fig:task1_large_spd_mnli}}
	\hspace{-0.1in}
	\subfigure[QQP dataset (large)]{
		\includegraphics[angle=0, width=0.24\linewidth]{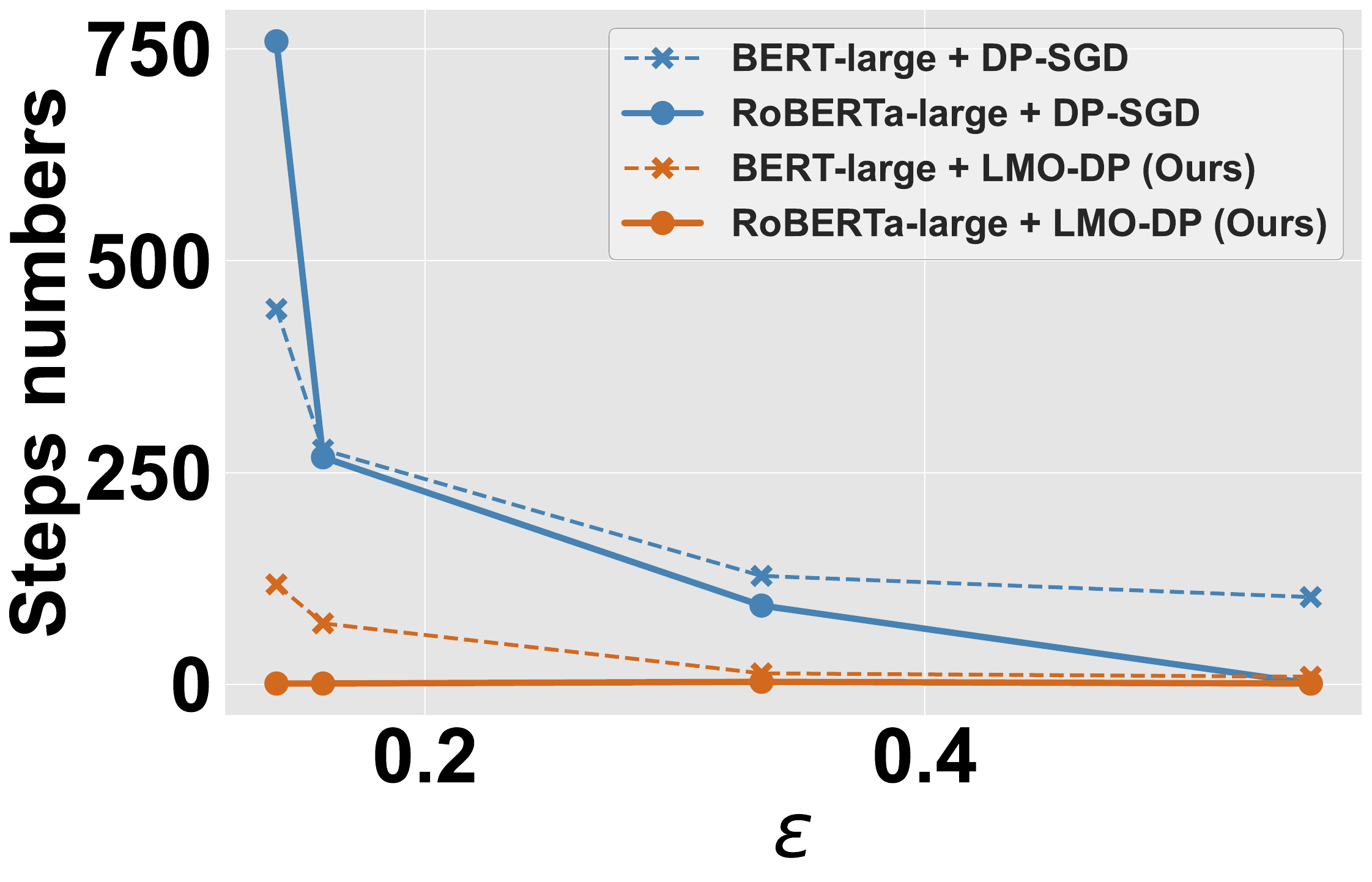}		\label{fig:task1_large_spd_qqp}}\vspace{-0.1in}
\caption[Optional caption for list of figures]
{$\epsilon$ vs. Steps of the sentence classification task for BERT families and RoBERTa families on MNLI-m and QQP datasets. For larger $\epsilon$, the results of baselines are approximating LMO-DP, and they are not plotted.}
\vspace{-0.05in}
\label{classification_step_largemodel}
\end{figure*}

In Figures \ref{classification_acc_basemodel} and \ref{classification_acc_largemodel}, the $x$-axis shows total privacy loss $\epsilon$ and the $y$-axis shows the accuracy under specific $\epsilon$ after $T$-step composition during training/fine-tuning. When $\epsilon>3$, we have accuracy that matches or surpasses that of the baseline; When $\epsilon\leq 3$, our mechanism largely outperforms the baseline, exhibiting close performance with non-private training.

\small
\begin{wraptable}{r}{0.5\textwidth}
\vspace{-8mm}
\scriptsize
\center
  \caption{Fine-tuning GPT-2 on the E2E dataset.}
  \addtolength{\tabcolsep}{-4.5pt}
    \begin{tabular}{ccccccc}
    \hline
    \multirow{2}{*}{Total $\epsilon$} & \multirow{2}{*}{Methods} & \multicolumn{5}{c}{Metrics} \\
\cmidrule{3-7}          &       & BLEU &  NIST & METEOR & ROUGE-L& CIDEr \\
    \hline
    \multirow{1}[6]{*}{0.046807} & DP-SGD & 21.36 & 2.3185 & 0.3575 & 55.99 & 0.66 \\
\cmidrule{2-7}          & \textbf{LMO-DP} & \textbf{30.82} & \textbf{3.2484}  & \textbf{0.3622} & \textbf{59.19}  & \textbf{1.3826}\\
    \midrule
    \multirow{1}[6]{*}{0.046870} & DP-SGD & 25.47 & 2.5943 & 0.4103 & 60.28 & 0.8656\\
\cmidrule{2-7}          & \textbf{LMO-DP} & \textbf{43.1} & \textbf{4.3517} & \textbf{0.441} & \textbf{69.7} & \textbf{2.0162}\\
    \midrule
    \multirow{1}[6]{*}{0.067623} & DP-SGD & 39.66 & 4.0486 & 0.433 & 67.56 & 1.836 \\
\cmidrule{2-7}          & \textbf{LMO-DP} & \textbf{49.91} & \textbf{5.3452} & \textbf{0.4495} & \textbf{68.94} & \textbf{3.073} \\
    \midrule
    \multirow{1}[6]{*}{0.176436} & DP-SGD & 44.32 & 4.471 & 0.4429 & 70.5 & 2.2172\\
\cmidrule{2-7}          & \textbf{LMO-DP} & \textbf{53.51} & \textbf{5.7178}  & \textbf{0.4489} & \textbf{68.87} & \textbf{3.3614} \\
\midrule 
    Non-private &  & 54.25 & 6.4832 & 0.4709 & 68.7 & 3.9951\\
    \bottomrule
    \end{tabular}%
  \label{tab:text}%
 \vspace{-0.15in}
\end{wraptable}
\normalsize

Besides, we observed that our method can reach the same accuracy with fewer steps compared to the DP-SGD baseline method \cite{li2021large}, especially under strong DP budgets. Considering baseline methods may not achieve the same high accuracy as the LMO-DP method, we report the steps that these two methods reach the same accuracy for each $\epsilon$. In particular, Specifically, we evaluate the steps that need to reach 48.51\% accuracy and 41.83\% accuracy for the BERT-base and RoBERTa-base on the MNLI-m dataset; the steps that need to reach 70.49\% accuracy and 71.09\% accuracy for the BERT-base and RoBERTa-base on the QQP dataset; the steps that need to reach 48.98\% accuracy and 45.67\% accuracy for the BERT-large and RoBERTa-large on the MNLI-m dataset; the steps that need to reach 63.18\% accuracy and 67.84\% accuracy for the BERT-large and RoBERTa-large on QQP dataset. From Figure \ref{classification_step_largemodel}, we conclude that we can reduce the $\sim 50\%$ training steps when $\epsilon\leq 3$.

Comparison in Appendix \ref{comapre_noises} shows a significantly smaller perturbation of LMO noises, the incorporation of LMO-DP noise in private training may intriguingly lead to faster convergence due to the dynamics of LMO-DP noise.
Hence, LMO noise effectively addresses two challenges (accuracy and convergence) simultaneously.

\subsection{Table-to-Text Generation Task}
\label{5.3}
We conduct the LMO-DP and DP-SGD baseline methods on a table-to-text generation task which generates the descriptions of table entries. We fine-tune the GPT-2 model \cite{yu2021differentially} on the E2E dataset \cite{novikova2017e2e} with $\delta=8\cdot10^{-6}$, evaluating five metrics in Table \ref{tab:text}.
We evaluate LMO-DP and DP-SGD \cite{li2021large} under a fixed $\delta=8\cdot10^{-6}$ (same setting as \cite{li2021large}). We also apply the same settings of privacy budget at each iteration, weights, and clipping threshold as the sentence classification. 
We only employ a batch size of $16$ which causes the total privacy budget to be less than $0.2$.

Table \ref{tab:text} presents the results with five different metrics by following \cite{yu2021differentially}. We observe that LMO-DP yields results that are more closely aligned with the non-private results (larger values of all these metrics exhibit more accurate generated texts). It is worth noting that the improvement can be up to $50\%$ on some metrics (e.g., CIDEr). The ROUGE-L of both LMO-DP and DP-SGD can be slightly higher than the original values (the last pair of results) since they are both fine-tuned based on the same LMs with rich vocabulary and downstream dataset and thus have a greater chance of generating approximate texts.

\subsection{Boosting Accuracy for the Existing Methods}
\label{subsec:compare}

\begin{wraptable}{r}{0.5\textwidth}
\vspace{-8mm}
\small
\center
\caption{Boosting accuracy (\%) with LMO-DP (w.l.o.g., for Ghost Clipping \cite{li2021large}). RoBERTa-large model on the SST-2 dataset for sentiment classification ($\delta=10^{-10}$).}
\addtolength{\tabcolsep}{-3.5pt}
\label{tab:comparison}
  \begin{tabular}{ccccccc}
    \toprule
    \multirow{2}[3]{*}{Method} & \multicolumn{6}{c}{Total $\epsilon$} \\
\cmidrule{2-7}          & 0.16   & 0.3     & 0.9     & 1.4     & 3          & $\infty$ \\
    \hline 
    \cite{li2021large} & \cellcolor{red!30}50.92     & \cellcolor{red!30}50.92   & \cellcolor{red!30}89.33     & \cellcolor{red!30}90.48 & \cellcolor{red!30}91.06 & 96.20  \\
    \textbf{LMO-DP} &  \cellcolor{green!30}\textbf{90.83}     &  \cellcolor{green!30}\textbf{92.20}    &    \cellcolor{green!30}\textbf{92.55}   &   \cellcolor{green!30}\textbf{93.00}    &  \cellcolor{green!30}\textbf{93.92}      & 96.20 \\
    \hline
    \cite{yu2021differentially} & -     & -     & 51.31    & 51.31     & 51.31    & 96.40 \\
    \cite{bu2022differentially} & 49.08     & 49.43     & 50.92     & 50.92 & 54.58 & 95.50 \\
    \cite{he2022exploring}& -     & -     & -     & - & 93.87$^*$  & 96.20$^*$ \\
    \cite{bu2023differentially} &49.08 &50.92& 87.72 & 90.02 & 90.14 & - \\
    \bottomrule
    \end{tabular}%
\vspace{-0.05in}
\end{wraptable}

\normalsize
Our LMO-DP mechanism can be implemented as a plug-and-play module to other orthogonal methods (e.g., memory reduction methods such as Ghost Clipping \cite{li2021large} or parameter efficiency \cite{yu2021differentially} during real-time private training. We take sentence classification as an example. In detail, we conduct sentence classification task with the SST-2 dataset on the RoBERTa-large model.

As $\epsilon>3$, the best accuracy of SOTA methods and LMO-DP would be close to each other (we validated this). Thus, the improvement for LMO-DP is marginal, and then we focus on the strong DP with smaller $\epsilon$ ($\delta$ is set to be close to 0, e.g., $10^{-10}$). The top two rows in Table \ref{tab:comparison} show that LMO-DP can significantly boost the accuracy for Ghost Clipping \cite{li2021large} from $\sim 50\%$ to $90\%+$. As illustrated in the last four rows of Table \ref{tab:comparison}, we also found that the accuracy of other SOTA methods \cite{yu2021differentially,bu2022differentially,bu2023differentially} have low accuracy in case of small $\epsilon$ (strong DP).\footnote{\cite{he2022exploring} is not open-sourced. We include its result ($^*$) for $\epsilon=3$, $\delta=1/n^{1.1}$ given training size $n$.} Given high noise reduction by LMO-DP compared to the Gaussian mechanism, we anticipate that LMO-DP can also drastically boost their accuracy to $\sim 90\%$ similar to Ghost Clipping \cite{li2021large}.

\subsection{Performance on the Llama 2 Model}
\label{sec:llama2}

\begin{wraptable}{r}{0.5\textwidth}
\vspace{-17mm}
\small
\center
\caption{Accuracy of LMO-DP private training on Llama2-7b-chat model on the SST-2 dataset for sentiment classification ($\delta=10^{-10}$).}
\addtolength{\tabcolsep}{-1pt}
\label{tab:llama_lmo}
  \begin{tabular}{ccccccc}
    \toprule
    \multirow{2}[3]{*}{Steps} & \multicolumn{6}{c}{Total $\epsilon$} \\
\cmidrule{2-7}          & 0.16   & 0.3     & 0.9     & 1.4     & 3          & $\infty$ \\
    \hline 
    7 &  54\%     & 56\%  & 60\%    & 75\% & 78\% & 85\% \\
    21 &  82\%     &    87\%    &   87\%     &  88\%  & 89\%    & 93\%  \\
    35 &  90\%     & 91\%     & 91\%    & 93\%     & 93\%    & 93\% \\
    841 & 93\%      & 93\%      & 93\%   & 93\%  & 93\% & 93\% \\
    \bottomrule
    \end{tabular}%
 \vspace{-2mm}
\end{wraptable}

\normalsize
Different from all existing works (e.g., \cite{yu2021differentially,bu2022differentially,bu2023differentially}), we take the first step to implement our new LMO noise to privately fine-tune the Llama2-7b-chat model on the SST-2 dataset for sentiment classification with strong $(\epsilon,\delta)$-DP guarantees. As demonstrated in Table \ref{tab:llama_lmo}, both non-private ($\infty$) and LMO-DP private fine-tuning can converge to high accuracy (non-private fine-tuning converges relatively faster). This confirms that LMO-DP also works effectively on LLMs (due to smaller impact on the overall model by each single data sample, we anticipate that such high accuracy can be maintained in other tasks). 

\section{Conclusion}
\label{Conclusion}

In this paper, we propose a Language Model-based Optimal Differential Privacy (LMO-DP) mechanism, allowing for accurate private fine-tuning (large) language models even in very strict privacy settings $(\epsilon<3, \delta=10^{-10})$. To our best knowledge, LMO-DP is the first non-Gaussian mechanism that can generate sub-optimal noise to ensure strong DP, and the first mechanism to support LLMs. It can also significantly boost the performance (e.g., accuracy and convergence) of DP-SGD and other variants with high noise reduction, as demonstrated in the experiments.

\appendix
\section{{Foundations and Frontiers / Scholarly Review and Key Definitions}}
\label{sec:r2dp}

{Our work is the instantiation and further exploration of the R$^2$DP framework within the Large Language Models (LLM) domain, originally proposed by Mohammady et al. \cite{mohammady2020r2dp}. To facilitate comprehension, we provide the necessary background information in \cite{mohammady2020r2dp}}

\subsection{\texorpdfstring{$R^2DP$}{R2DP} Mechanism}

\begin{theorem}
\label{theorem:mechanism-probability}
For a given query function $q: \mathcal{D} \rightarrow \mathbb{R}$, any measurable subset $S \subset \mathbb{R}$, and a dataset $D \in \mathcal{D}$, consider a randomized mechanism $\mathcal{M}_q(D,b): \mathcal{D} \times \Omega \to \mathbb{R}$ defined as $\mathcal{M}_q(d) = q(d) + w$, where $w \sim \text{Lap}(b)$ and $1/b \sim f \in \mathcal{F}$. Then, the probability that $\mathcal{M}_q(D,b)$ falls within subset $S$ can be expressed as:
\small
\begin{equation}
\label{eqn11}
\mathbb{P}(\mathcal{M}_q(D,b)\in S) = \frac{1}{2} \cdot \left[-M_{1/b}(-|x-q(D)|) \cdot \mathds{1}_{\{S\geq q(D)\}} 
+ M_{1/b}(-|x-q(D)|) \cdot \mathds{1}_{\{S< q(D)\}}\right]
\end{equation}
\normalsize
where $M_{f}(t)$ denotes the moment-generating function (MGF) of the random variable $f$, and $\mathds{1}$ is the indicator function (Proof in \cite{mohammady2020r2dp}, Appendix C).
\end{theorem}

\begin{theorem}
\label{theorem:epsilon_delta_DP}
The $R^2DP$ mechanism $\mathcal{M}_q(D,b)$ is $\ln\left [\frac{\mathbb{E}\left (\frac{1}{b}\right )}{\frac{dM_{\frac{1}{b}(t)}}{dt}|_{t=-\Delta q}}\right ]$-differentially private (proof in \cite{mohammady2020r2dp}, Appendix C).
\end{theorem}

\subsection{Moment Generating Function and Linear Combination of MGFs}
\label{sec:mgfdef}
\begin{definition}
\label{MGFdefn}
    The moment-generating function of a random variable $x$ is $M_{X}(t):=\mathbb E \left[e^{tX}\right], t\in \mathbb {R}$ wherever this expectation exists. The moment-generating function is the expectation of the random variable $e^{tX}$~ \cite{Walker1965ProbabilityTA}. 
\end{definition}

\begin{theorem}[MGF of Linear Combination of RVs]
\label{thm:lin}
 If $x_1, \cdots, x_n$ are $n$ independent random variables (RVs) with MGFs $M_{x_i}(t)=\mathbb E (e^{t x_i})$ for $i = 1,\cdots , n$, then the MGF of the linear combination $Y=\sum\limits_{i=1}^{n}a_ix_i$ is $\prod \limits_{i=1}^{n} M_{x_i} (a_it)$.
\end{theorem}

\section{Proofs}
\label{sec:Proofs}

\subsection{Proof of Rényi DP of LMO Noise}
\label{q_not_1}

\normalsize
\begin{proof}
First, we define the Laplace distribution $P \sim \Lambda (0, b)$ and $Q \sim \Lambda (C, b)$ $(C>0, b>0)$ with the density are $p(x)=\frac{1}{2b} \exp \left (-|x|)/b\right )$ and $q(x)=\frac{1}{2b} \exp \left (-|x-C|)/b\right )$, respectively; For any $\alpha>1$, we define the integral of them over the intervals $(-\infty, 0], \left [0, C \right ]$, $\left [C, +\infty \right ]$ as follows.

\small
\begin{align}
    & \int^{\infty}_{-\infty} p(x)^\alpha q(x)^{1-\alpha} dx \nonumber \\
    & = \frac{1}{2b}\int^{0}_{-\infty} \exp\left (\alpha x/ b + (1-\alpha) (x-1) /b \right )dx \nonumber\\
    &+ \frac{1}{2b}\int^{C}_{0} \exp\left (-\alpha x/ b + (1-\alpha) (x-1) /b \right )dx \nonumber \\
    &+\frac{1}{2b}\int^{\infty}_{C} \exp\left (-\alpha x/ b - (1-\alpha) (x-1) /b \right )dx \nonumber\\
    & = \frac{1}{2} \exp \left (\frac{\alpha-1}{b}\right ) + \frac{1}{2(1-2\alpha)}\exp \left (\frac{\alpha-1}{b} \right ) \left [ \exp \left (\frac{1-2\alpha}{b} C\right ) -1\right ] + \frac{1}{2} \exp \left (\frac{1-\alpha-C}{b}\right ) \nonumber \\
    & = \frac{\alpha}{2\alpha-1} \exp \left (\frac{\alpha-1}{b}\right )  + \frac{1}{2}\exp\left (\frac{1-C-\alpha}{b}\right) + \frac{1}{2(1-2\alpha)}\exp \left (\frac{\left ( 1-2C\right )\alpha + \left ( C-1 \right )}{b}\right )
\end{align}

\normalsize
With the definition of Rényi DP of two continuous distributions, we have

\small
\begin{align}
    & D_\alpha(P||Q) = \frac{1}{\alpha-1} \log \int^{\infty}_{-\infty} p(x)^\alpha q(x)^{1-\alpha} dx \nonumber \\
    & = \frac{1}{\alpha-1} \log \left \{ \frac{\alpha}{2\alpha-1} \exp \left (\frac{\alpha-1}{b}\right )  + \frac{1}{2}\exp\left (\frac{1-C-\alpha}{b}\right) \right.\nonumber \\
    & + \left. \frac{1}{2(1-2\alpha)}\exp \left (\frac{\left ( 1-2C\right )\alpha + \left ( C-1 \right )}{b}\right )\right \}
\end{align}
\normalsize
when we using the substitution $\exp(t/b)\to M(t)$, we can have the $R\acute{e}nyi$-DP for LMO Noise as claimed. Specifically, when we have LMO mechanism $P$ and $Q$, we have
\begin{align}
    \epsilon^{\text{LMO-DP}}_\alpha = \frac{1}{\alpha-1} \log \left \{ \frac{\alpha}{2\alpha-1} M_{\text{LMO}} \left (\alpha-1\right )  + \frac{1}{2}M_{\text{LMO}}\left (1-C-\alpha\right)  \right. \nonumber \\
    + \left. \frac{1}{2(1-2\alpha)}M_{\text{LMO}} \left (\left ( 1-2C\right )\alpha + \left ( C-1 \right )\right )\right \}
\end{align}
\end{proof}

\section{Quantifying the LMO Search Space: A Proposed Approach}
\label{sec:proposed-approach} 

In our quest to understand the $\mathcal{S}_\text{LMO}$ space, we introduce the ``Comprehensiveness Explorer''. This algorithm probes the $\mathcal{S}_\text{LMO}$ space's diversity and capability to represent various probability density functions.
\subsection{The Key Tool: Simulated Search Space}
At the core of the Comprehensiveness Explorer is the simulated search space, constructed using the Multinomial Probability Density Function (PDF). It enables us to quantitatively assess the richness of the $\mathcal{S}_\text{LMO}$ space. The Multinomial PDF, denoted as $\text{Multinomial}(N, k, p=1/k)$, models the scenario of distributing a unit probability mass uniformly among $k$ classes, with precision controlled by parameter $q$. We sample from this space to capture probability distributions under varying quantization rates $q$ and domain sizes $k$.

\subsubsection{Quantifying Comprehensiveness}\label{sec:quantify}

Algorithm \ref{alg:quantify-lmo-space} (Quantification of LMO Search Space) is employed to assess the comprehensiveness of this search space in comparison to a universally simulated space, introducing a novel quantification test. We utilize three distance metrics (both probabilistic and deterministic): KL divergence, $\ell_2$, and EMD metrics to measure the distance between these two spaces. Our results demonstrate that, for any given noise `n', there exists an LMO noise that remains close to `n'. Please note for the EMD metric, we experience a small divergence (scaled by $10^{-3}$) when the domain of noise increases which necessitates a smaller quantization rate to achieve even better results.

\subsubsection{Adaptive LMO Sample Generation}
To ensure LMO samples resemble the simulated space, we adjust LMO distribution parameters based on simulated space statistics. Specifically, results generated in Figure~\ref{nearoptim} are for $q$ and $k$ finetuned using the following steps: 1) estimate $\mu_{\text{sim}}$ and $\sigma_{\text{sim}}^2$, 2) adjust $\mu_{\text{LMO}}$ to match $\mu_{\text{sim}}$, 3) adjust $\sigma_{\text{LMO}}$ based on $\sigma_{\text{sim}}^2$ and 4) generate LMO samples with adjusted parameters. This adaptive approach ensures LMO samples match the simulated space's statistical properties.

\begin{algorithm}[tb]
\caption{Quantification of LMO Search Space $\mathcal{S}_\text{LMO}$}
\label{alg:quantify-lmo-space}
\begin{algorithmic}[1]
\REQUIRE $\mathcal{Q}$: quantization values; $\mathcal{K}$: domain size; $M$: sampling times; $\mathcal{S}$: searched ranges of distribution parameters and their weights; $\ell$: distance metric  (e.g., KL-divergence, $\ell_2$ distance) \color{blue}\COMMENT{$\mathcal{Q}=\left[10^{-1},\cdots, 10^{-q}\right]$; $q$, $\mathcal{K}$, $N$, $M$ are large numbers.}\color{black}

\ENSURE Distance $\mathcal{D}$ between the universal search space and $\mathcal{S}_\text{LMO}$ space 

\color{blue}\COMMENT{Simulating universal search space}\color{black}

  \FOR{$q$ in $\mathcal{Q}$}
    \FOR{$k$ in $\mathcal{K}$}
    \FOR{$i \in [1, M]$}
       \STATE $N =1/q$, $p = 1/k$      
        \STATE $x_i \sim \text{Multinomial}(N, k, p)$ 
    \ENDFOR       
    \FOR[\color{blue}Generating LMO search space $\mathcal{S}_\text{LMO}$ \color{black}]{$j \in [1, M]$}       
         \STATE $y_j \sim \text{Lap}(0, b(\mathcal{S}), k)$
    \ENDFOR

    \STATE $\mathcal{D}_{q,k} = \ell (x, y)/M$
    \ENDFOR
    \ENDFOR

\STATE \textbf{Return} $\mathcal{D}$
\end{algorithmic}
\end{algorithm}

\begin{figure}[!tbh] \centering	\includegraphics[width=0.8\linewidth, viewport=60 330 555 490,clip]{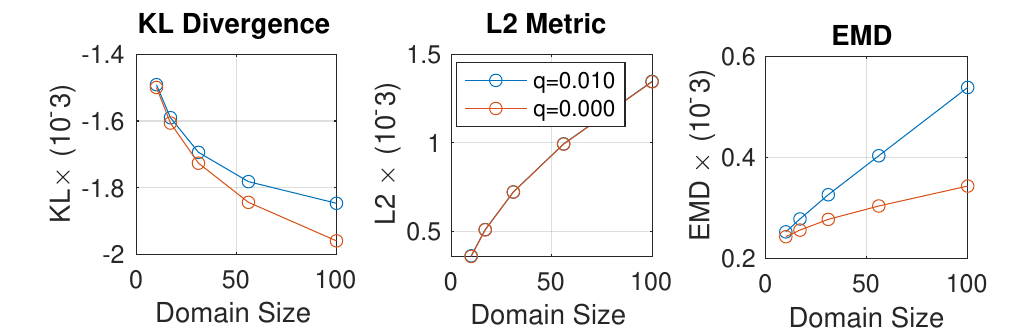} \caption[]{Generated noise using the $\mathcal{S}_\text{LMO}$ space exhibits a remarkable level of comprehensiveness concerning three distance metrics. These results are derived from the quantification of the $\mathcal{S}_\text{LMO}$ space, as outlined in Algorithm~\ref{alg:quantify-lmo-space}.}
\label{nearoptim}
\end{figure}

\section{Algorithm}
\label{sec:algmapp}

We illustrate the algorithm to return the optimal noise parameter in Algorithm \ref{alg:rdp-optimization}.

\begin{algorithm}[tb]
\caption{R\'enyi Accountant Optimization $F_1$}
\label{alg:rdp-optimization}
\begin{algorithmic}[1]
\REQUIRE $A_1$: privacy budget $\left \{\epsilon, \delta \right \}$, $A_3$\color{blue}\COMMENT{$A_3$=\{(\text{Gamma}, \text{Exponential}, \text{Uniform}), $\alpha_{max}$, $\mathcal{S}$\}}\color{black}

\ENSURE LMO parameters $\mathcal{S}_{\text{opt}}$\color{blue}\COMMENT{Best LMO mechanism}\color{black}

\IF[\color{blue}Step 1: Defining the MGF of distributions\color{black}]{\text{Gamma} $\in A_3$}
    \STATE $M_{Y_1}(t)\gets {\displaystyle (1-t\theta )^{-k},~t<{\tfrac {1}{\theta }}}$
\ELSE
   \STATE $M_{Y_1}(t)\gets 0$
\ENDIF

\IF{\text{Exponential} $\in A_3$}
  \STATE $M_{Y_2}(t)\gets {\displaystyle \left(1-t\lambda ^{-1}\right)^{-1},~t<\lambda }$
\ELSE
  \STATE $M_{Y_2}2(t)\gets 0$
\ENDIF

\IF{\text{Uniform} $\in A_3$}
 \STATE $M_{Y_3}(t)\gets {\displaystyle {\frac {e^{tb}-e^{ta}}{t(b-a)}}}$
\ELSE
 \STATE $M_{Y_3}(t)\gets 0$
\ENDIF

\STATE $M_Y(t) \gets a_1 \cdot M_{Y_1}(t) + a_2 \cdot M_{Y_1}(t) + a_3 \cdot M_{Y_3}(t)$

\FOR[\color{blue}Step 2: Finding the optimal $S_{opt}$ by grid search\color{black}]{$S \in \mathcal{S}$}   
\STATE $\alpha_1, \alpha_2, \alpha_3, k, \theta, \lambda, b, a = S$ \color{blue} \COMMENT{Theorem F.4 in \cite{mohammady2020r2dp}}\color{black}

\FOR{$\alpha \in [2, \alpha_{max}]$}              
     
    \STATE $\epsilon_{\text{R\'enyi}, \alpha}= \frac{1}{\alpha-1} \log \left [ \frac{\alpha M_Y(\alpha-1)+(\alpha-1)M_Y(-\alpha)}{2\alpha-1}\right ] $ 
    
    \STATE $\epsilon'$.append($(\mathcal{H}(\alpha, \epsilon_{\text{R\'enyi}, \alpha}, \delta)$) \color{blue}\COMMENT{$\mathcal{H}$: Converting R\'enyi-DP to DP}\color{black}
\ENDFOR
    
\IF{max($\epsilon'$) $< \epsilon_0$}
    \STATE $S_{\text{opt}} \gets S$
\ENDIF

\ENDFOR

\STATE \textbf{Return} $S_{\text{opt}}$
\end{algorithmic}
\end{algorithm}

\section{Additional Experiments}

\subsection{Experimental Setting}\label{setting}
We conducted our experiments on two servers: Intel(R) Xeon(R) Platinum 8336C CPU @ 2.30GHz, 2T RAM, and 8$\times$NVIDIA A100 SXM4 80G GPUs, and AMD Ryzen Threadripper PRO 5975WX 32-Cores CPUs, 500G RAM, 3$\times$NVIDIA Quadro RTX A6000 48GB GPUs.

\textbf{Hyperparameter Setting.} LMO-DP ensures $\epsilon$-DP with $\delta=10^{-10}$, and thus we set $\delta=10^{-10}$ for the baseline DP-SGD \cite{li2021large} for relatively fair comparisons. We first select different privacy parameters at each iteration of the fine-tuning: $\{0.3, 0.7, 2, 3\}$, and the total privacy loss $\epsilon$ will be derived with composition. 
The optimal weight values for three PDFs $a_1$, $a_2$, and $a_3$ are in the range $[0.1, 0.9]$ (varying on different $\epsilon$). Additionally, we set the clipping threshold for gradients as $1$, across all methods. We employ a batch size of $2048$ and $6$ training epochs. Consequently, the sampling rate for the training data is calculated as $\frac{2048}{|D|}$, where $|D|$ denotes the size of the dataset. 
\subsection{LMO noise v.s. Gaussian noise}
\label{comapre_noises}

\begin{figure*}[htbp]
\begin{minipage}[t]{0.24\textwidth}
\centering
\includegraphics[scale=0.22]{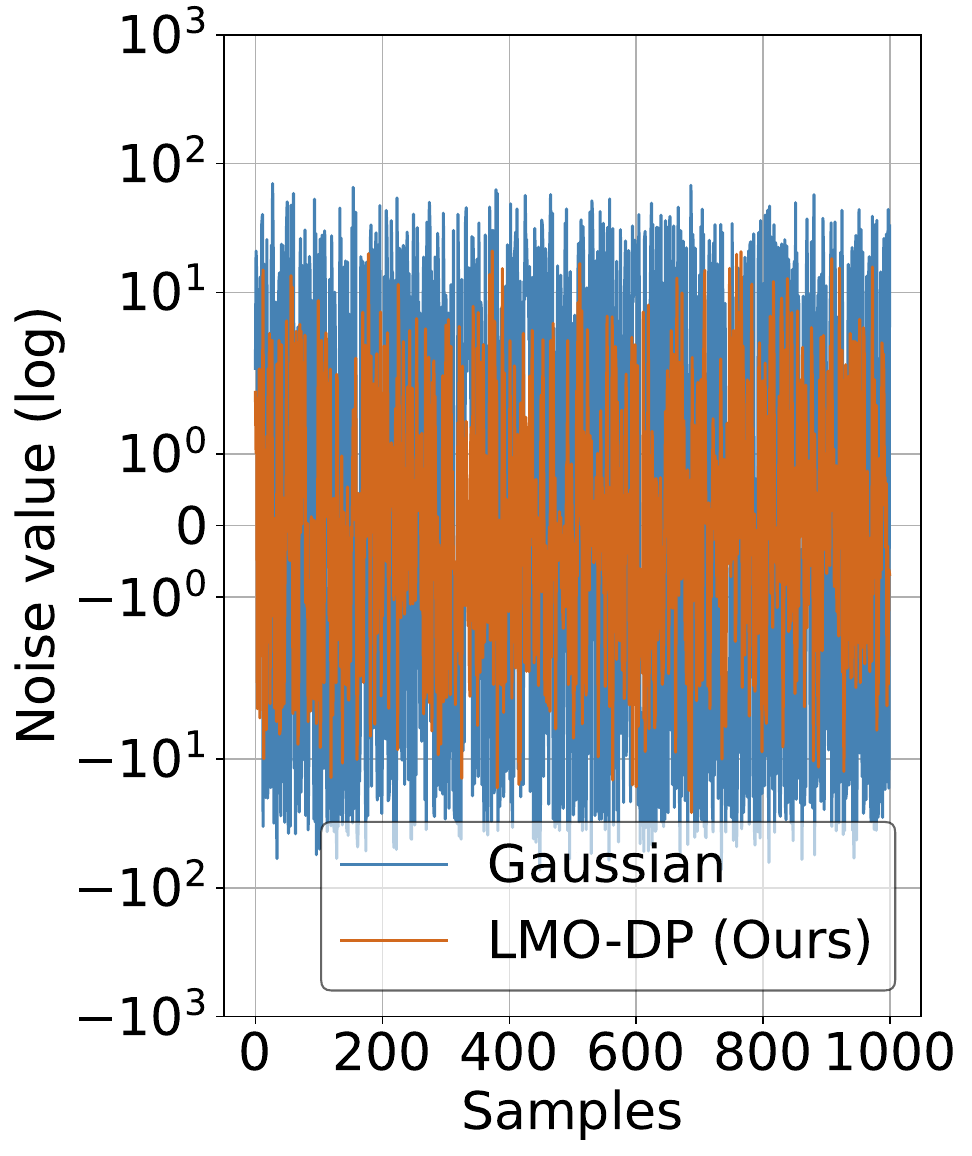}
\text{(a) $\epsilon=0.3$}
\end{minipage}
\begin{minipage}[t]{0.24\textwidth}
\centering
\includegraphics[scale=0.22]{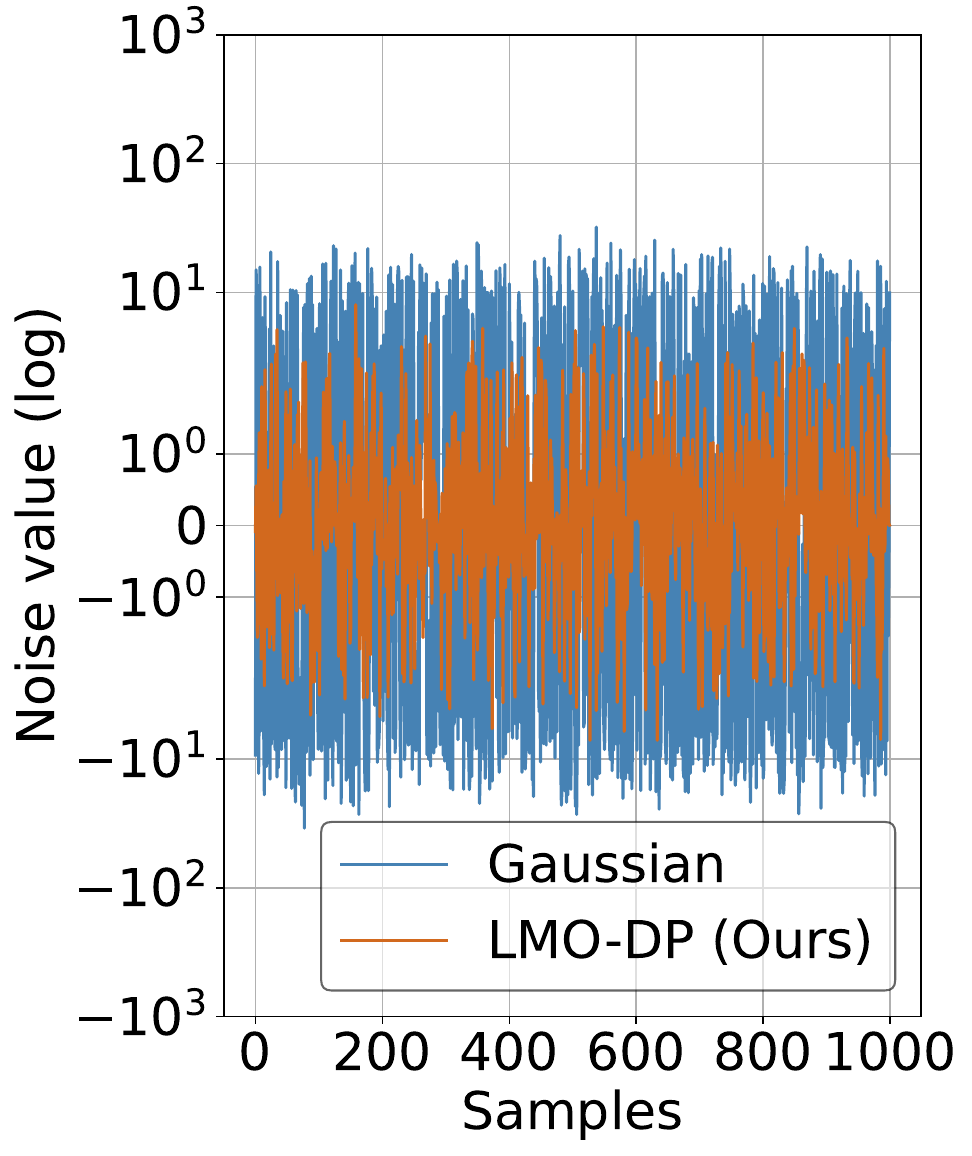}
\text{(b) $\epsilon=0.7$}
\end{minipage}
\begin{minipage}[t]{0.24\textwidth}
\centering
\includegraphics[scale=0.22]{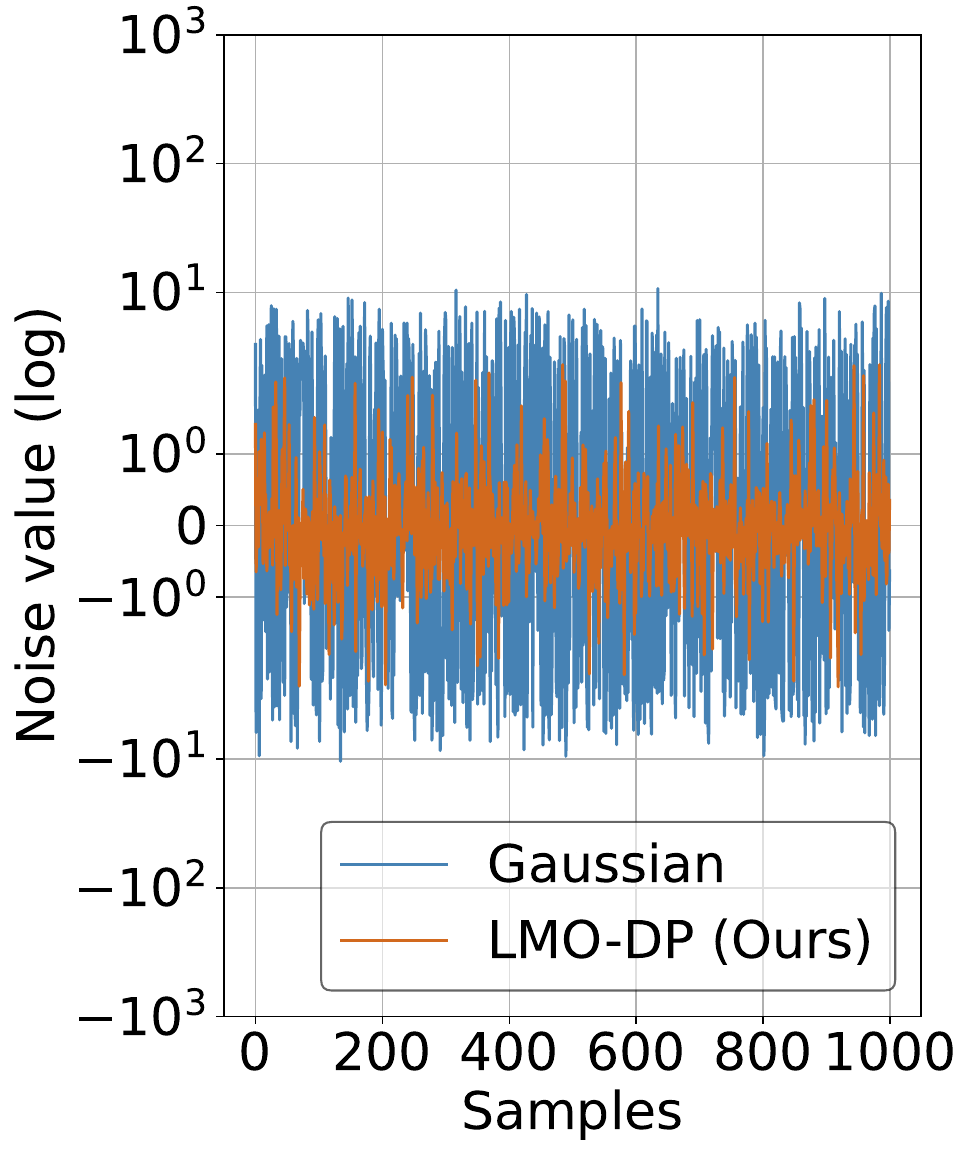}
\text{(c) $\epsilon=2$}
\end{minipage}
\begin{minipage}[t]{0.24\textwidth}
\centering
\includegraphics[scale=0.22]{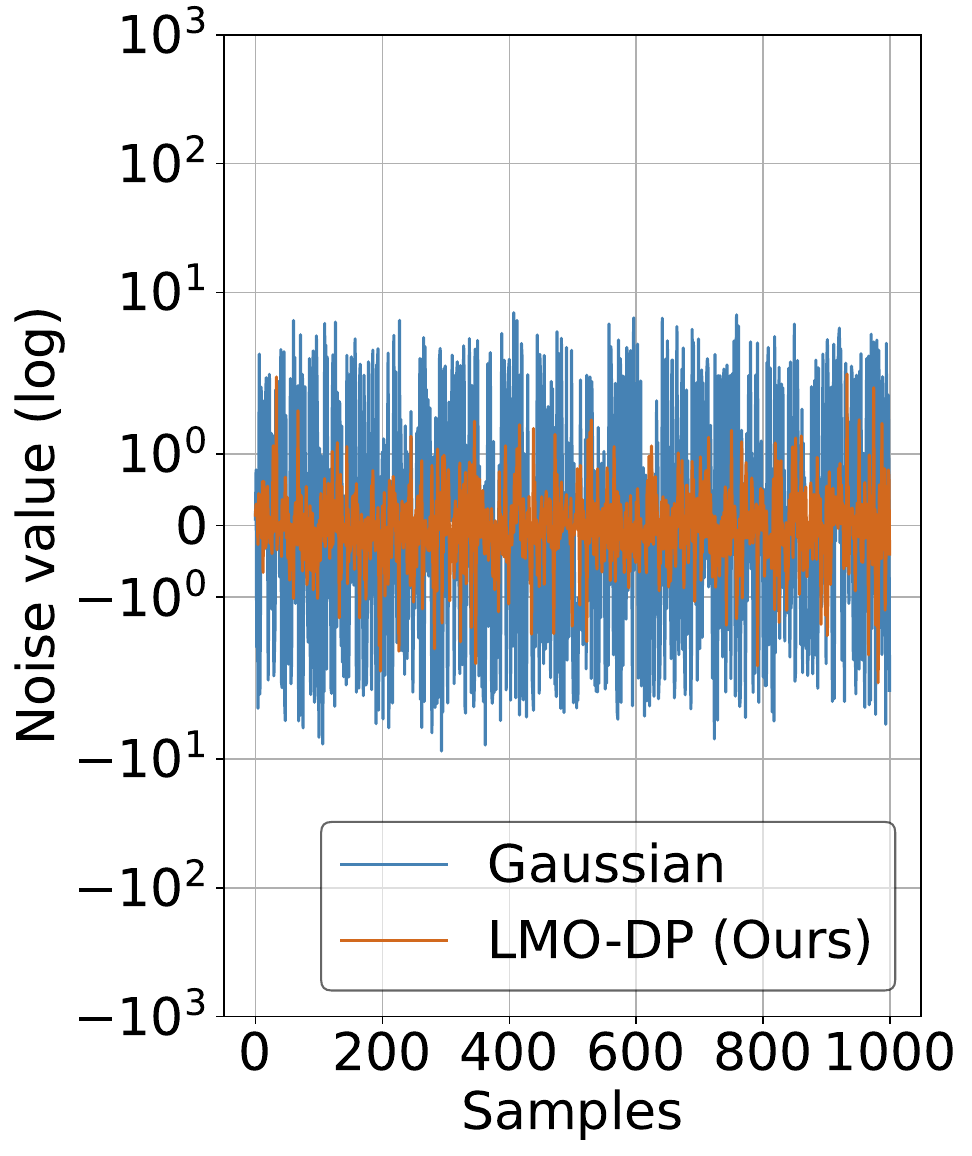}
\text{(d) $\epsilon=3$}
\end{minipage}
\caption{LMO-DP vs Gaussian.
(a) average reduction rate $95.13\%$. (b) average reduction rate $92.19\%$. (c) average reduction rate $87.71\%$. (d) average reduction rate $87.31\%$. The results demonstrate that the LMO-DP noise significantly outperforms the Gaussian noise; LMO-DP performs even better for smaller $\epsilon$ since the average reduction rate slightly declines as $\epsilon$ increases.}\vspace{-0.1in}
\label{fig_compare_noises2}
\end{figure*}

\noindent\textbf{LMO-DP Noise vs Gaussian Noise}. LMO-DP noise shows a significantly smaller amplitude as shown in Figure \ref{fig_compare_noises2} (given the same DP guarantee and shown in logarithmic scale). The reduction rate of LMO-DP noise compared to Gaussian noise is as high 95.13\% for $\epsilon=0.3$, and then slightly reduced to 87.31\% as $\epsilon$ increases to $3$. This also shows that LMO-DP has superior performance for strong privacy guarantees (small $\epsilon$).

In addition, Figure \ref{fig0.0} plots the simple entropy and Figure \ref{fig0.1} plots the variance of the Gaussian and LMO-DP noises. First, we generated the Gaussian noise and LMO-DP noises which have the exact privacy cost for a single noise (same $\epsilon$ value). Then, we computed the histogram and probability density function of these sampled noises. Finally, we plot the simple entropy and variance of the Gaussian and LMO-DP noises for specific $\epsilon$ values. These comparisons exhibit that our LMO-DP noises have lower entropy and variance when compared to Gaussian noises with the same privacy budget. Moreover, in Appendix~\ref{sec:proposed-approach}, we propose a novel approach to quantify the extent of the LMO search space via first simulating the entire search space with different quantization rates and domain sizes and then measuring the worst-case distance of LMO search space to the simulated space, using both probabilistic ``KL-divergence'' and deterministic ``$\ell_2$'' distance metrics.

\begin{figure}[!ht]
	\centering
	\subfigure[Entropy]{
    \includegraphics[angle=0, width=0.45\linewidth]{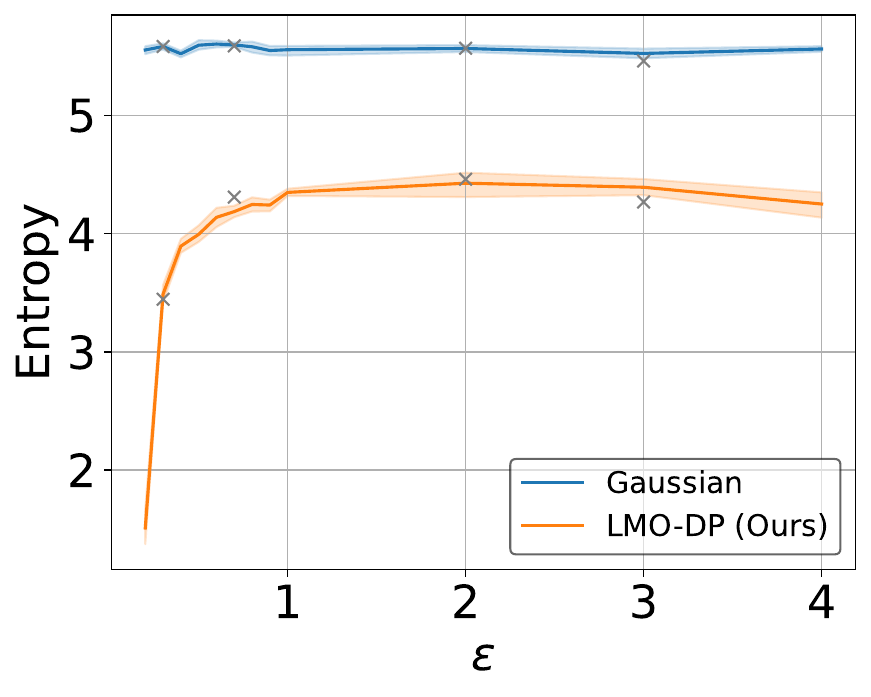}
\label{fig0.0}}
    \subfigure[Variance]{
    \includegraphics[angle=0, width=0.49\linewidth]{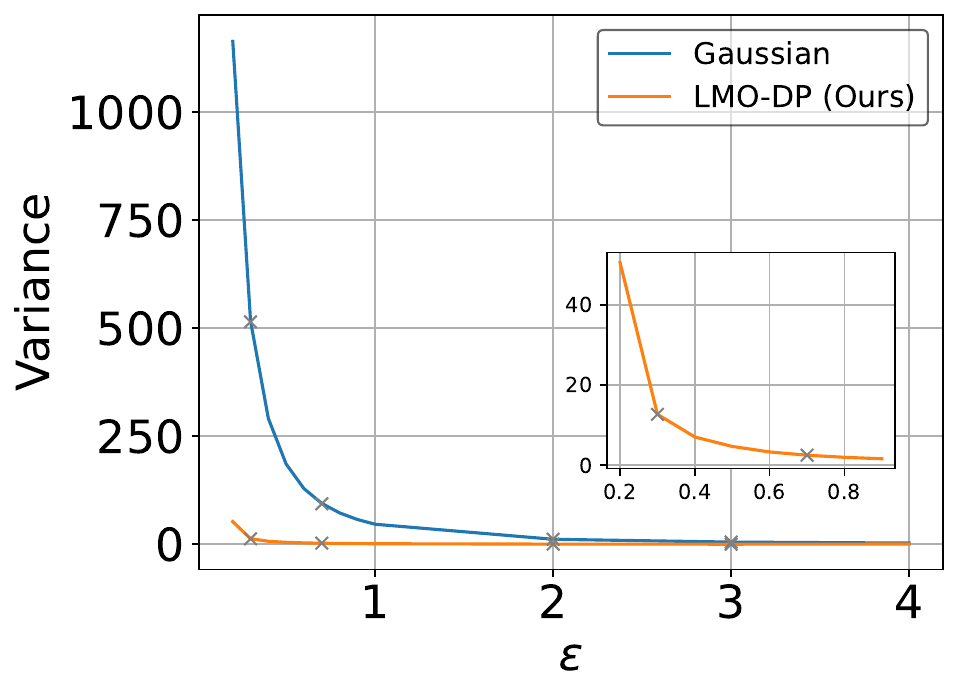}
\label{fig0.1}}\vspace{-0.1in}
\caption[Optional caption for list of figures]
{(a) The simple entropy comparison of LMO-DP and Gaussian noises. (b) The variance comparison of LMO-DP and Gaussian noises.}
\label{fig_entropy_variance}
\end{figure}

\noindent
\textbf{Ablation Study of LMO-DP Noise}. Since the LMO-DP noise is a Laplace-based two-fold noise (the first-fold is the Laplace distribution while the second-fold is a mixture distribution of three PDFs), we conducted an ablation study for it. Specifically, the ``inverse'' of ``scale parameter'' of the Laplace distribution (1/b) is subject to the linear combination of Gamma, Exponential, and Uniform distributions. Our ablation study in Appendix \ref{sec:ablation} shows that the Uniform distribution (in the second-fold mixture distribution) contributes most to the sub-optimal performance of LMO-DP noise.

\subsection{On Quality of LMO-DP Noise during Fine-tuning}
\label{sec:vsguassian}
Figure \ref{base_trainingprocess} and \ref{fig:large_trainingprocess} show the fine-tuning process of sentence classification tasks for different models with different privacy parameters.

\begin{figure*}[!ht]
\begin{minipage}[t]{0.245\textwidth}
\centering
\includegraphics[width=\textwidth]{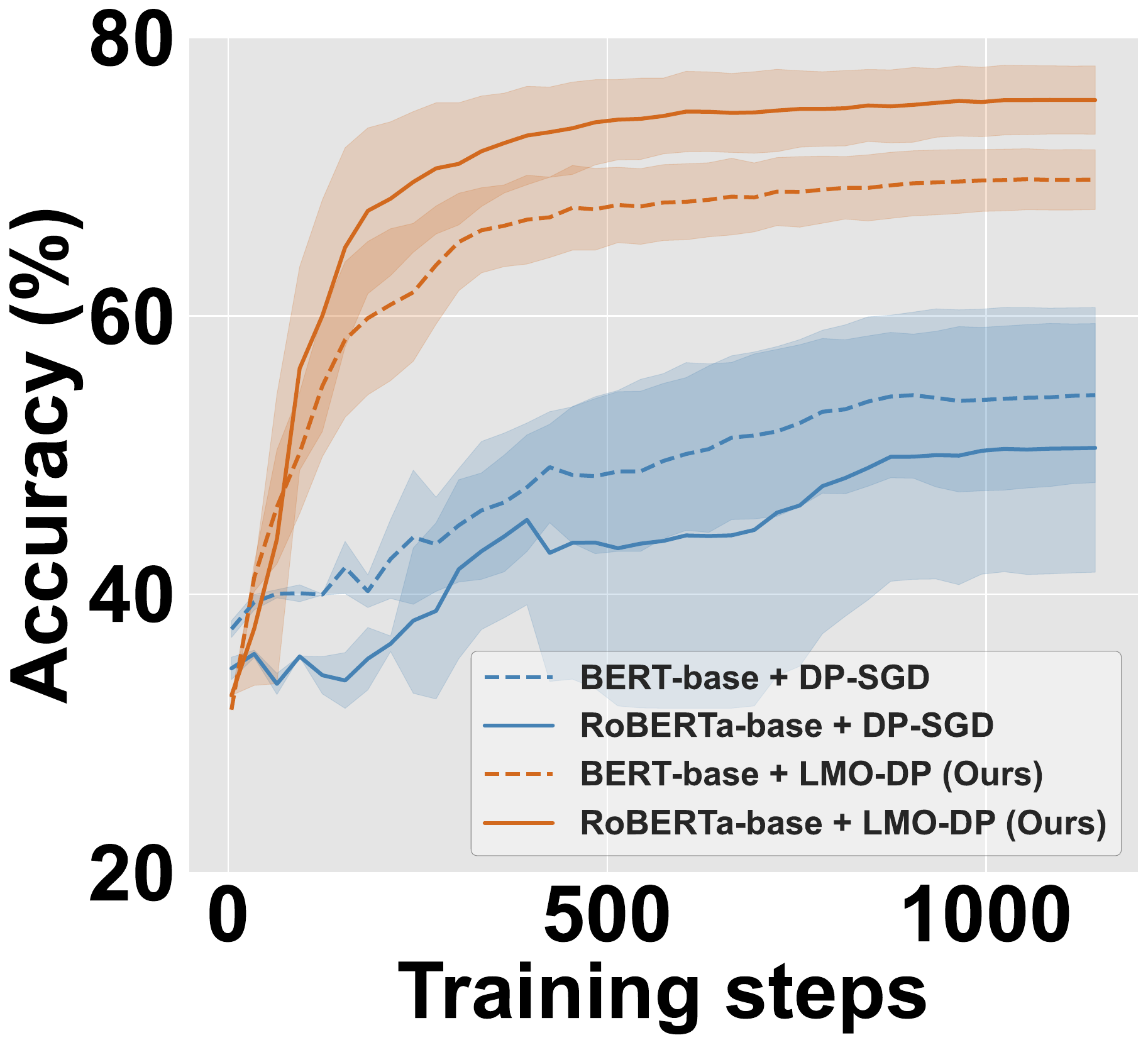}
\text{(a)}
\label{fig1_1}
\end{minipage}
\begin{minipage}[t]{0.245\textwidth}
\centering
\includegraphics[width=\textwidth]{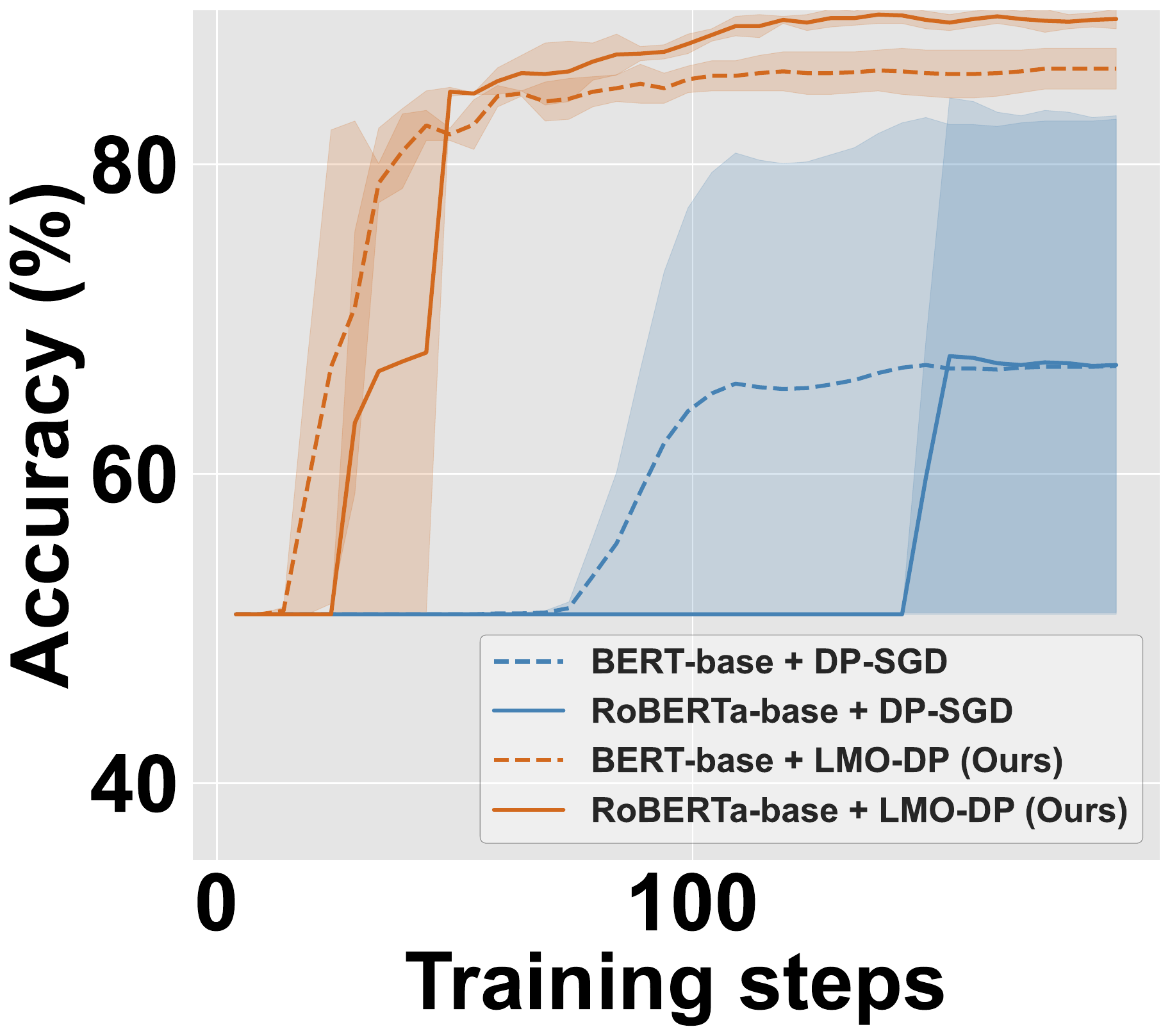}
\text{(b)}
\end{minipage}
\begin{minipage}[t]{0.245\textwidth}
\centering
\includegraphics[width=\textwidth]{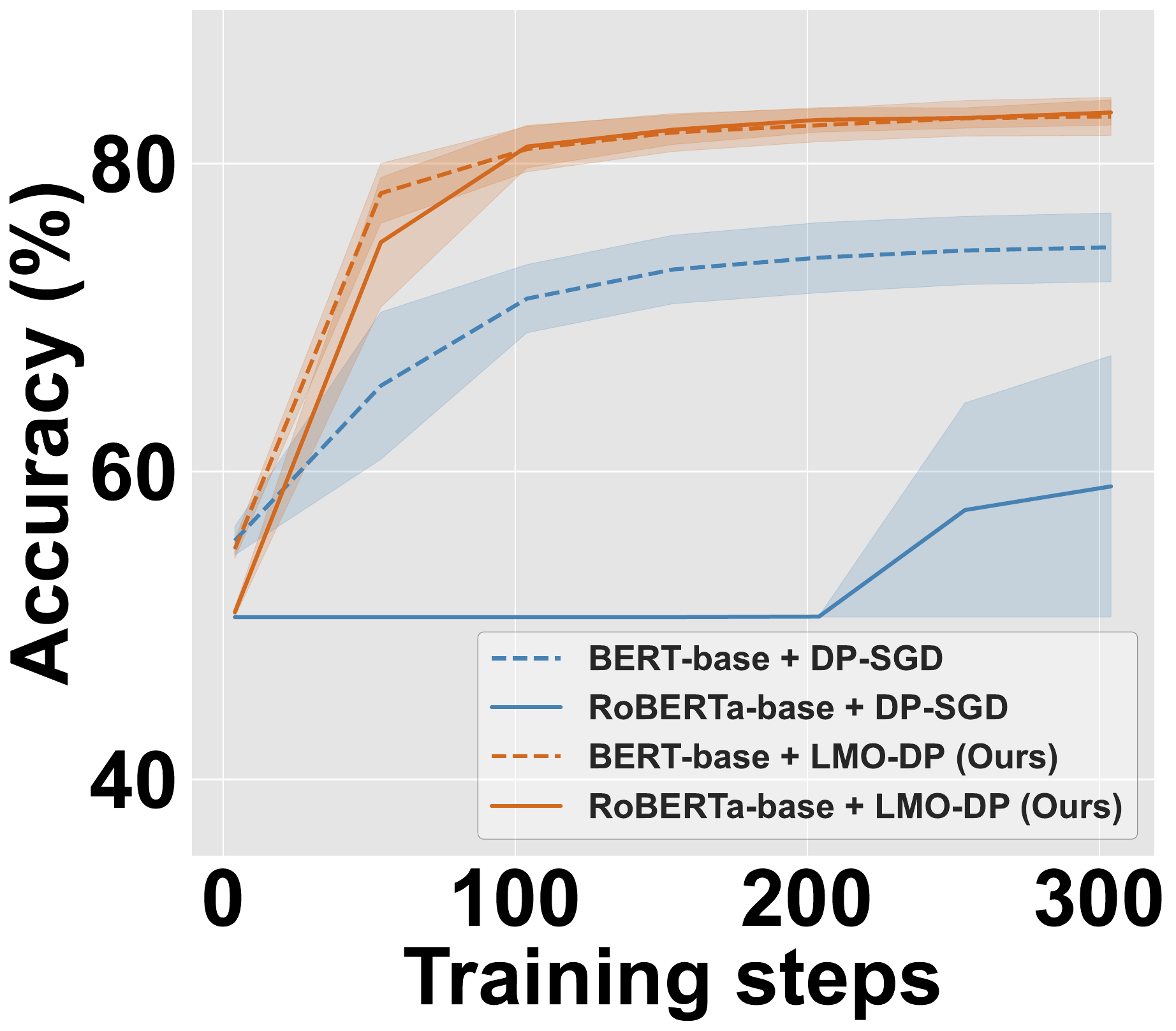}
\text{(c)}
\end{minipage}
\begin{minipage}[t]{0.245\textwidth}
\centering
\includegraphics[width=\textwidth]{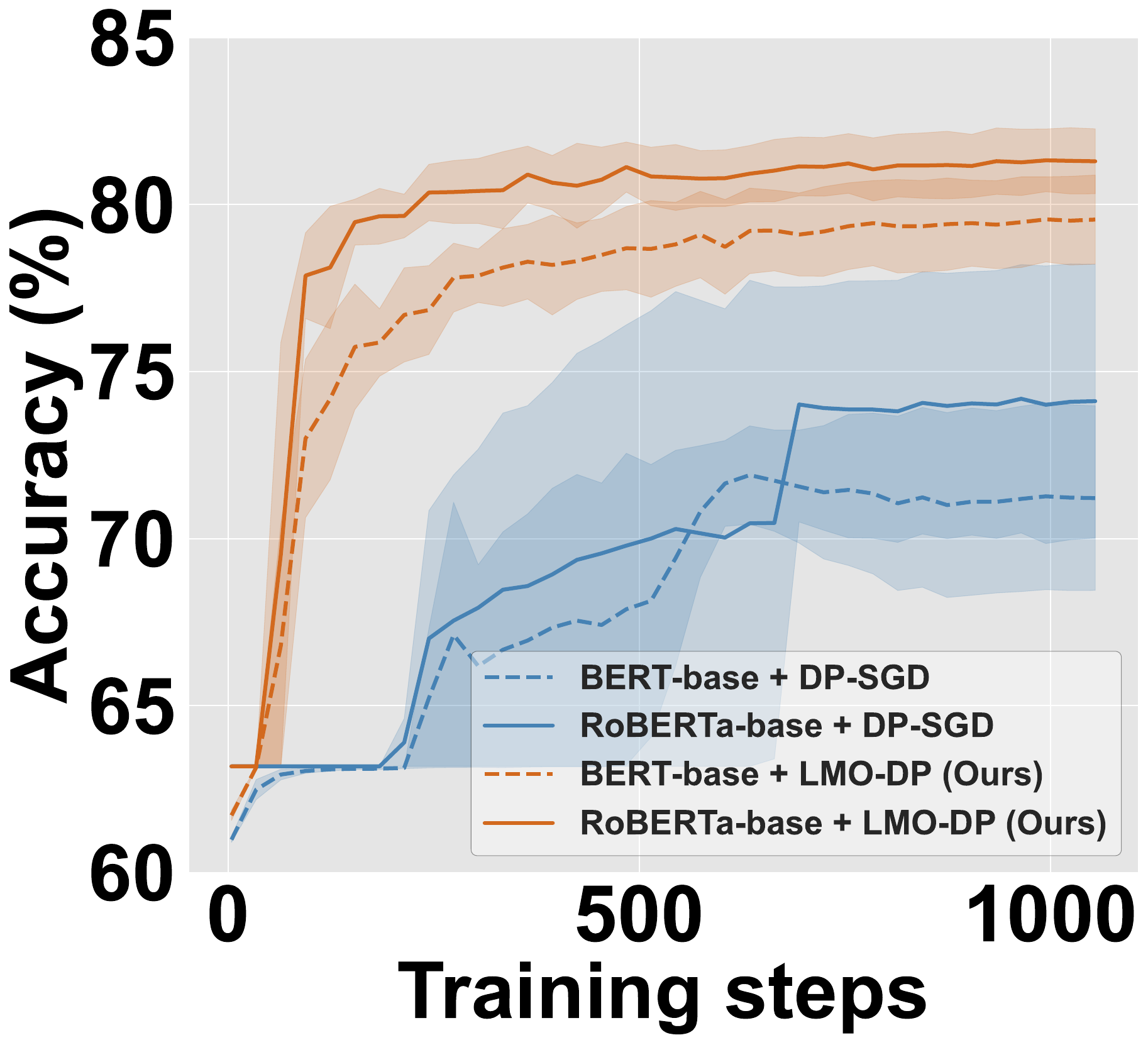}
\text{(d)}
\end{minipage}
\vspace{-0.1in}
\caption{The fine-tuning process of sentence classification task for BERT-base and RoBERTa-base (using 100M parameters) with small privacy budget ($\epsilon$=0.2 or $\epsilon$=0.3). (a) MNLI-m dataset; (b) SST-2 dataset; (c) QNLI dataset; (d) QQP dataset.}\vspace{-0.1in}
\label{base_trainingprocess}
\end{figure*}

\begin{figure*}[!ht]
\begin{minipage}[t]{0.245\textwidth}
\centering
\includegraphics[width=\textwidth]{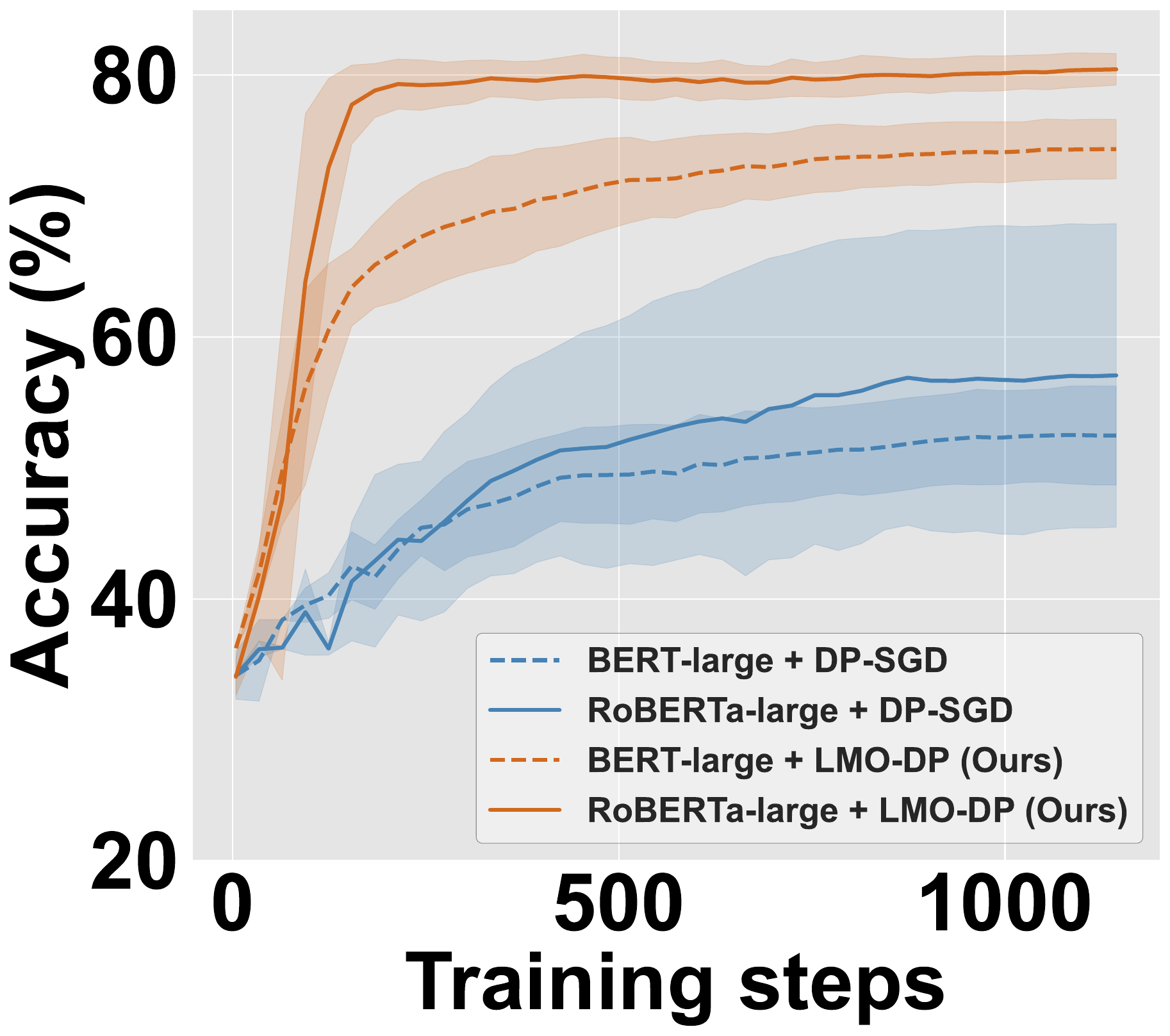}
\text{(a)}
\label{fig1}
\end{minipage}
\begin{minipage}[t]{0.245\textwidth}
\centering
\includegraphics[width=\textwidth]{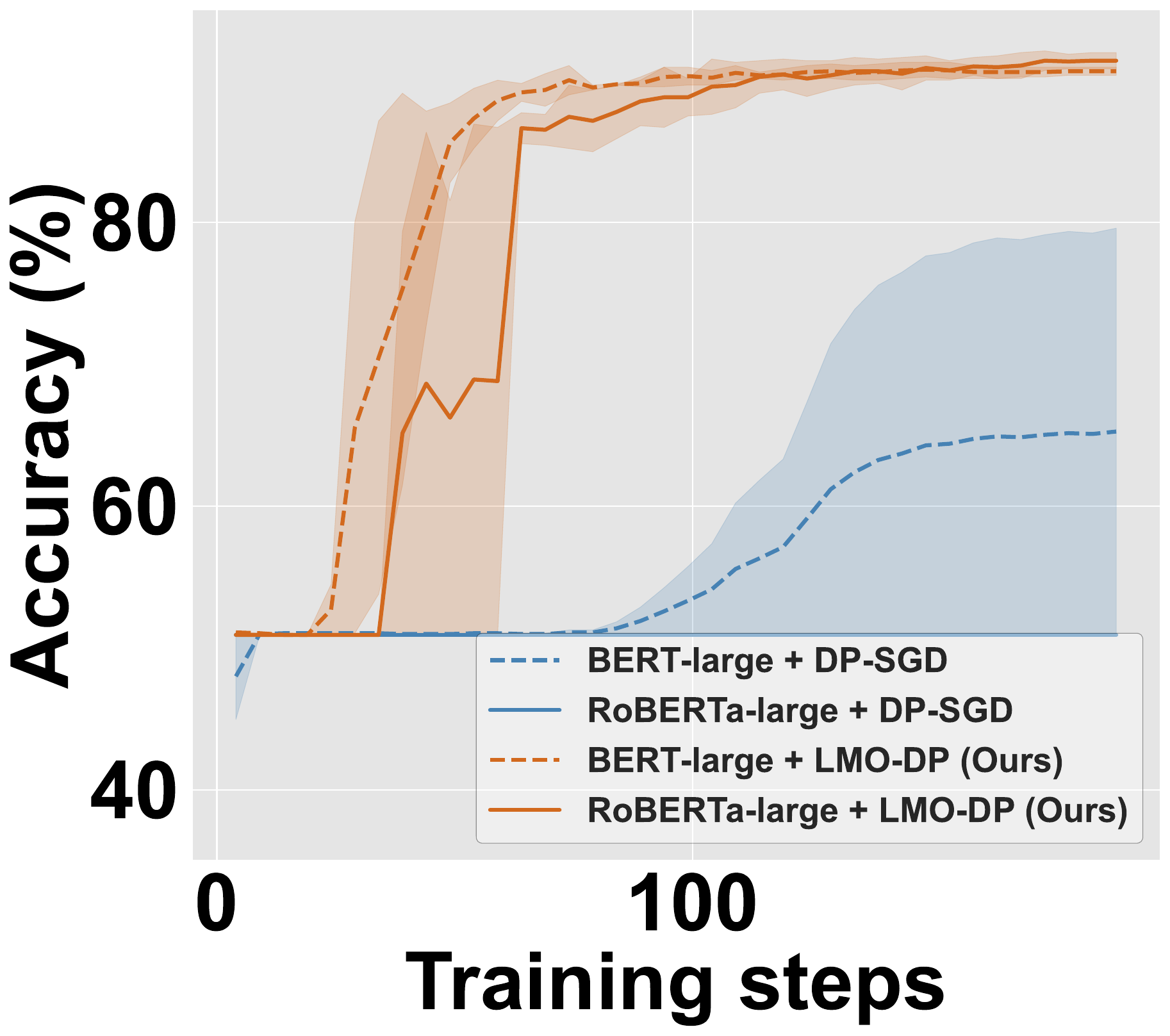}
\text{(b)}
\end{minipage}
\begin{minipage}[t]{0.245\textwidth}
\centering
\includegraphics[width=\textwidth]{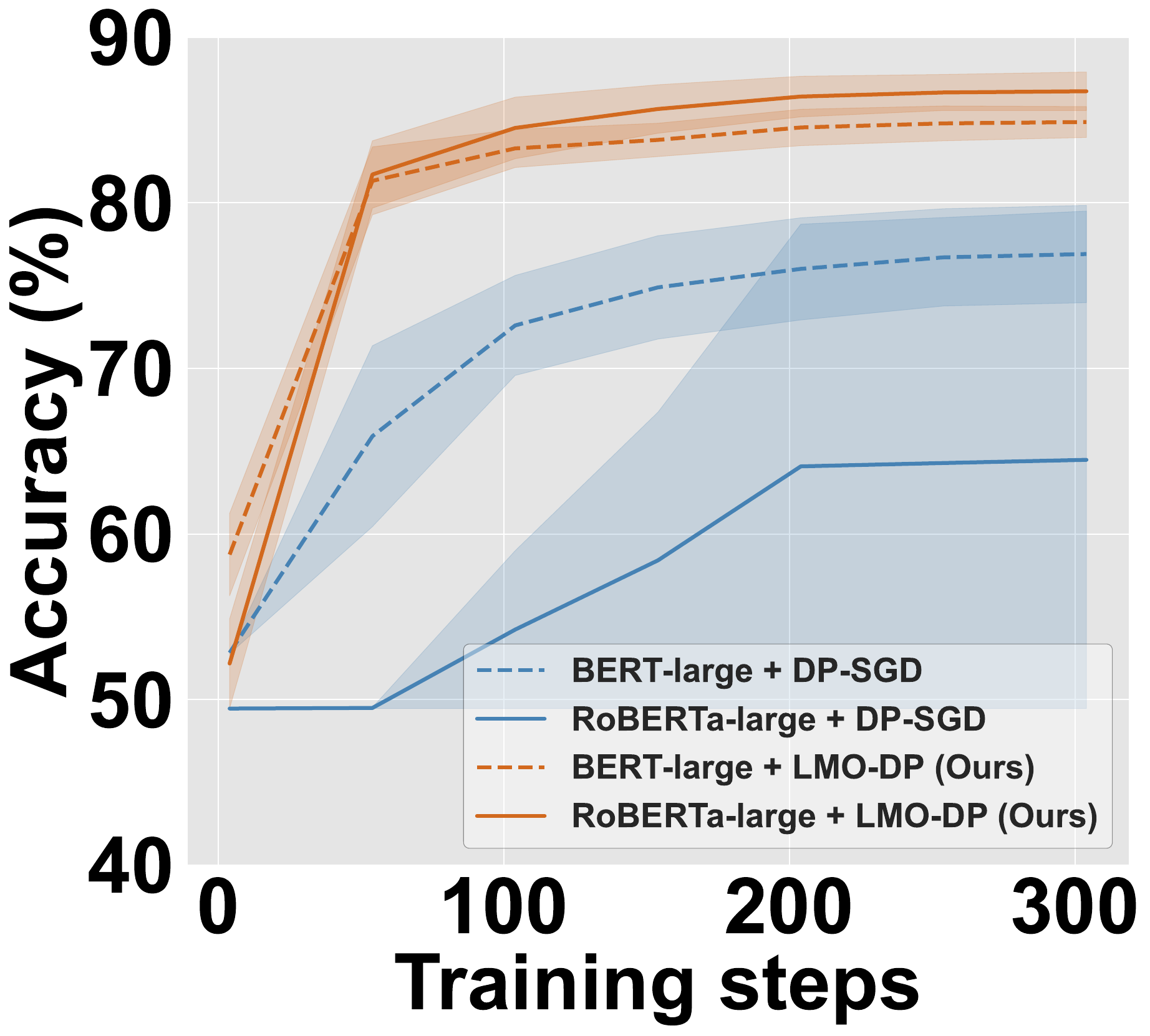}
\text{(c)}
\end{minipage}
\begin{minipage}[t]{0.245\textwidth}
\centering
\includegraphics[width=\textwidth]{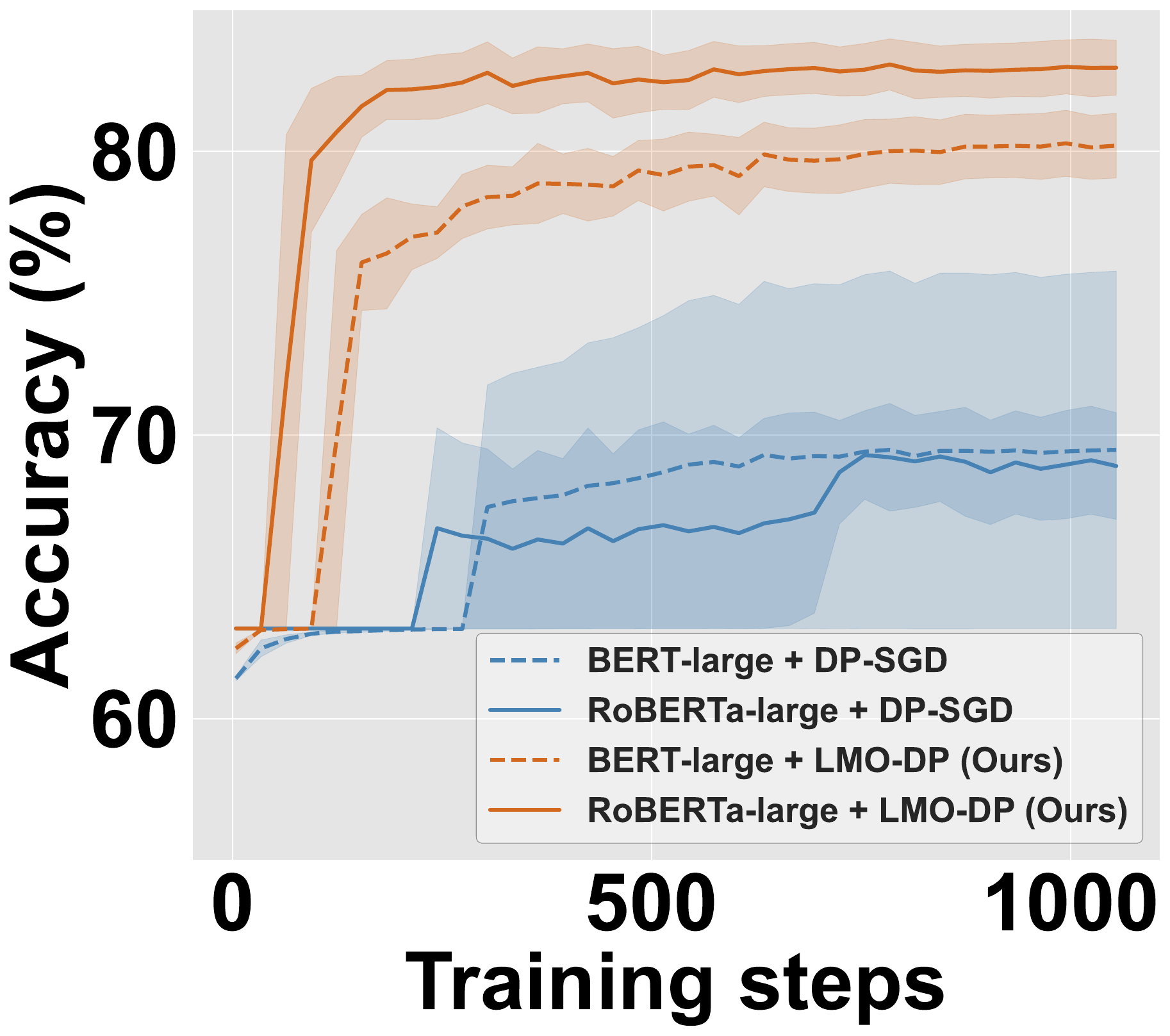}
\text{(d)}
\end{minipage}
\caption{The fine-tuning process of the sentence classification task for BERT-large and RoBERTa-large (using 300M parameters) with small privacy budget ($\epsilon$=0.2 or $\epsilon$=0.3). (a) MNLI-m dataset; (b) SST-2 dataset; (c) QNLI dataset; (d) QQP dataset.}
\label{fig:large_trainingprocess}
\end{figure*}

\newpage

\subsection{Ablation Study}
\label{sec:ablation}

\subsubsection{One Distribution vs. Mixture Distribution in LMO-DP}

\begin{figure*}[!ht]
\begin{minipage}[t]{0.245\textwidth}
\centering
\includegraphics[width=\textwidth]{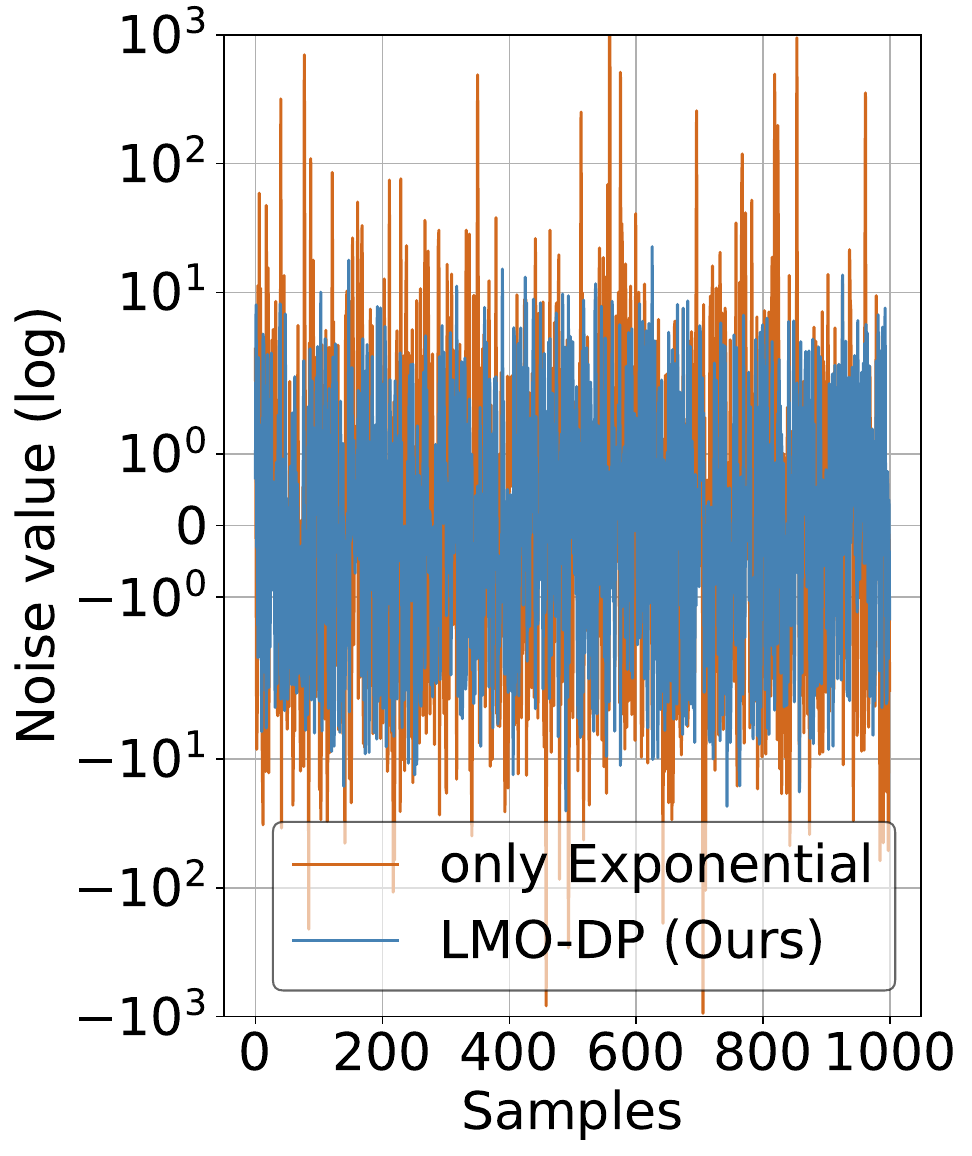}
\text{(a) $\epsilon=0.3$}
\end{minipage}
\begin{minipage}[t]{0.245\textwidth}
\centering
\includegraphics[width=\textwidth]{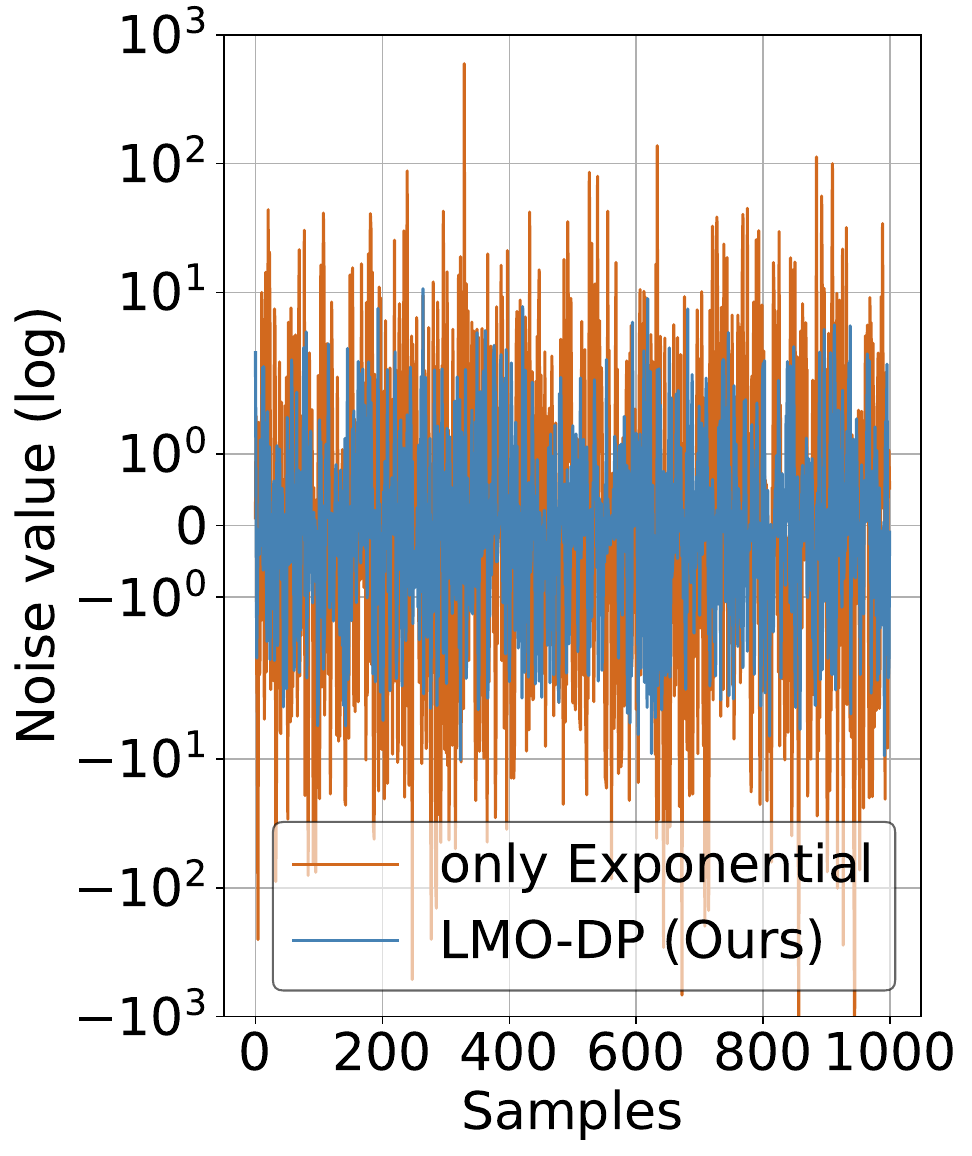}
\text{(b) $\epsilon=0.7$}
\end{minipage}
\label{sec:exp_noise6}
\begin{minipage}[t]{0.245\textwidth}
\centering
\includegraphics[width=\textwidth]{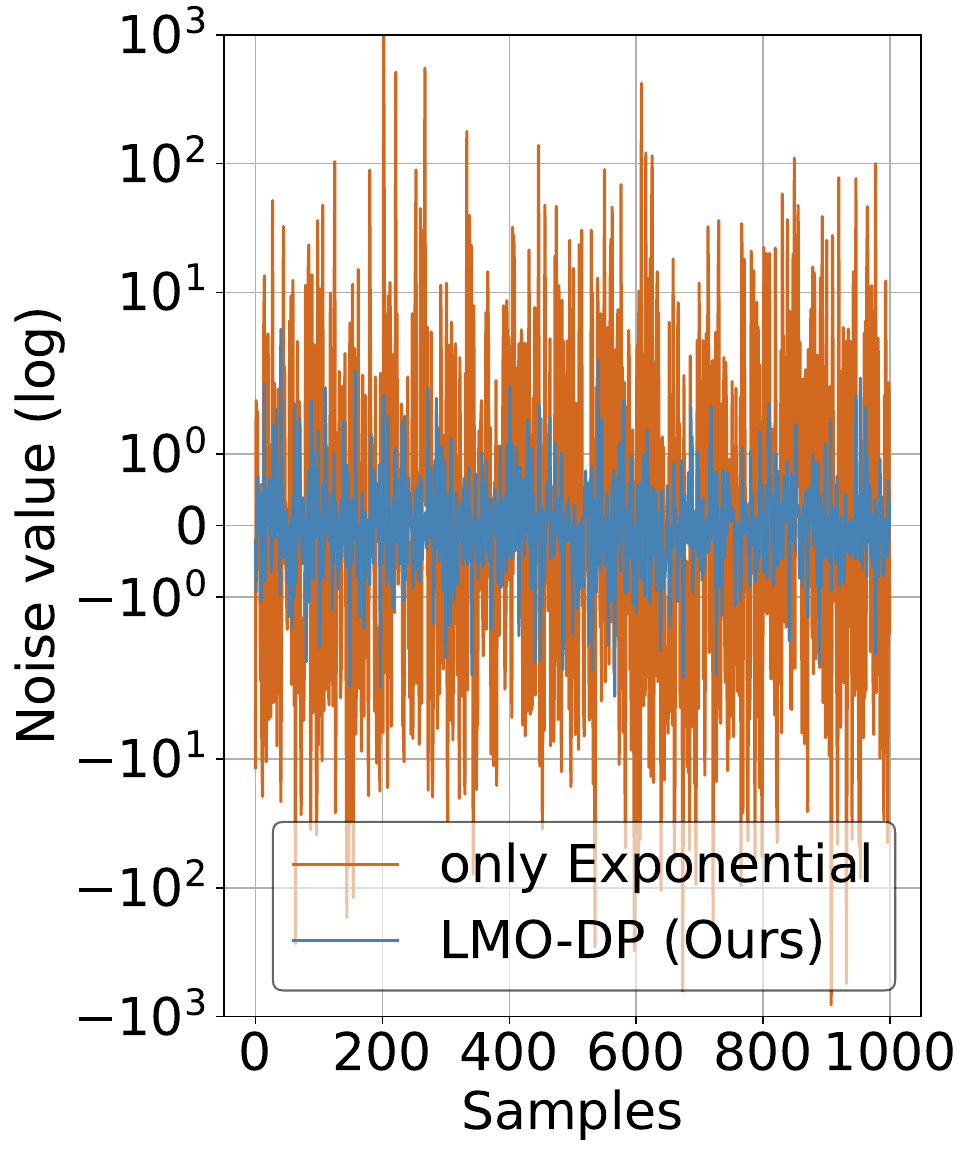}
\text{(c)  $\epsilon=2$}
\end{minipage}
\begin{minipage}[t]{0.245\textwidth}
\centering
\includegraphics[width=\textwidth]{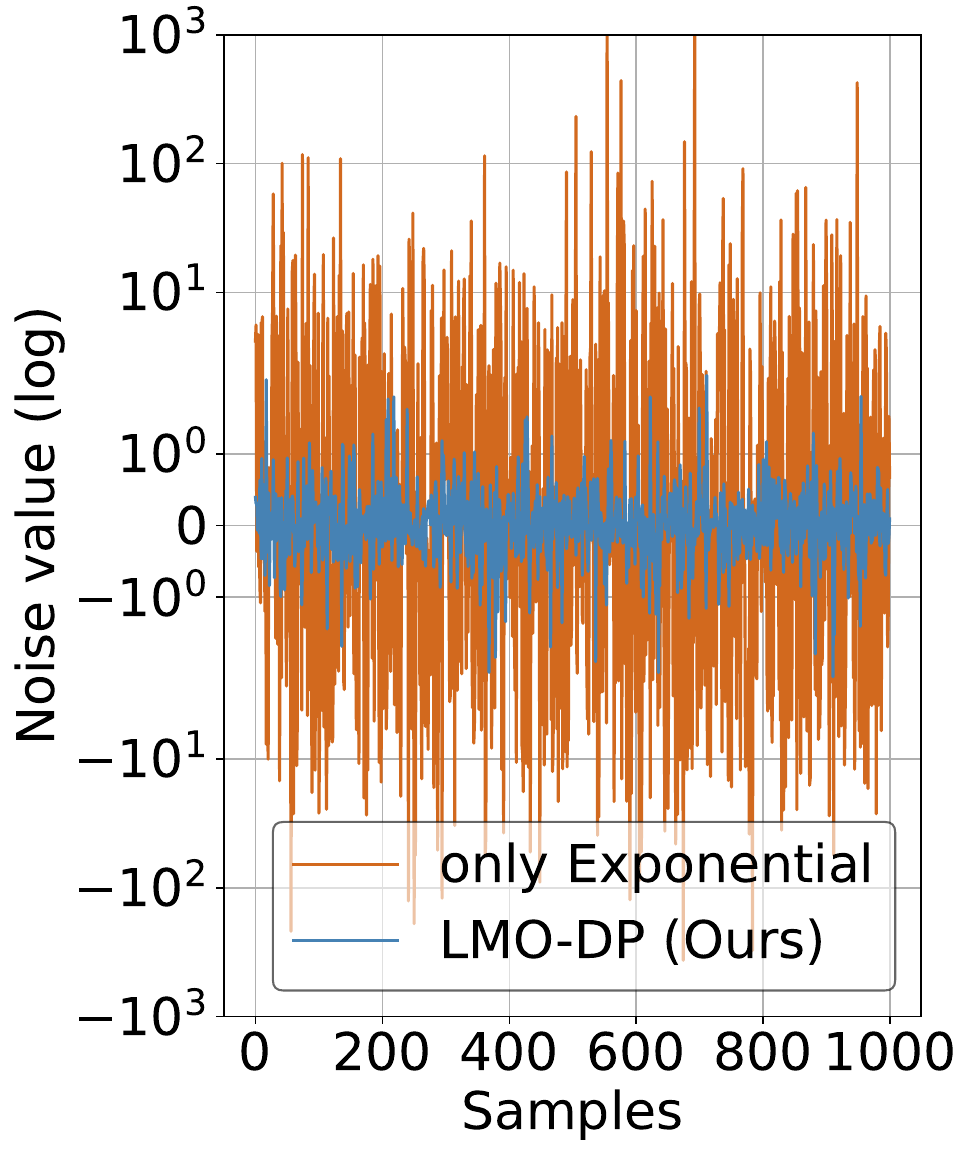}
\text{(d)  $\epsilon=3$}
\end{minipage}
\vspace{-0.1in}
\caption{Exponential distribution vs mixture distribution (with the same remaining setting). The noise generated by the mixture distribution (as the second-fold) in LMO-DP is significantly smaller than that replaces the mixture distribution with the Exponential distribution, especially $\epsilon = 2$ or $3$.}  \vspace{-0.1in}
\label{fig_compare_noises1}
\end{figure*}

\vspace{-0.15in}
\begin{figure*}[!ht]
\begin{minipage}[t]{0.245\textwidth}
\centering
\includegraphics[width=\textwidth]{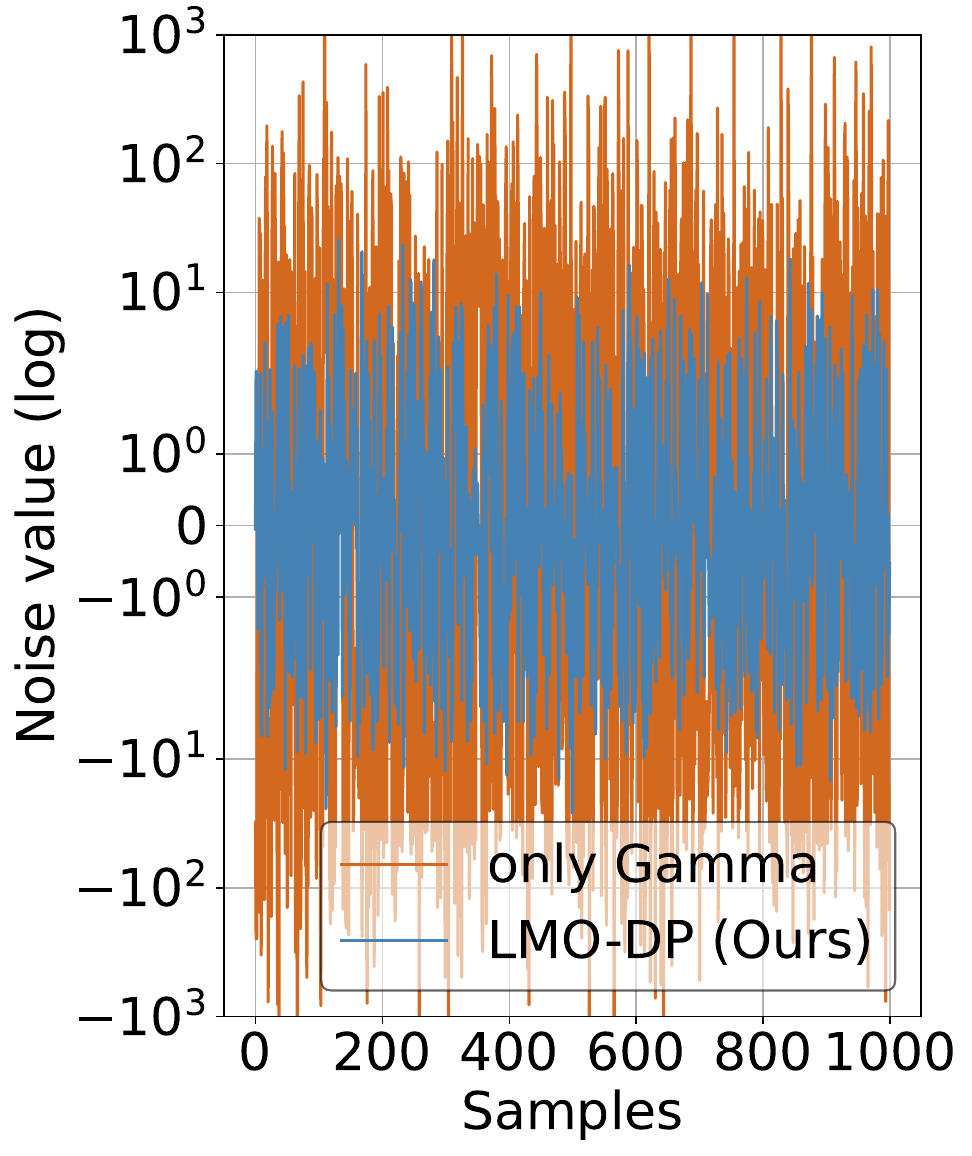}
\text{(a) $\epsilon=0.3$}
\end{minipage}
\begin{minipage}[t]{0.245\textwidth}
\centering
\includegraphics[width=\textwidth]{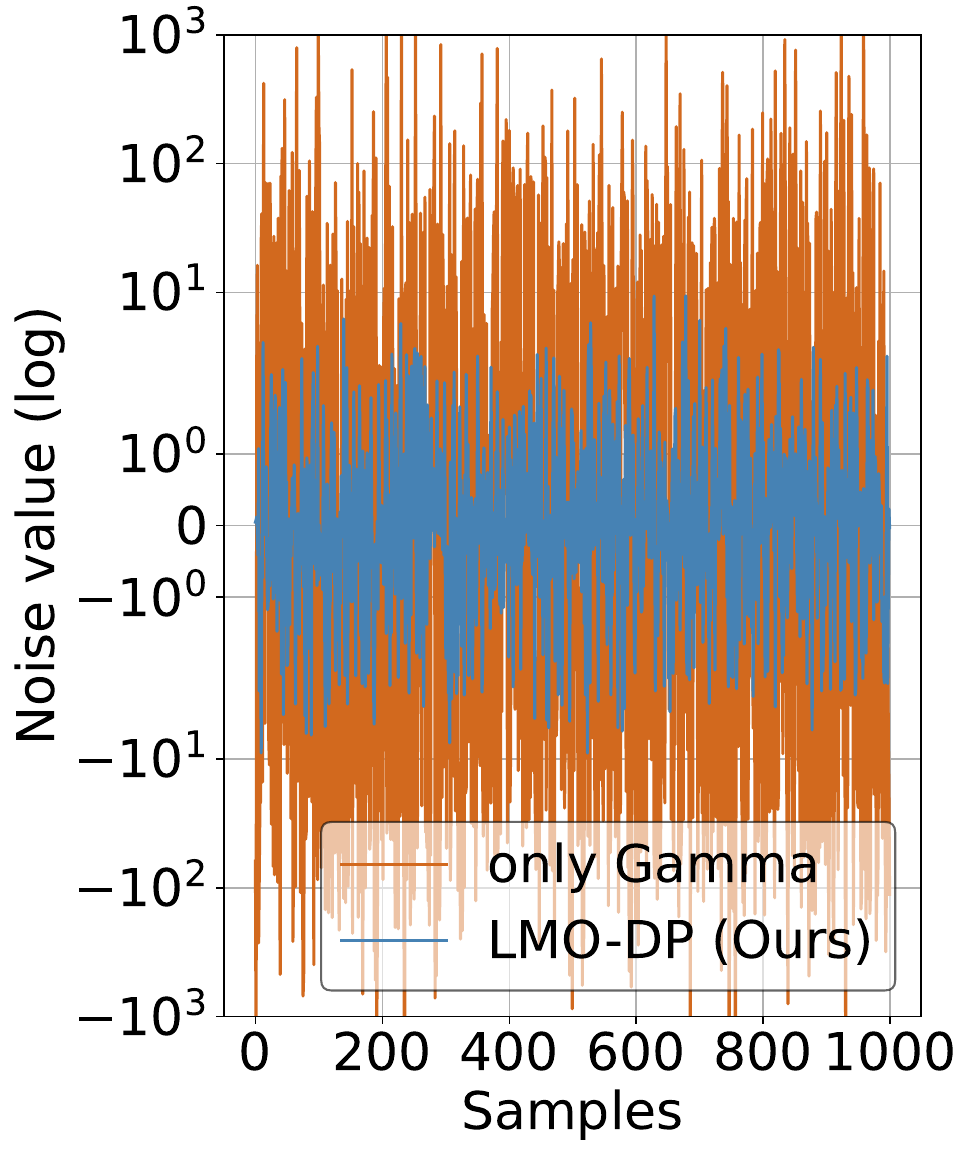}
\text{(b) $\epsilon=0.7$}
\end{minipage}
\label{sec:exp_noise5}
\begin{minipage}[t]{0.245\textwidth}
\centering
\includegraphics[width=\textwidth]{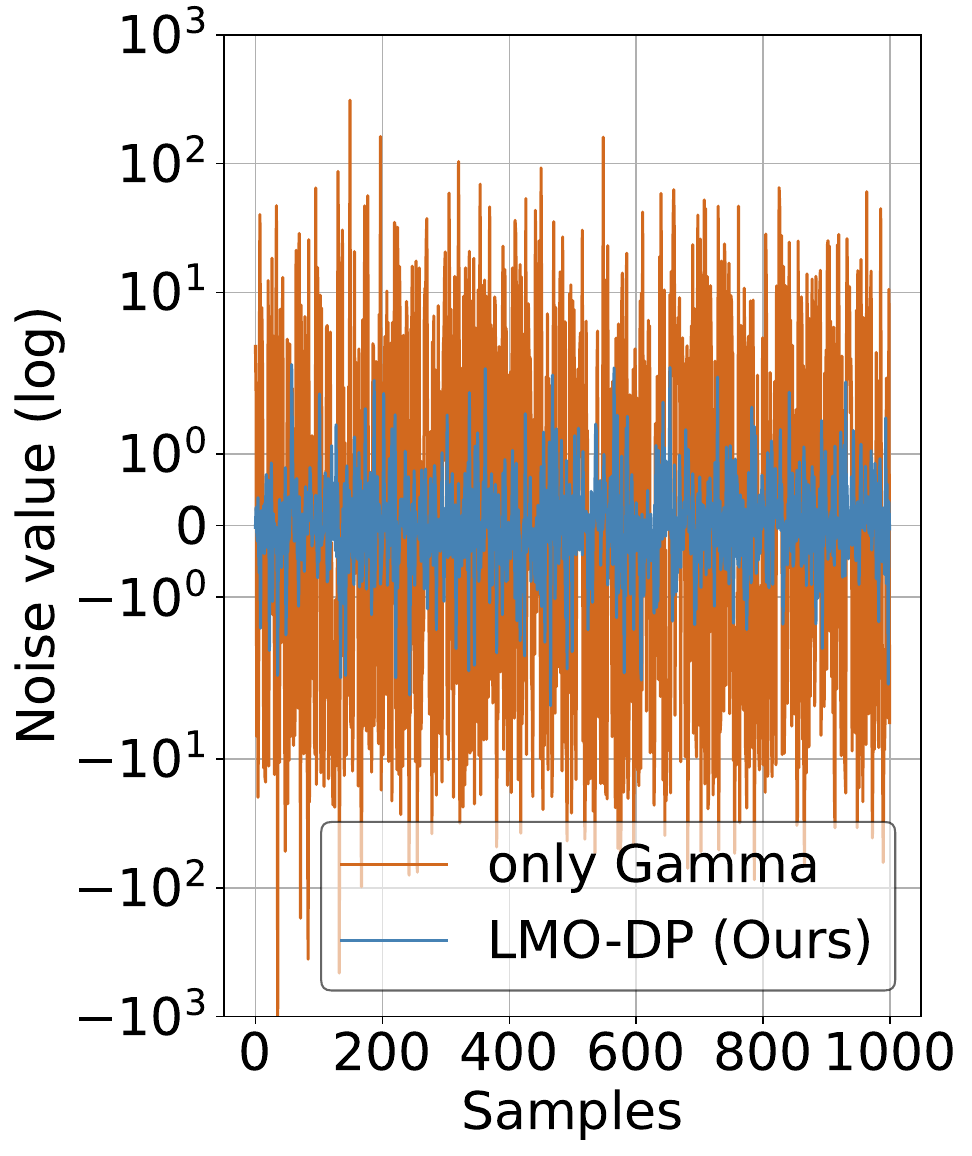}
\text{(c)  $\epsilon=2$}
\end{minipage}
\begin{minipage}[t]{0.245\textwidth}
\centering
\includegraphics[width=\textwidth]{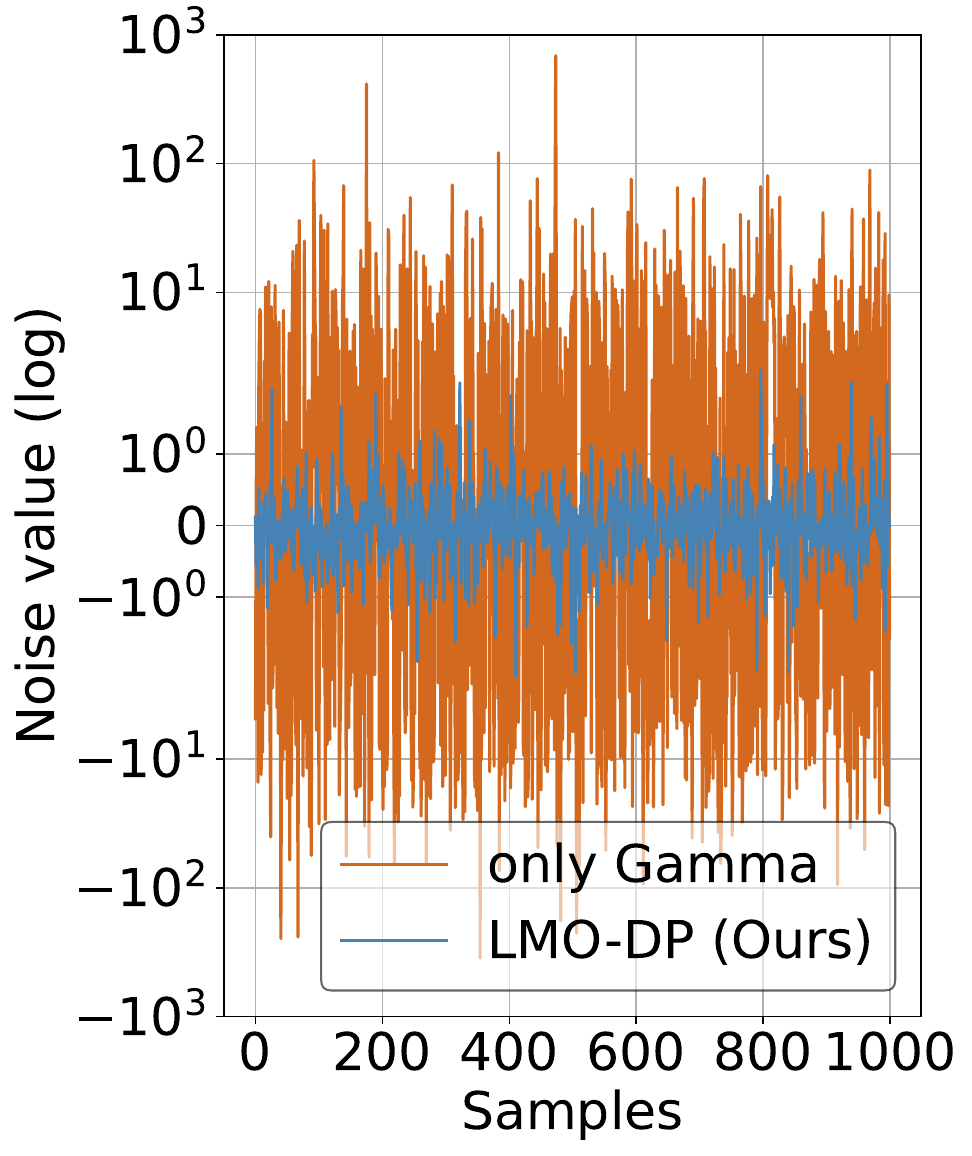}
\text{(d) $\epsilon=3$}
\end{minipage}
\vspace{-0.1in}
\caption{Gamma distribution vs mixture distribution (with the same remaining setting). The noise generated by the mixture distribution (as the second-fold) in LMO-DP is significantly smaller than that replacing the mixture distribution with the Gamma distribution for all $\epsilon$.}  \vspace{-0.1in}
\label{fig_compare_noises22}
\end{figure*}

\begin{figure*}[htbp]
\begin{minipage}[t]{0.245\textwidth}
\centering
\includegraphics[width=\textwidth]{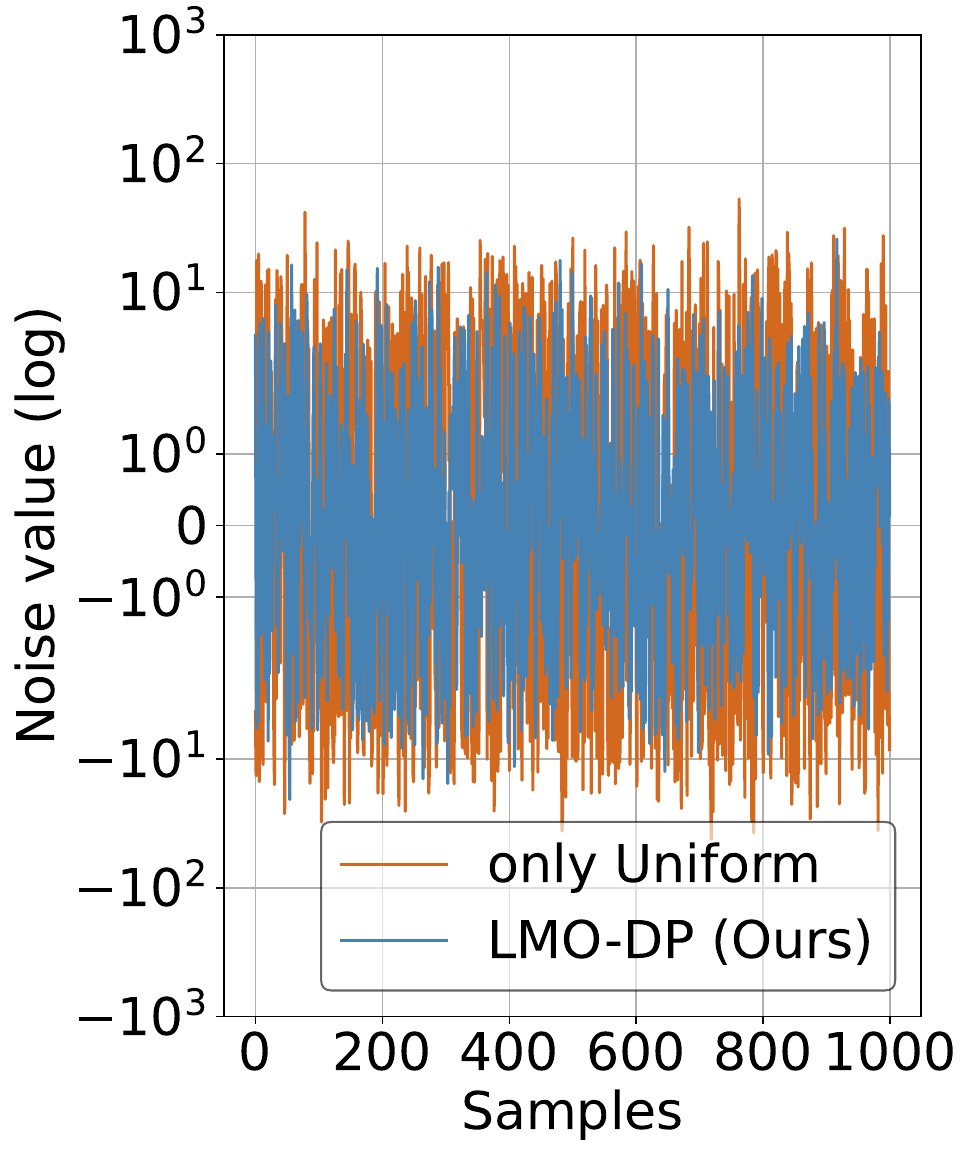}
\text{(a) $\epsilon=0.3$}
\end{minipage}
\begin{minipage}[t]{0.245\textwidth}
\centering
\includegraphics[width=\textwidth]{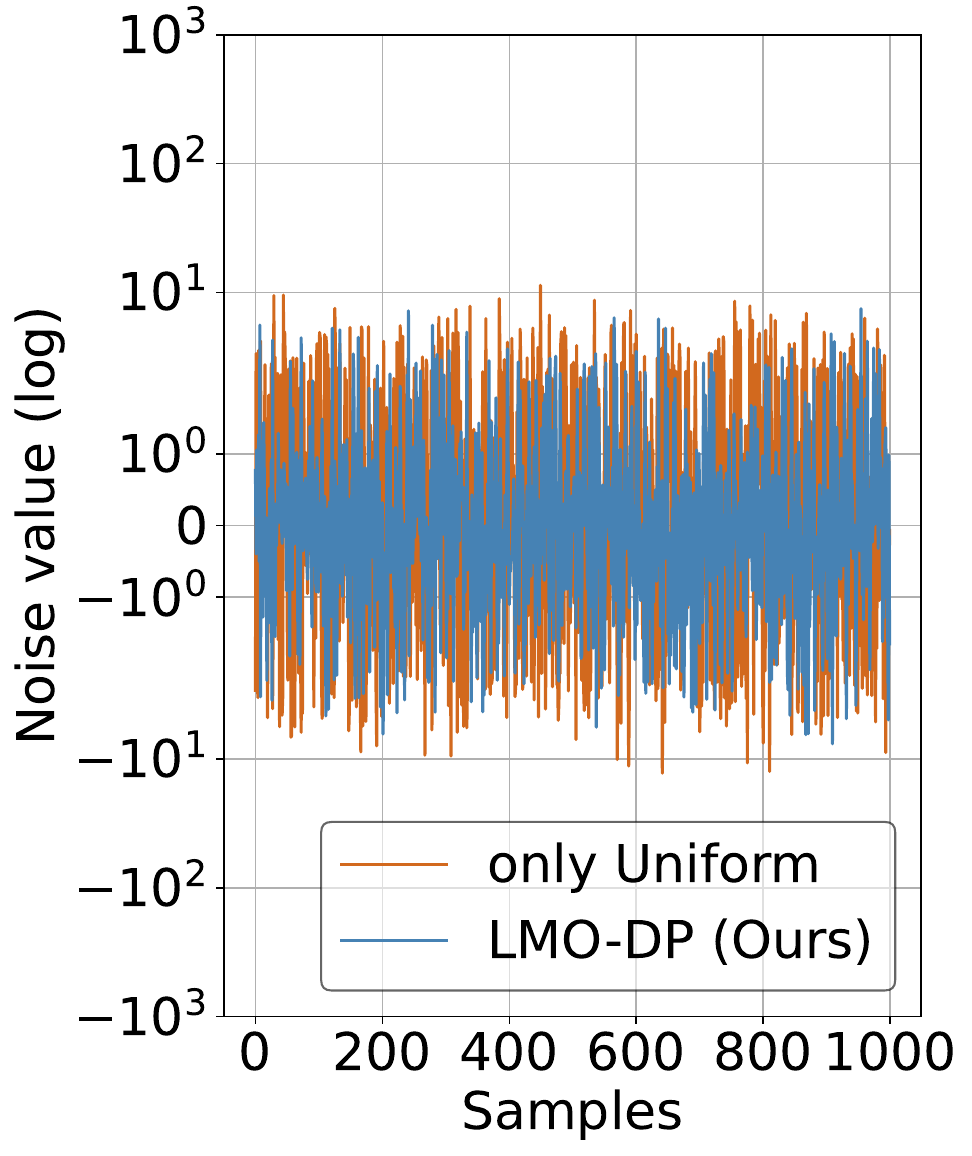}
\text{(b) $\epsilon=0.7$}
\end{minipage}
\label{sec:exp_noise4}
\begin{minipage}[t]{0.245\textwidth}
\centering
\includegraphics[width=\textwidth]{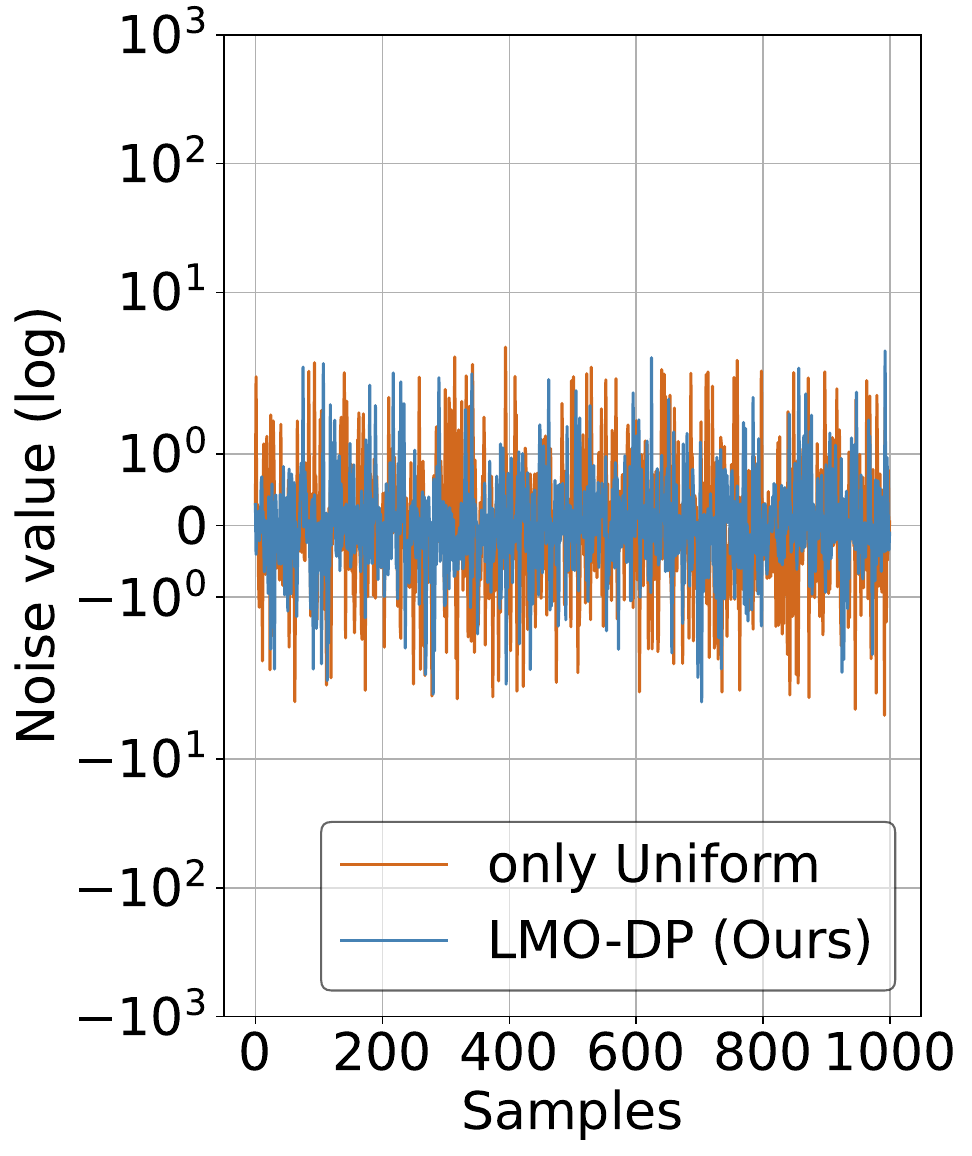}
\text{(c) $\epsilon=2$}
\end{minipage}
\begin{minipage}[t]{0.245\textwidth}
\centering
\includegraphics[width=\textwidth]{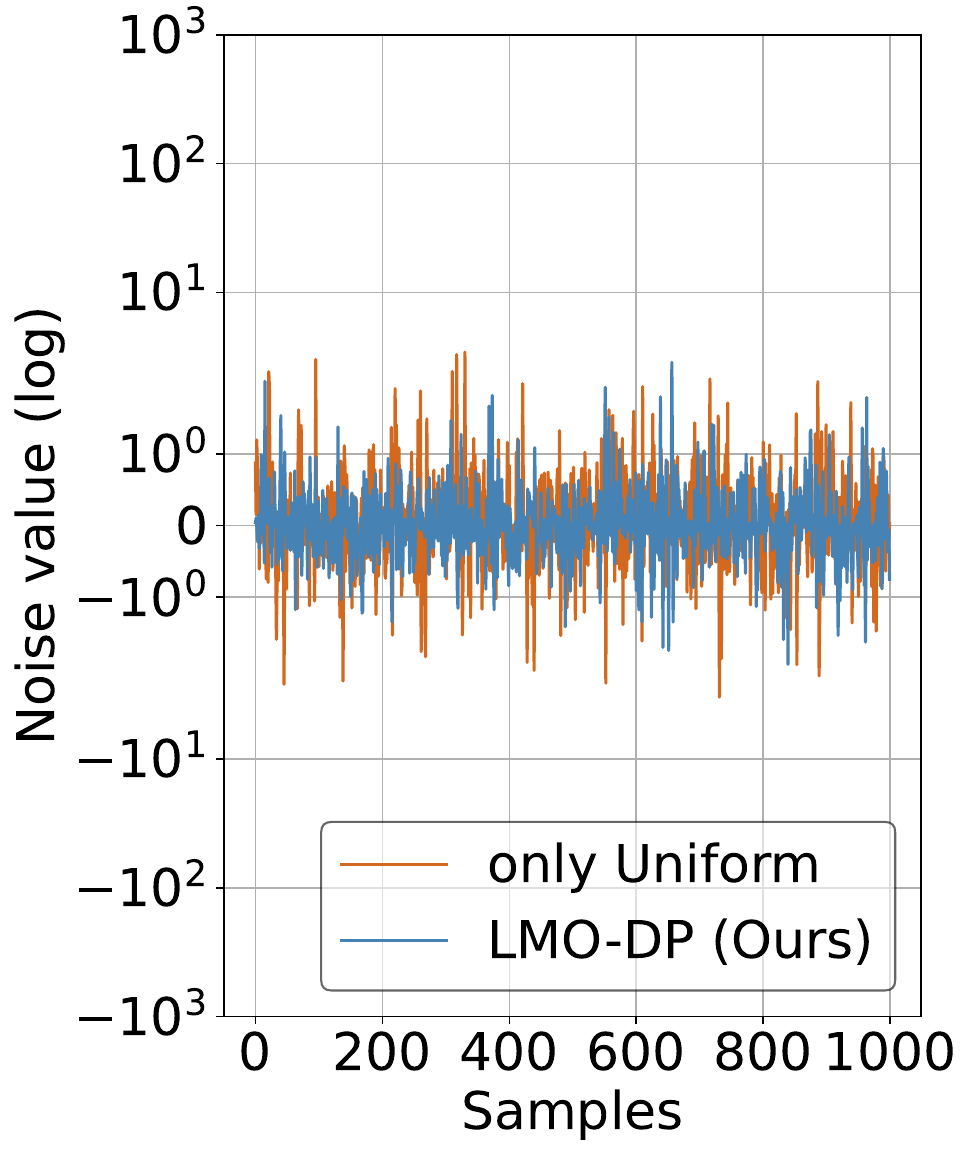}
\text{(d) $\epsilon=3$}
\end{minipage}
\vspace{-0.1in}
\caption{Uniform distribution vs mixture distribution (with the same remaining setting). The noise generated by the mixture distribution (as the second-fold) in LMO-DP is slightly smaller than that replaces the mixture distribution with the Uniform distribution. \textbf{The results demonstrate that Uniform distribution contributes more to the sub-optimal noise}.}  \vspace{-0.1in}
\label{fig_compare_noises3}
\end{figure*}

\newpage

\subsubsection{Mixture of Two Distributions vs. Mixture of Three Distribution in LMO-DP}

\begin{figure*}[htbp]
\begin{minipage}[t]{0.245\textwidth}
\centering
\includegraphics[width=\textwidth]{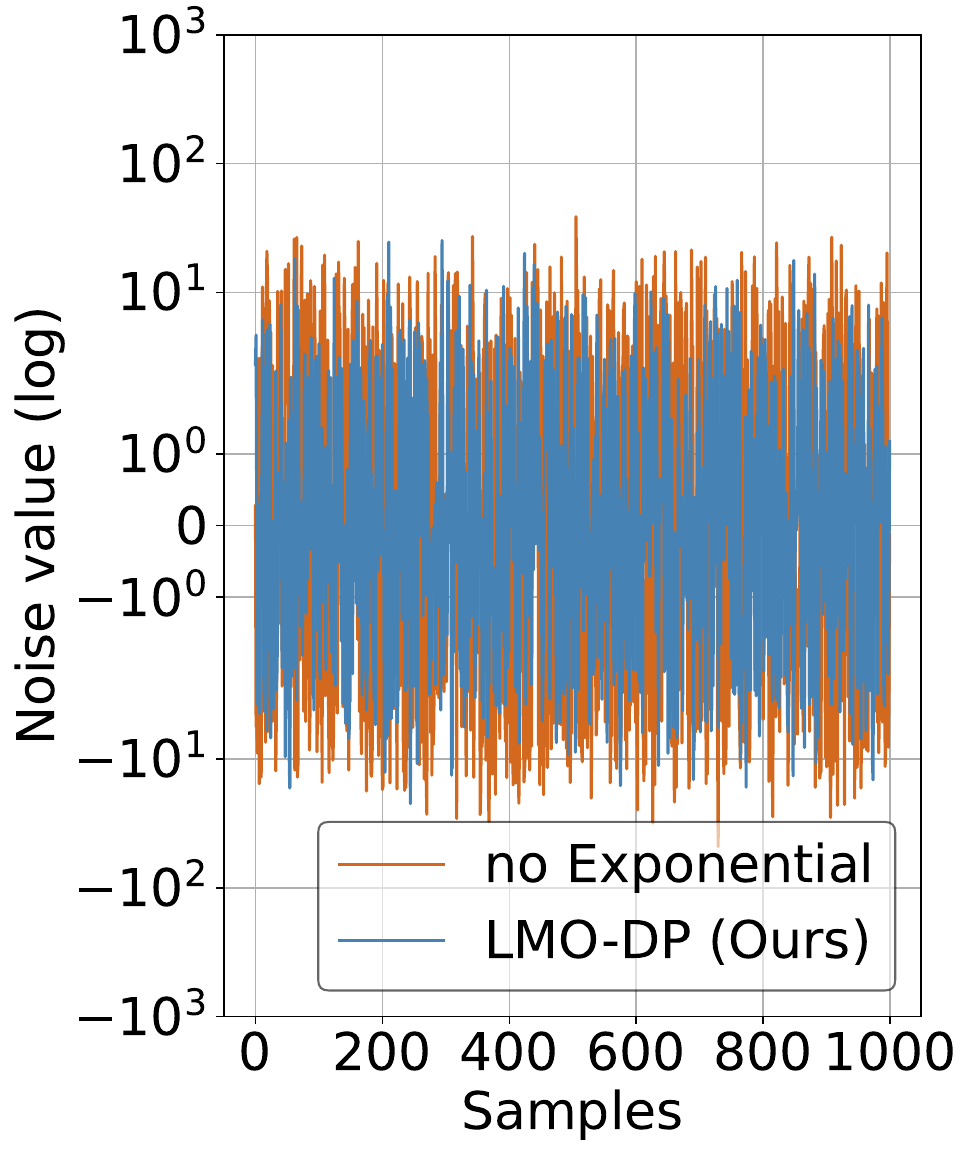}
\text{(a) $\epsilon=0.3$}
\end{minipage}
\begin{minipage}[t]{0.245\textwidth}
\centering
\includegraphics[width=\textwidth]{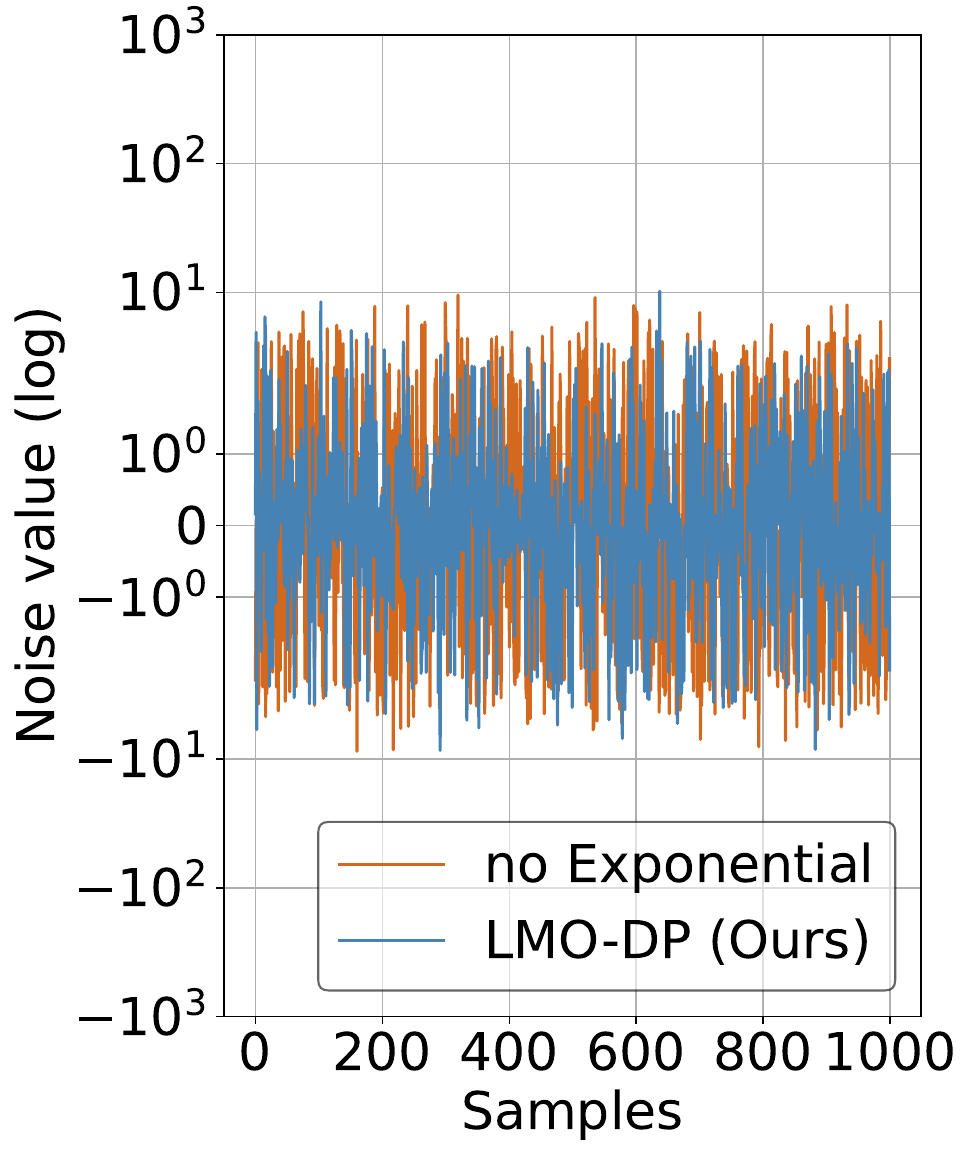}
\text{(b)  $\epsilon=0.7$}
\end{minipage}
\label{sec:exp_noise3}
\begin{minipage}[t]{0.245\textwidth}
\centering
\includegraphics[width=\textwidth]{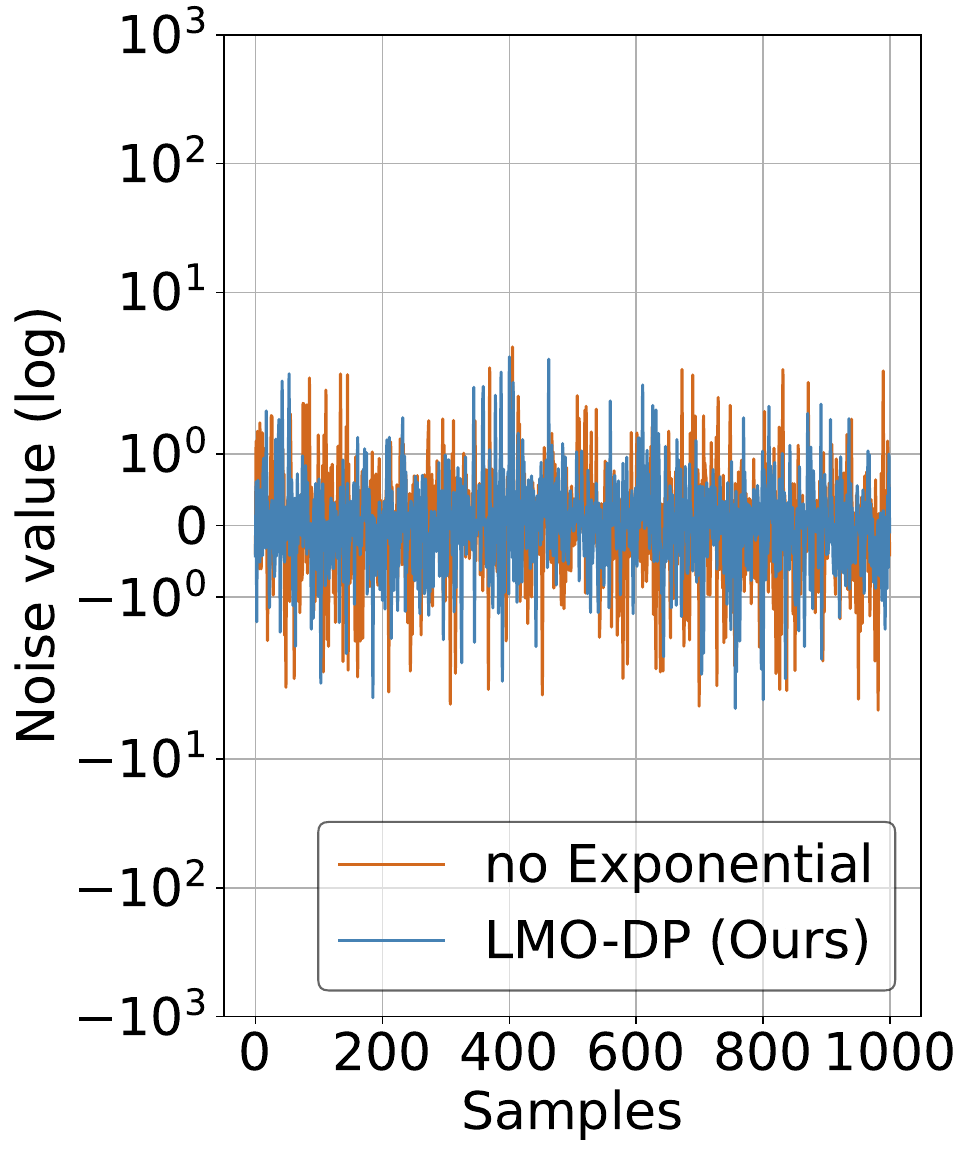}
\text{(c)  $\epsilon=2$}
\end{minipage}
\begin{minipage}[t]{0.245\textwidth}
\centering
\includegraphics[width=\textwidth]{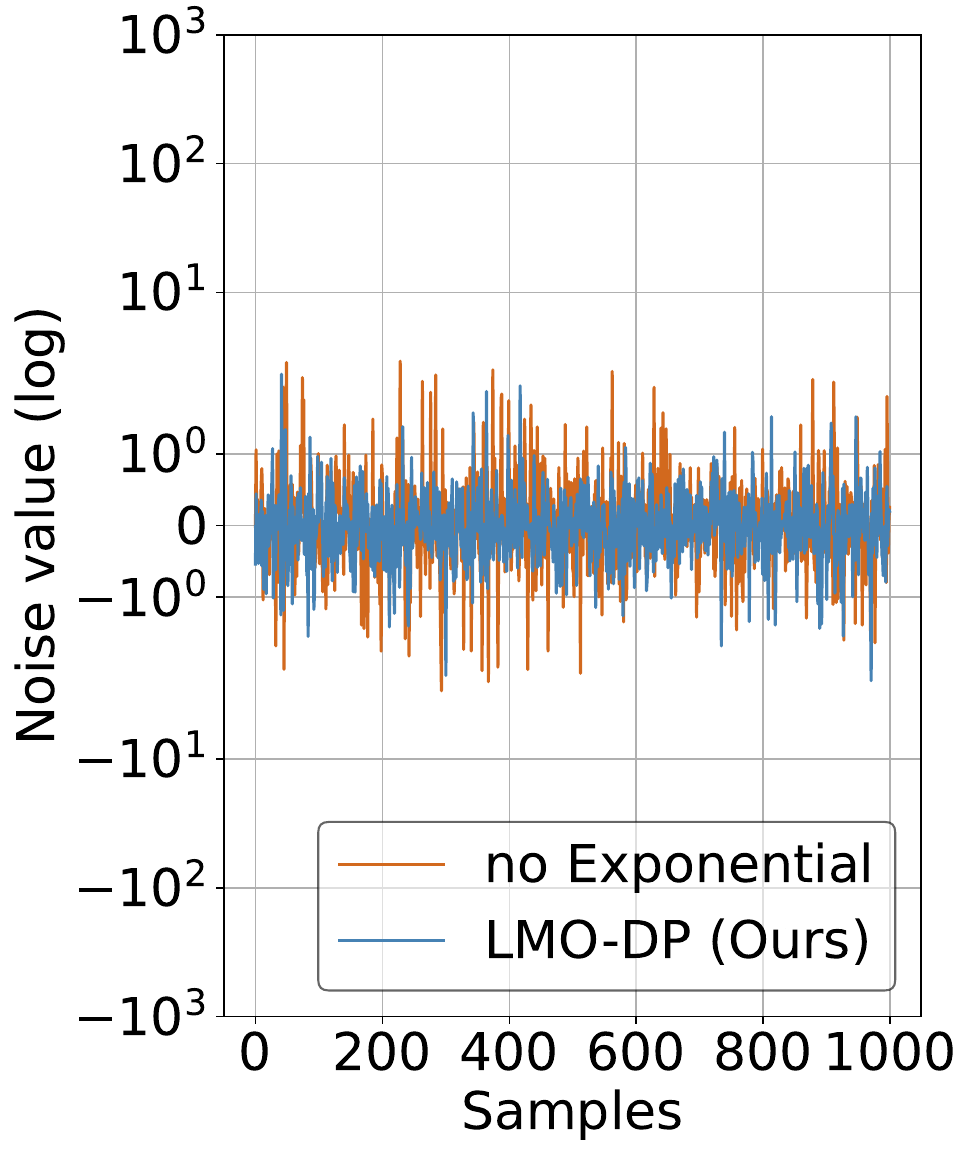}
\text{(d)  $\epsilon=3$}
\end{minipage}
\vspace{-0.1in}
\caption{Mixture of Gamma and Uniform distributions vs mixture of three distribution (with the same remaining setting). The noise generated by the mixture of three distributions (as the second-fold) in LMO-DP is slightly smaller than that removes the Exponential distribution.}  \vspace{-0.1in}
\label{fig_compare_noises4}
\end{figure*}

\begin{figure*}[htbp]
\begin{minipage}[t]{0.245\textwidth}
\centering
\includegraphics[width=\textwidth]{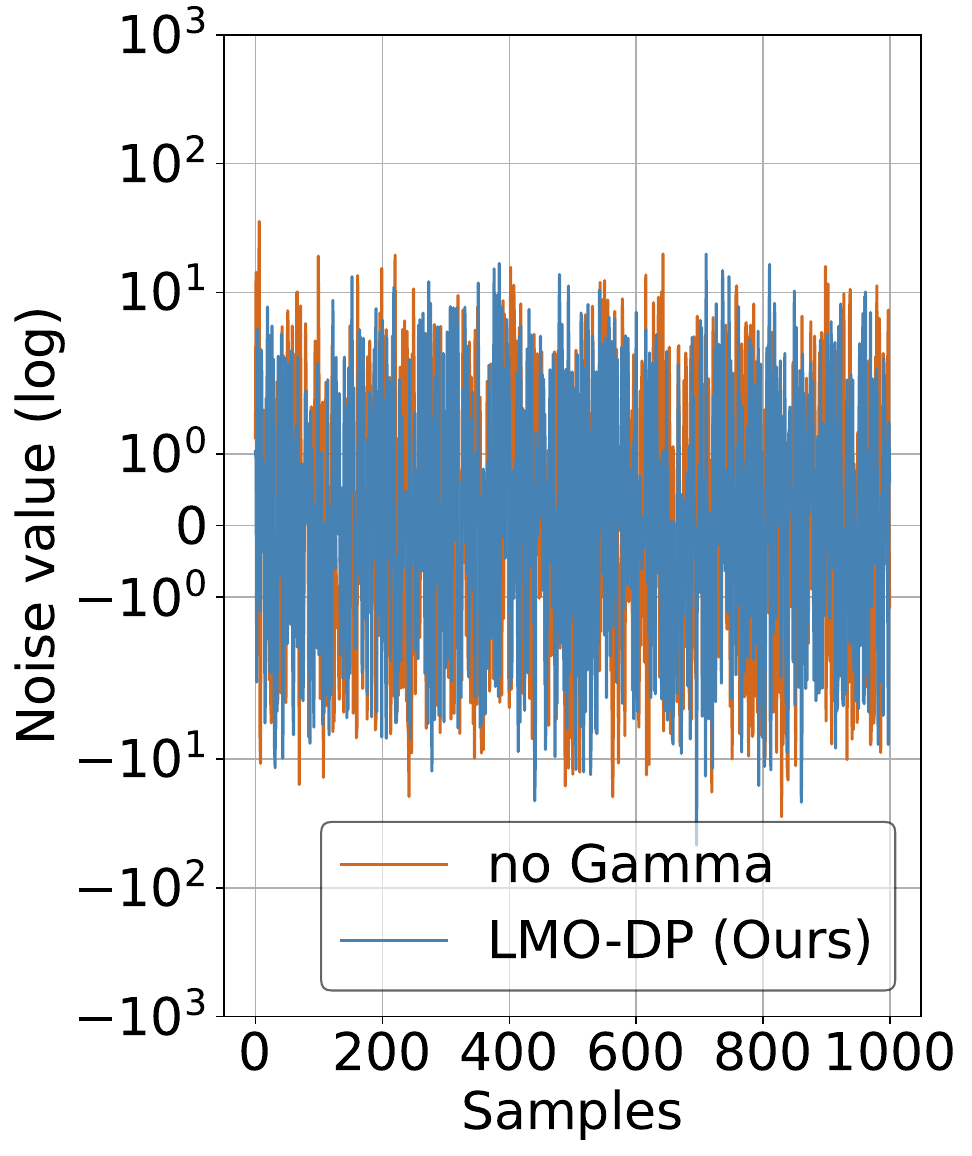}
\text{(a)  $\epsilon=0.3$}
\end{minipage}
\begin{minipage}[t]{0.245\textwidth}
\centering
\includegraphics[width=\textwidth]{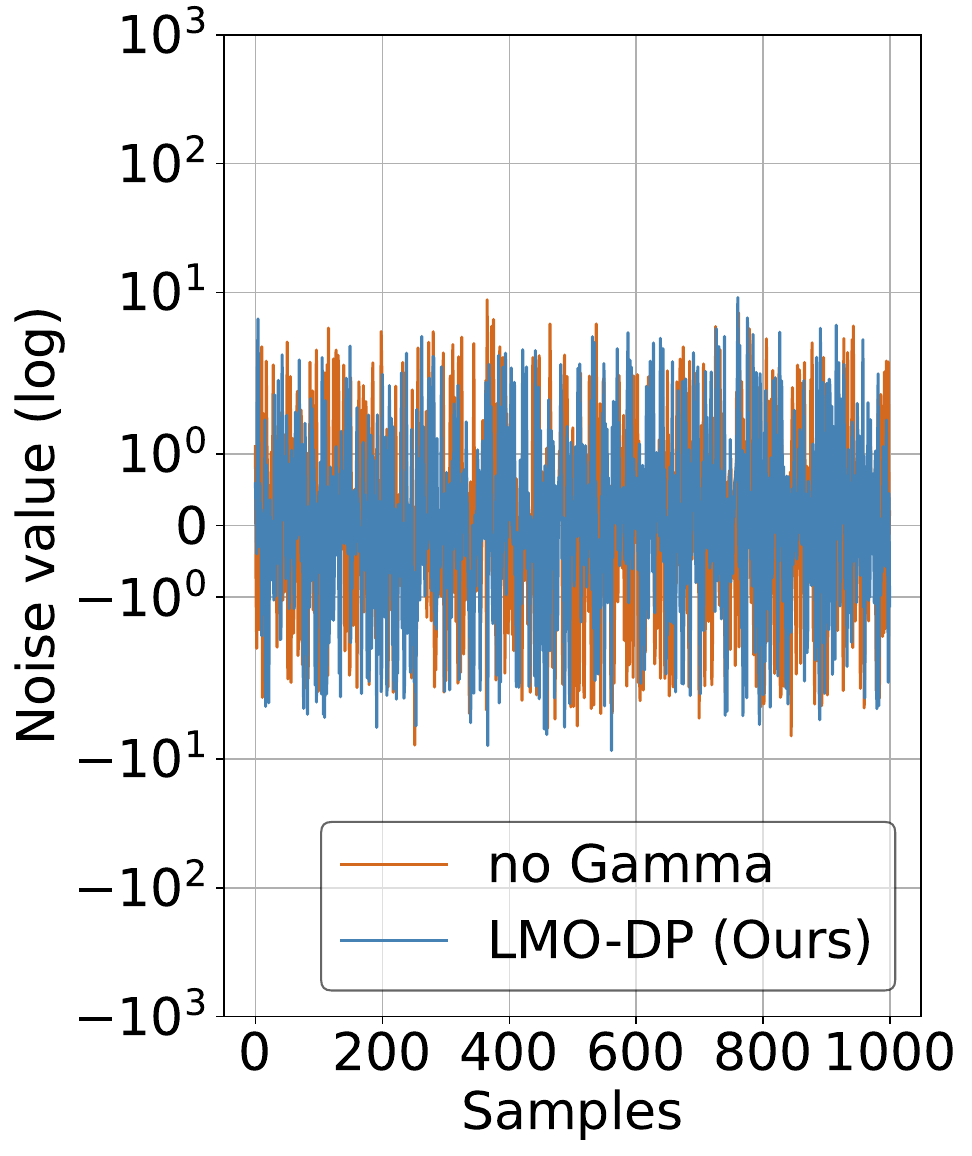}
\text{(b) $\epsilon=0.7$}
\end{minipage}
\label{sec:exp_noise2}
\begin{minipage}[t]{0.245\textwidth}
\centering
\includegraphics[width=\textwidth]{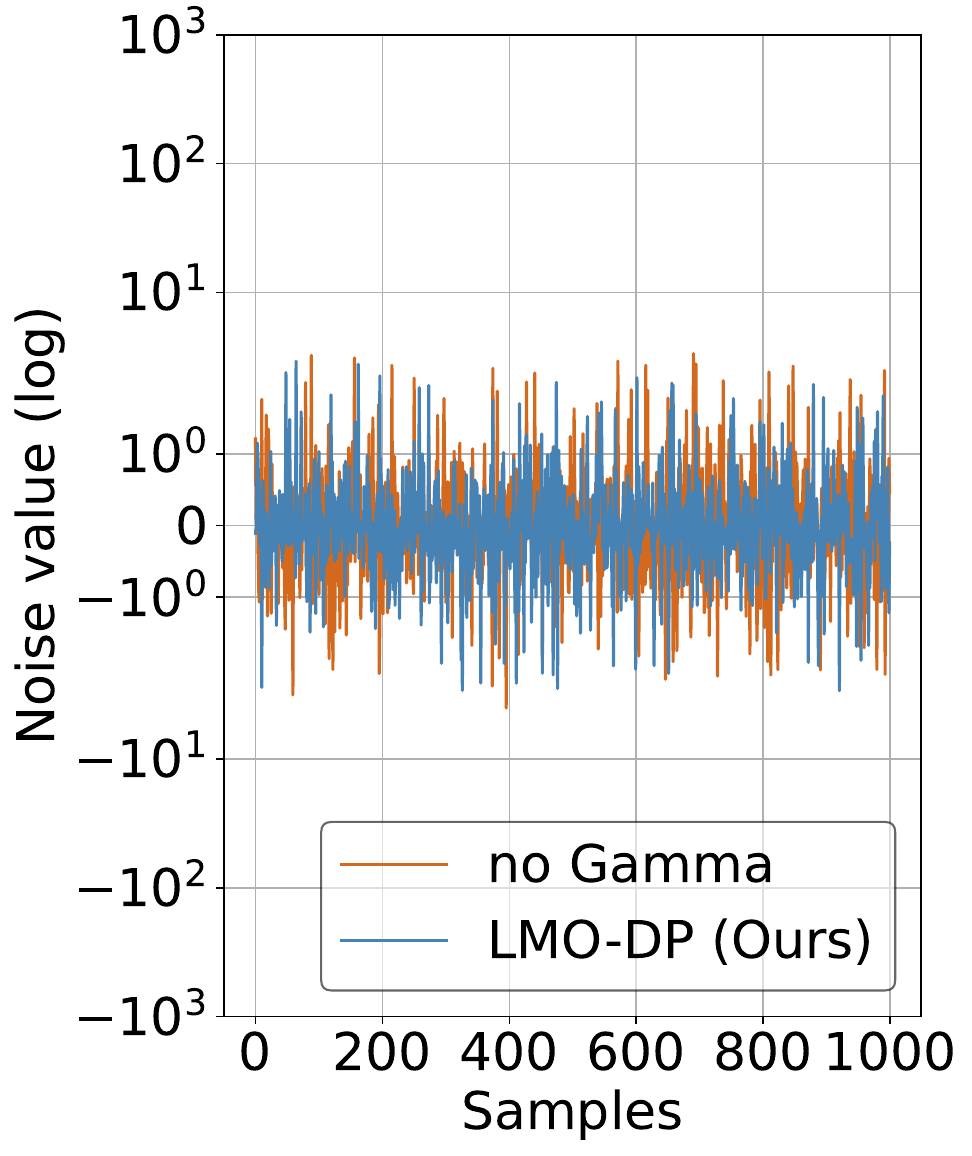}
\text{(c) $\epsilon=2$}
\end{minipage}
\begin{minipage}[t]{0.245\textwidth}
\centering
\includegraphics[width=\textwidth]{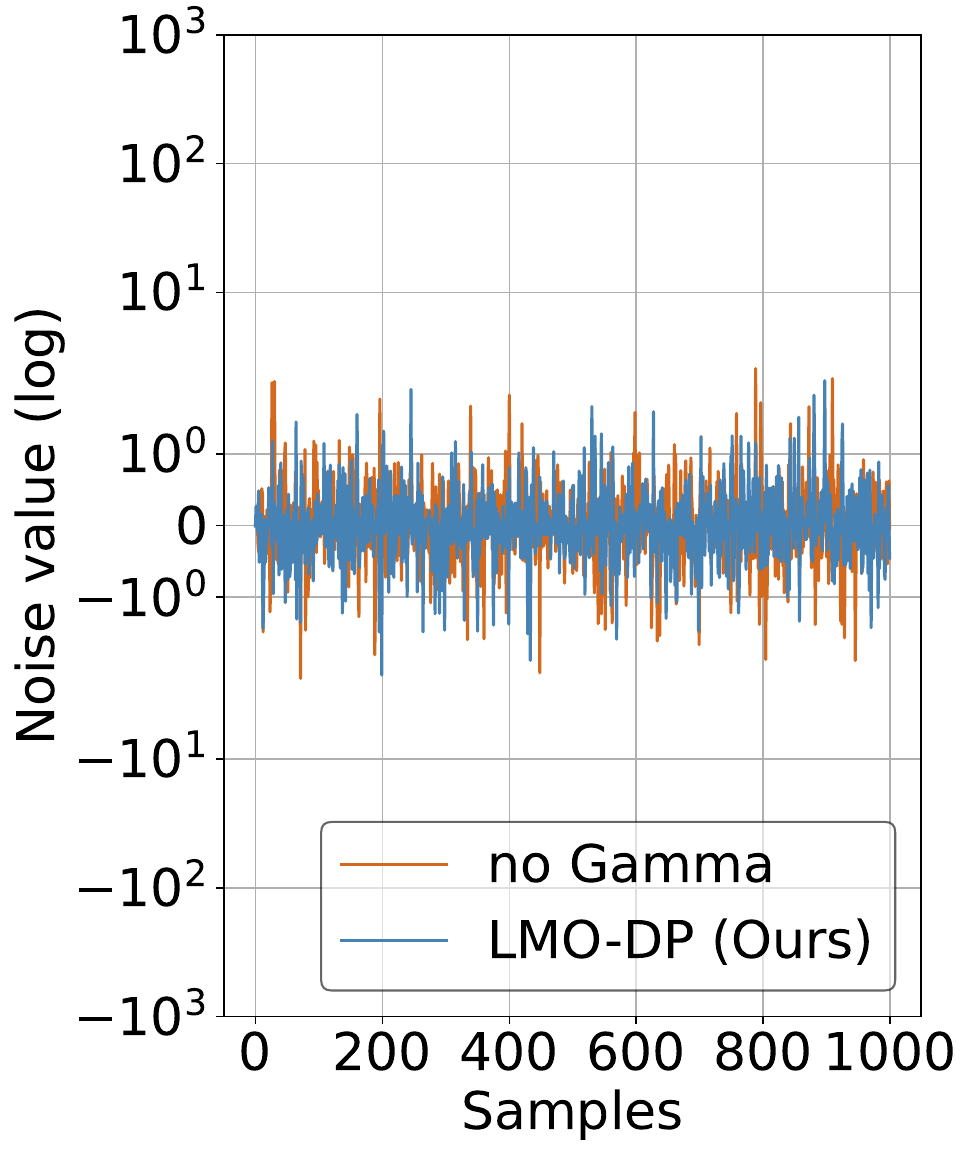}
\text{(d) $\epsilon=3$}
\end{minipage}
\vspace{-0.1in}
\caption{Mixture of Exponential and Uniform distributions vs mixture of three distribution (with the same remaining setting). The noise generated by the mixture of three distributions (as the second-fold) in LMO-DP is slightly smaller than that removes the Gamma distribution.}  \vspace{-0.1in}
\label{fig_compare_noises5}
\end{figure*}
\thispagestyle{empty} 
\begin{figure*}[htbp]
\begin{minipage}[t]{0.245\textwidth}
\centering
\includegraphics[width=\textwidth]{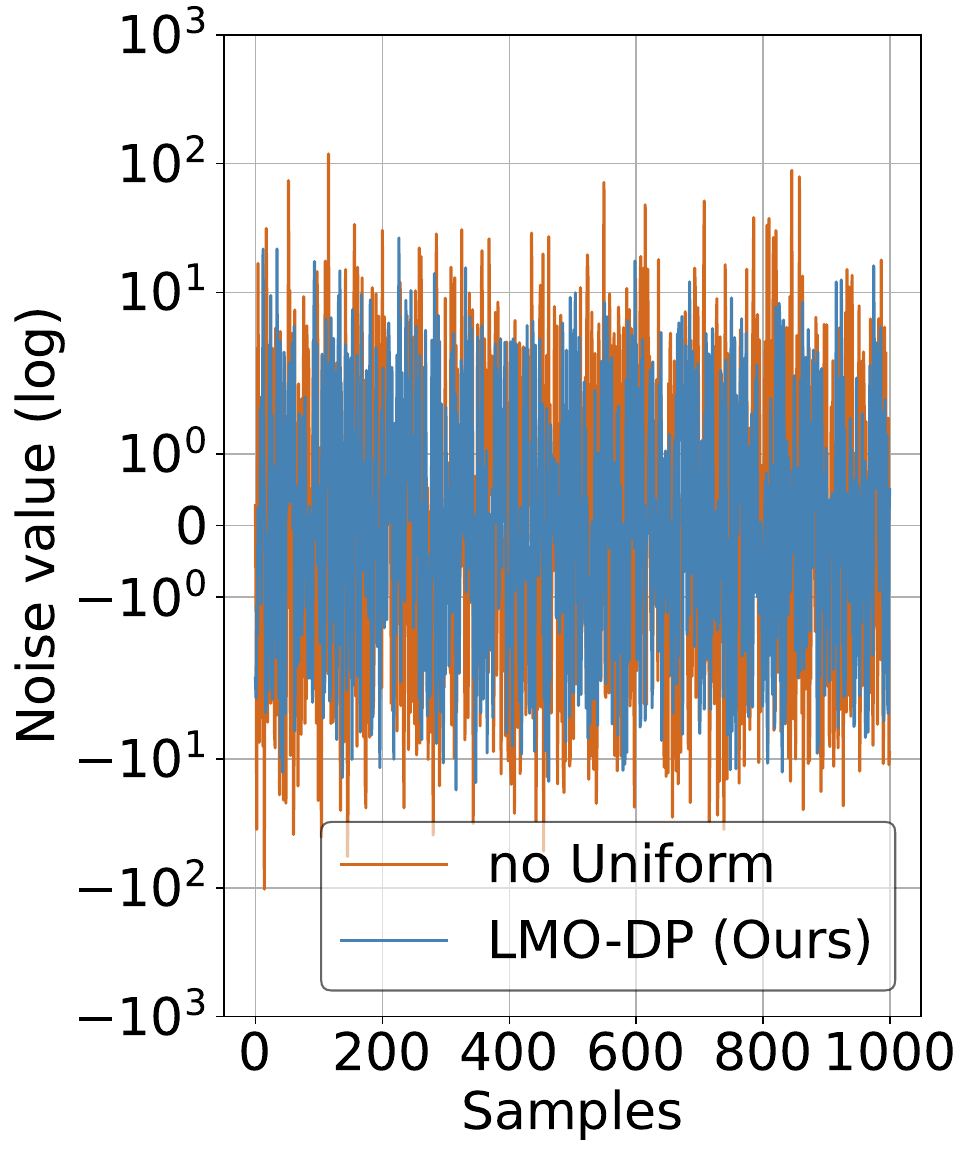}
\text{(a) $\epsilon=0.3$}
\end{minipage}
\begin{minipage}[t]{0.245\textwidth}
\centering
\includegraphics[width=\textwidth]{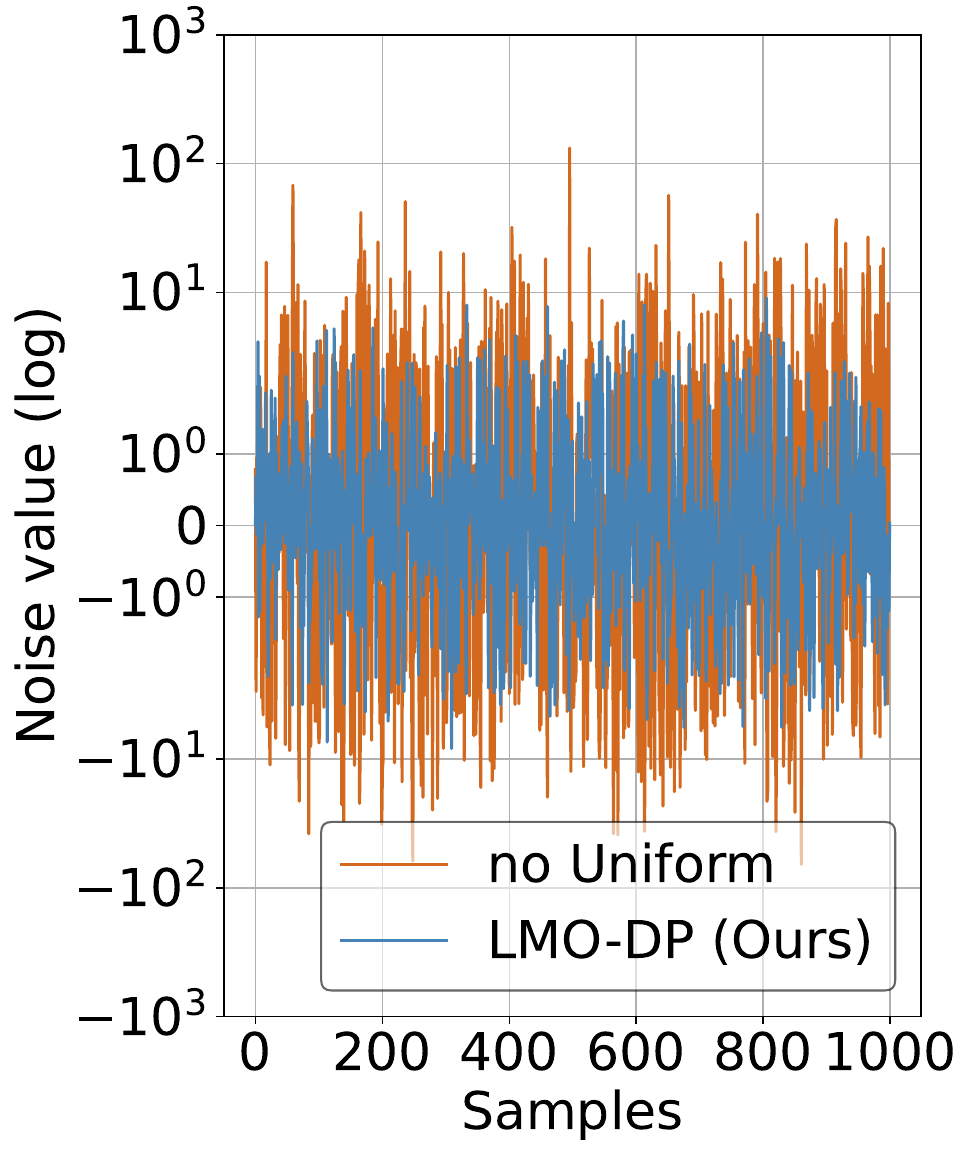}
\text{(b) $\epsilon=0.7$}
\end{minipage}
\label{sec:exp_noise1}
\begin{minipage}[t]{0.245\textwidth}
\centering
\includegraphics[width=\textwidth]{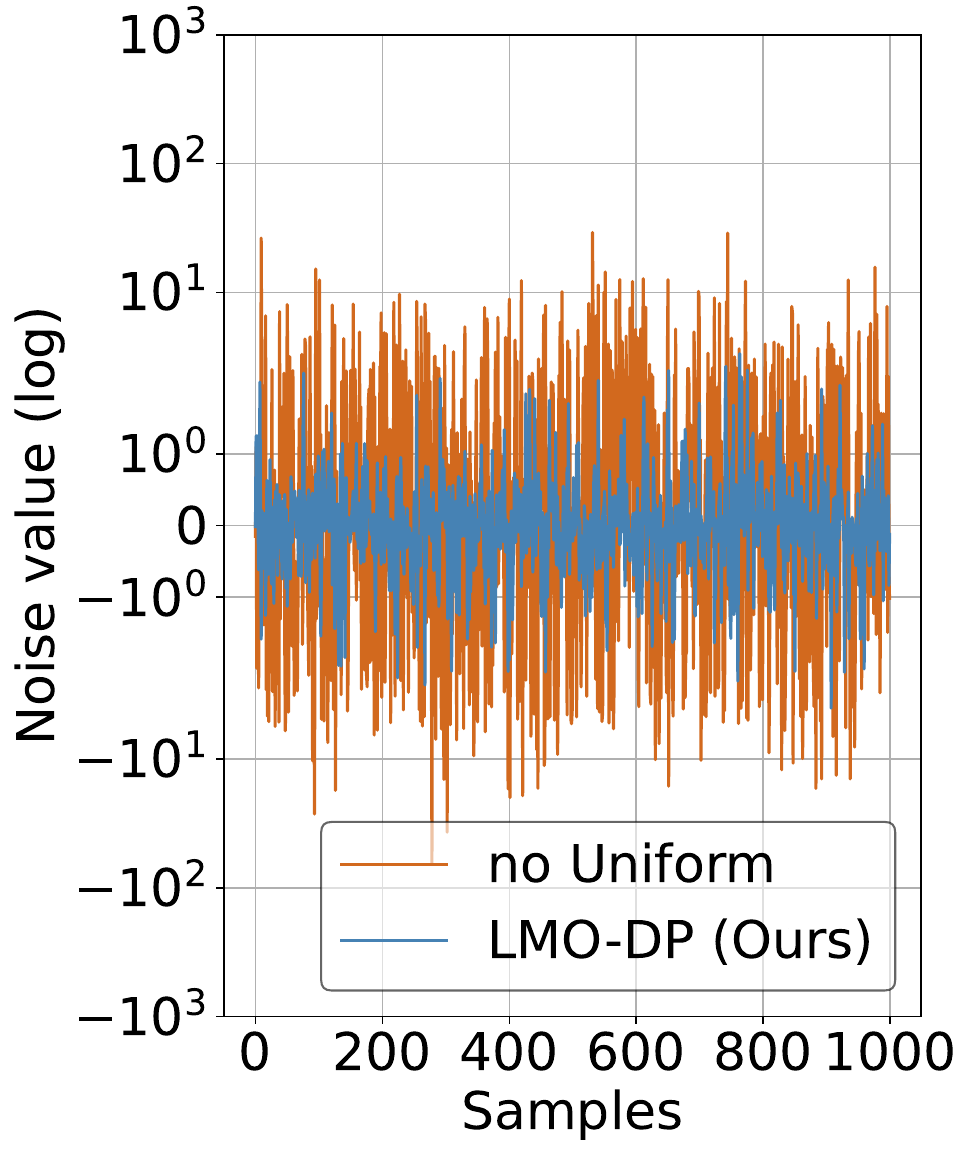}
\text{(c) $\epsilon=2$}
\end{minipage}
\begin{minipage}[t]{0.245\textwidth}
\centering
\includegraphics[width=\textwidth]{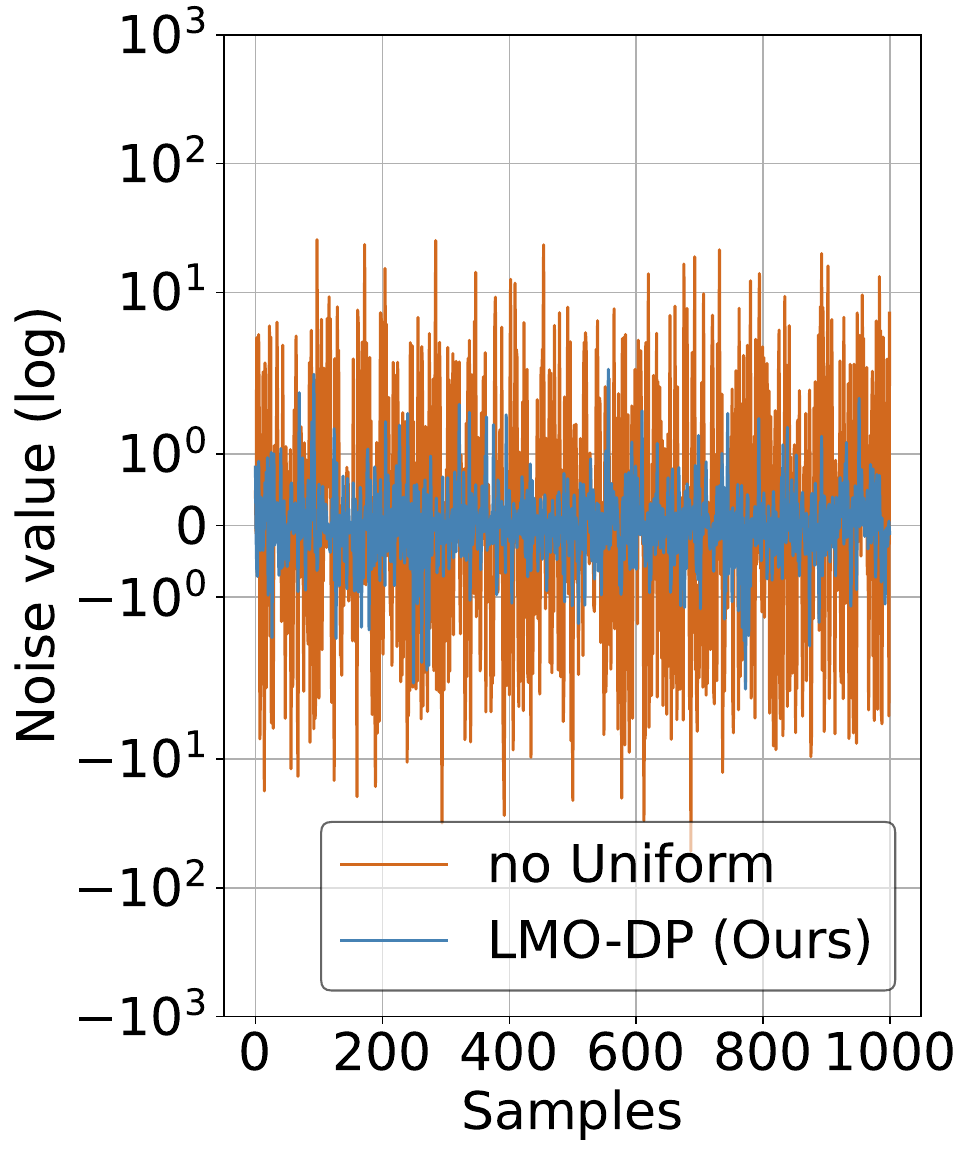}
\text{(d) $\epsilon=3$}
\end{minipage}
\vspace{-0.1in}
\caption{Mixture of Gamma and Exponential distributions vs mixture of three distribution (with the same remaining setting). The noise generated by the mixture of three distributions (as the second-fold) in LMO-DP is smaller than that removes the Uniform distribution, especially for large $\epsilon$. \textbf{The results again demonstrate that Uniform distribution contributes more to the sub-optimal noise}.} \vspace{-0.1in}
\label{fig_compare_noises6}
\end{figure*}

\end{document}